%% file: diser.tex
\documentclass[12pt]{report}
\usepackage{epsfig}
\usepackage{cite}
\usepackage{ODUthesis}
\begin{document}

\title{SOME ASPECTS OF\\ DEEPLY VIRTUAL COMPTON SCATTERING}

\author{Elena Kuchina}
\principaladviser{Ian Balitsky}
\member{Anatoly Radyushkin (Member)}
\member{John A. Adam (Member)}
\member{Charles I. Sukenik (Member)}
\member{Rocco Schiavilla (Member)}

\degrees{
         M.S. June 1987, Kazakh State University\\ 
         Ph.D. June 1993, Kazakh State University\\
         M.S. May 2001, Old Dominion University}
\dept{Physics}

\submitdate{May 2004}
\figurespagefalse
\tablespagefalse
\copyrightfalse
\vita{
M.S. in Physics, Kazakh State University, June 1987\\ 
Ph.D. in Physical and Mathematical Science, Kazakh State University, June 1993\\
M.S. in Physics, Old Dominion University, May 2001
}

\abstract{
We consider different aspects of the virtual Compton amplitude in QCD 
on two examples: small-x physics accessible in the Regge regime and twist-3 
approximation in the description of DVCS through the general parton 
distributions.
Using this model, we give an estimate for the cross section of deeply 
virtual Compton scattering for the kinematics of CEBAF at Jefferson Lab.}

\beforepreface

\prefacesection{Acknowledgments}
       I am grateful to many people for their help and support during all the years I have been a graduate student.

I would particularly like to thank my  thesis advisor Ian Balitsky 
for assigning me such a challenging and rewarding thesis project.
I express my deep gratitude to Professor Anatoly Radyushkin for his continuing support during my graduate student career and for giving me many opportunities to learn good physics.

It has been my privilege to collaborate with an amazing group of people, the Theory group at Jefferson Lab, who are equal parts teachers, scientists, and tireless workers. Igor Musatov, my colleague and friend, provided invaluable help in nearly every project I worked on.

I would like to thank my family and friends for supporting me through what has been a difficult road. 

I thank my husband Sergey for his willingness to help and my mom for keeping my children safe and sound for me, for their patience, permanent spiritual support and encouragement.

I would also like to extend my gratitude to the members of my committee, for taking time to read this manuscript and for their helpful comments and suggestions. 

\afterpreface

\chapter{Introduction}
\input{intro/intro.tex}

\chapter{DVCS at small-$x$}
\input{dvcs/dvcs.tex}

\chapter{Generalized parton distributions}
\input{GPD/GPD.tex}

\chapter{Conclusion\label{Summary}}
\input{conclu/conclu.tex}

\appendix
\input{GPD/GPD-appendix.tex}

\newpage
\vitapage
\end{document}

%% file: intro/intro.tex
One of the inexhaustible studies in physics is the study of hadron structure
\nnfootnote{The style specifications used in this thesis follow
those of {\it Physical Review D}}.
It is believed that hadrons and nuclei are built from quarks and gluons, 
but it is still under investigation how those building blocks reveal themselves
in nuclear reactions and how we could get information about those 
blocks from modern experiments.

Theory offers Quantum Chromodynamics (QCD) as the gauge theory of strong 
interactions describing
the dynamics between colored quarks and gluons.
According to QCD, the interaction between the quarks becomes weak at very short
distances. This phenomenon, known as asymptotic freedom, allows
us to use perturbation theory to describe high energy strong interaction
processes. At low energy, quarks and gluons are confined into colorless mesons and
baryons which are the real world particles. The QCD coupling constant
grows at these energies plus the
 chiral symmetry of QCD is spontaneously broken by almost massless quarks. 
It brings us to the necessity of using non-perturbative methods at this regime.

In search of ways describing the complexity of hadrons and hadronic systems 
one should perform precise low-energy experiments to gain information about 
the dynamics inside the hadron. However, one should also look at the specific high energy
kinematical regimes where the factorization theorems have been proven. Those experiments
could reveal some new non-perturbative hadron structure information with the help of an
accurate perturbative QCD description of the reaction dynamics \cite{Aip01,Sau00,Adl01,Ste01,SteBurEloGar01,Chen00,Nowak01}.

The experiments performed with electromagnetic probes (Compton scattering) are the best
ones. 
In Compton scattering, a real or virtual photon, emitted by a lepton,  interacts
with the nucleon with initial momentum $p$ and as a result a real photon is emitted.
Due to the new generation 
of electron accelerators and  high precision, large acceptance detectors, 
very high precision Compton scattering experiments have become a reality. 
Despite the small cross sections of the process, it serves as a clean probe 
of hadron structure.

It is convenient to examine this process in a parton model \cite{feynman}. 
In this model, the nucleon consists of {\it (a)} three valence quarks, {\it (b)}
quark-antiquark sea, and {\it (c)} gluons, the carriers of quark interactions. 
The sea and gluons is created and dissipated through time and is
expected to have almost no influence on the process' outcome. 
The leading role in the nucleon belongs to the valence quarks. 
The typical time between interactions should be $1/\Lambda$ 
because $\Lambda\approx200$MeV is the only genuine scale in light quark QCD.
The  virtual photon, emitted by a lepton and carrying the momentum $q$ ($q^2=-Q^2$), 
has a lifetime $1/Q$, which varies inversely with the momentum transferred by the photon. 
The momentum $Q$ defines the virtuality of the photon.
The higher virtuality, the shorter is a photon's life.
At some point one may view the photon as being absorbed instantaneously 
by one of the quarks in the hadron. 

Suppose the quark that absorbs the photon has longitudinal momentum $r$. 
Upon absorbing the virtual photon, the struck quark becomes 
highly virtual with a lifetime $r/Q^2$. Since this time is much 
shorter than the normal interaction time between quarks in the proton, 
the struck quark must reemit the photon before any interactions 
with the other quarks and gluons in the proton take place.

Finally, since the transverse momentum of the absorbed photon 
is $\mid q \mid =Q$ the photon must be absorbed and re-emitted
over a transverse coordinate region having $|\Delta x| \approx 1/Q$. 
The quark which absorbs the virtual photon is point-like (bare) down to a transverse 
size $|\Delta x|\approx 1/Q$.

Thus, the scattering by the virtual photon takes place essentially 
instantaneously and over a very small, almost point-like, spatial region. 
Since the photon interacts only with a single quark and the transverse momenta 
$r_{\perp }$ of partons inside, the moving proton is assumed to be independent of 
photon virtuality $Q^2$. It is expected that the amplitude 
of the process should be given in terms of the number density of quarks 
in the proton multiplied by the scattering amplitude of an individual quark.
Charge and energy
conservation brings us to the following sum rules for the parton distributions $n_i(x)$
inside the proton

\begin{equation}
1=\sum_ie_i\int_0^1dx\,(n_i(x)-n_{\overline{i}}(x))\,,\,\,1=\sum_i\int_0^1dx
\,x\,n_i(x),
\end{equation}
while the cross-section $\sigma _L$ for the longitudinally polarized 
virtual photon is
zero 
\begin{equation}
\sigma_L=0,
\end{equation}
and the cross-section $\sigma _T$ for the transversely polarized
photon is expressed in the impulse approximation as a sum of
photon-quark cross-sections averaged over the distributions of quarks 
$n_i(x)$ and anti-quarks $n_{\overline{i}}(x)$ in the proton:
\begin{equation}
\sigma _T=\sum_ie_i^2\frac{4\pi ^2\alpha }{Q^2}\,x\,(n_i(x)+n_{\overline{i}
}(x))\,,
\end{equation}
where $e_i$ is the $i$-quark charge measured in the units of the electron charge
$e$ and $\alpha =e^2/4\pi$. 

\begin{figure}[htb]
\mbox{
   \epsfxsize=13cm
 \epsfysize=5cm
 \hspace{1cm}  \epsffile{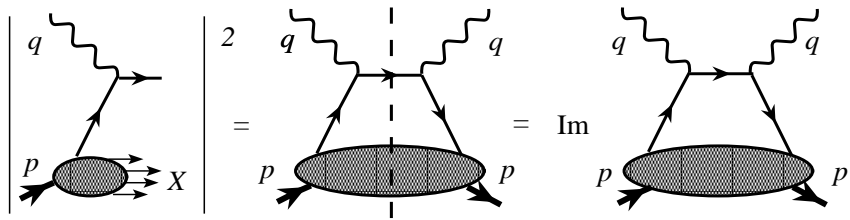}  }
{\caption{\label{dis-dia} The diagram for deep inelastic scattering 
as an imaginary part of the virtual Compton amplitude.
   }}
\end{figure}

Most of the internal structure of the nucleon has been revealed during
the last few decades through Deep Inelastic Scattering (DIS). 
DIS is a type of inclusive scattering of high energy leptons on the nucleon in the 
Bjorken regime,
which means that the photon virtuality $Q^{2}$ is very large,
 $x_B=Q^2/2p \cdot q$ is finite, and the total center-mass energy of the 
photon-nucleon system $s=(p+q)^2$ (one of the Mandelstam variables) is 
above the resonance region.

The range of $x_B$
\[ 0 \leq x_B \leq 1 \]
is given by the fact that the invariant mass of the unobserved final hadronic state is larger than the nucleon mass $M$
\[ (p+q)^2= q^2+2 p \cdot q +M^2 \geq M^2\]
and for the elastic scattering $x_B=1$.

In the process of deep inelastic scattering, the photon emitted by the lepton 
and absorbed by the proton creates an infinite number of possible final states. 
Summing up all those possible outcomes gives the intermediate state 
of elastic lepton-hadron scattering. 
This means that deep inelastic scattering could be described using the
optical theorem by the imaginary part of forward Compton scattering.

Elastic scattering has the same initial and final momentum for both 
of the participating particles. 
The amplitude of this process would be dominated by short distance
interactions in the Bjorken regime defined above.

The leading contribution comes from so-called handbag diagram. 
The factorization theorem states that in the large-$Q^2$ limit the 
perturbatively calculable hard quark
propagators completely factorize from  parton distribution
functions $n_i(x)$ ($i=u,d,s, \ldots$), 
which describe non-perturbative long-distance information about hadronic structure.

It was this particular process to which
the perturbative quantum chromodynamics (QCD) was traditionally 
and successfully applied  and where the approximate scaling behavior 
of the structure functions (independence of hadronic structure functions from 
the virtuality of the photon) was discovered.
Unpolarized DIS experiments have
mapped out the quark and gluon distributions in the nucleon, while
polarized DIS experiments have shown that quarks carry only a small fraction of
the nucleon spin.  As a result, new investigations to understand the nucleon spin, both experimental and theoretical, became necessary. 

Yet, even more complex and intriguing information could be extracted from
Deeply Virtual Compton Scattering (DVCS) \cite{ji,compton}. 
As well as in DIS, the virtual photon carries large negative momentum $q^2=-Q^2$, 
but now the final momentum of the particles differs from its initial values.
One should make sure that the momentum transfer $t$ is as 
small as possible \cite{ji,ji2}.
Again, the hard short-distance part factorizes from the non-perturbative 
long-distance part, general parton distributions (GPD), 
which now would be much more complex. 

\begin{figure}[htb]
\mbox{
   \epsfxsize=13cm
 \epsfysize=5cm
 \hspace{1cm}  \epsffile{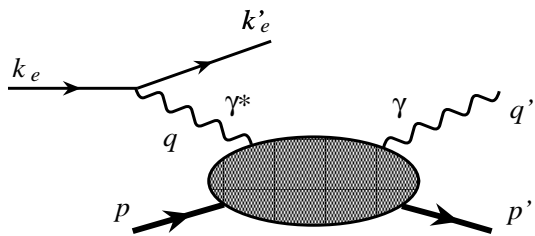}  }
{\caption{\label{dvcs-dia} The diagram for virtual Compton scattering.
   }}
\end{figure}

In addition to usual variable $x$, the fraction of the nucleon momentum carried by the 
photon absorbing quark, which runs from 0 to 1 for the quark and from -1 to 0 
for antiquark, depends on two other variables. 
One of them is the momentum transfer $t=(q-q')^2=(p'-p)^2$ which is  
an independent  Mandelstam variable, that should be small and negative.
The other one is the so-called skewedness $\xi$, defined as a fraction of the average nucleon 
momentum carried by the overall momentum transfer in the process.
These two extra degrees of freedom, $t$ and $\xi$, make the dynamics of deeply 
virtual Compton scattering rich and diverse \cite{npd,jios2,cofre}.

\begin{figure}[htb]
\mbox{
   \epsfxsize=13cm
 \epsfysize=5cm
 \hspace{1cm}  \epsffile{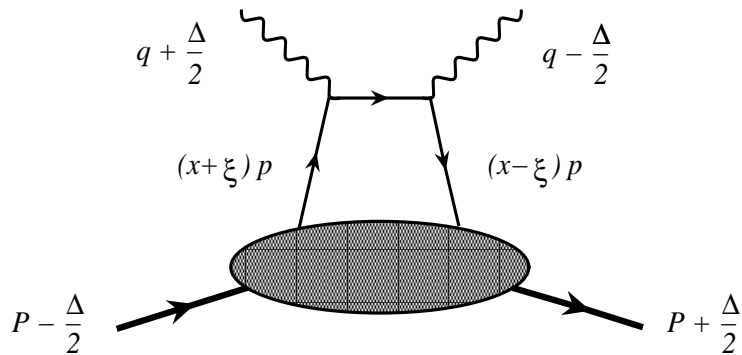}  }
{\caption{\label{dvcs-var} Parton picture for DVCS. 
Here $P=(p+p')/2$ is an average nucleon momenta, $\Delta =(p'-p)$ 
is the overall momentum transfer.
   }}
\end{figure}

Let us look at the handbag diagram of deeply virtual Compton scattering. 
From this figure it is easy to see that, depending on values of $x$ and $\xi$, 
the non-forward parton distribution functions can represent either the correlation 
between two quarks (when $x > \xi$), two antiquarks ($x < -\xi$), or between 
a quark and antiquark ($-\xi < x < \xi$). 
In the last case, GPDs behave like a meson distribution and 
uncover completely new information about nucleon structure which is 
inaccessible in DIS (which corresponds to the limit $\xi \rightarrow 0$).
That is why DVCS is so compelling and intriguing for physicists nowadays.

\begin{figure}[htb]
\mbox{
   \epsfxsize=13cm
 \epsfysize=5cm
 \hspace{0.5cm}  \epsffile{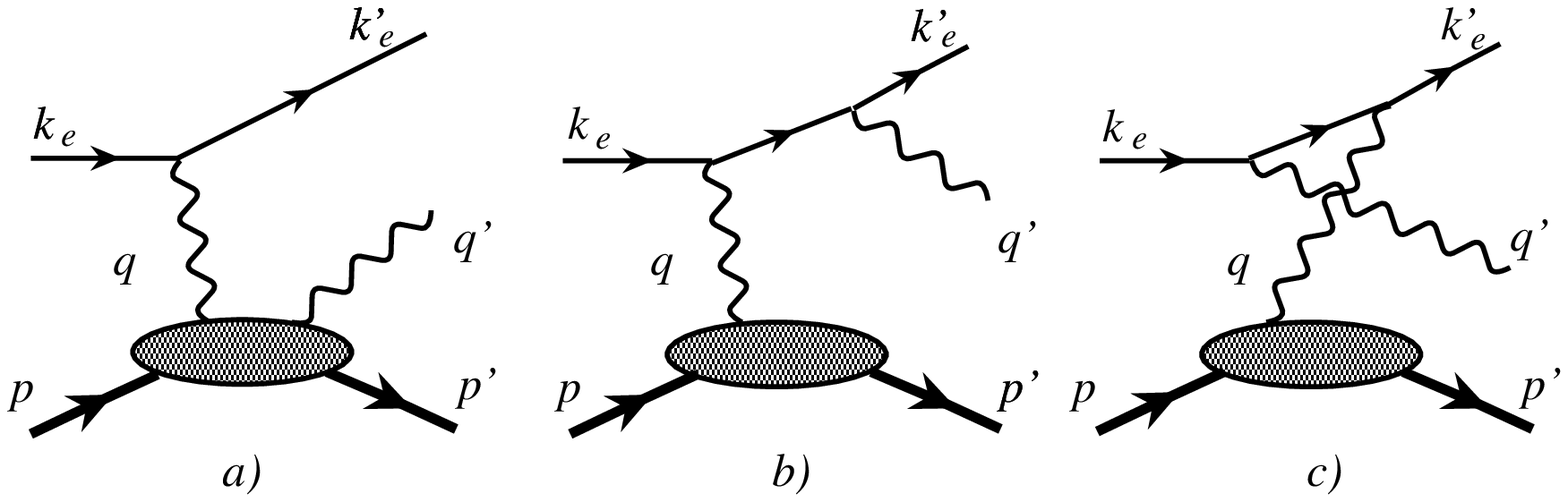}  }
{\caption{\label{dvcs-BH-dia} $a)$ the DVCS part of the amplitude;
$b),c)$ the Bethe-Heitler part.
   }}
\end{figure}

There is only one disadvantage of measuring the virtual Compton scattering 
amplitude, namely that the final 
photon can be emitted not only by the proton,
the process we are most interested in, 
but also by an electron, which is referred to as the
Bethe-Heitler (BH) process. 
This process is well known and it can be calculated exactly in quantum electrodynamics.
The only thing which must be known is the elastic form factors of the nucleon. 
Unfortunately, light
particles such as electrons radiate much more than the heavy proton.
Therefore the BH process generally dominates the DVCS amplitude, especially at small $t$. 

One way to minimize this problem is to find the kinematical regions where the BH
process is suppressed (or at least is comparable with the DVCS amplitude). 
One of those cases will be discussed in Chapter 2.

Another way is to exploit the interference between the DVCS 
and the calculable  Bethe-Heitler process, by independently measuring both the real and 
imaginary parts of the amplitude. This approach requires the 
theoretical models which will allow one to calculate the interference term 
with high precision and better understanding. In Chapter 3, the leading 
(twist-2) approximation
for the DVCS process will be discussed as well as the influence of the next 
to leading (twist-3) approximation on the final result.

There are other promising non-forward hard exclusive processes such as 
longitudinal electroproduction of vector and pseudoscalar mesons at large $Q^{2}$, 
but they are not discussed here.

%% file: dvcs/dvcs.tex
\section{Introduction}

In this chapter, the specific kinematical regime for the DVCS is considered 
when the energy of an incoming virtual photon $E_e$ is very large  $p\cdot q \rightarrow \infty$ 
in comparison to its virtuality $Q^2\rightarrow \infty$, 
while the Bjorken variable $x_B=Q^2/2 M E_e$ is finite and small.
It is the so called small-$x$ region. To be specific, we calculate the DVCS amplitude in the region
\begin{equation}
s\gg Q^2\gg -t\gg M^2,
\label{fla1}
\end{equation}
where $s=(p+q)^2 \simeq 2 M E_e$ is the Mandelstam variable corresponding to the center-of-mass energy squared 
of the photon-hadron system, $M$ is the nucleon mass, $t=(q-q')^2$ is another independent Mandelstam variable 
equal to the momentum transfer. The mass of the lepton is neglected.

The first study of the small-$x$ DVCS was undertaken in  Ref. \cite{Bar82}.
The DVCS in this region is a semihard process in which quark ladders are dominated by gluon ladders well-known as the
Balitsky-Fadin-Kuraev-Lipatov (BFKL) pomeron. 
It turns out that at large momentum transfer 
the coupling of the BFKL pomeron to the nucleon is essentially equal to the 
Dirac form factor of the nucleon $F_1(t)$. Thus, the DVCS amplitude in this
region (\ref{fla1}) can be calculated without any model assumptions.
Since there are only model predictions for the small-$x$ DVCS in the current literature
\cite{strikfurt1}, even the approximate calculations of the
cross section in QCD are very timely.
The results obtained in this region
can be used for the estimates of the
amplitude at experimentally accessible energies  where one or more conditions
in Eq.~(\ref{fla1}) are relaxed.

To proceed further, at high energies it is convenient to  use the Sudakov variables. Let us 
define
\begin{equation}
\begin{array}{rcl} \displaystyle
%q&=&p_1(1-\frac{{r_\perp}^2}{s})-x p_2-r_\perp \nonumber\\
q'&=&p_1,\nonumber\\
%p&=&p_2(1+x)+\frac{M^2+{r_\perp}^2}{} p_1+r_\perp \nonumber\\
p'&=&p_2+\frac{M^2}{\textstyle s} p_1, \nonumber\\
\end{array}\label{sudakov_variables}
\end{equation}
where $(p_1,p_2)$ is the light-cone basis of the final particles momenta plane
\begin{equation}
p_1^2=p_2^2=0, ~~~ 2p_1p_2=s \rightarrow \infty.
\end{equation}
One advantage of these coordinates is their simple
scaling properties when one takes the high energy limit.

All the other momenta are introduced in this basis as $k=\alpha_k p_1+
\beta_k p_2+k_\perp$ with $p_1 k_\perp=p_2 k_\perp=0$ by definition. 
For example:
\[ r=\alpha_r p_1+\beta_r p_2+r_\perp. \]

From $p^2=(p'-r)^2=M^2$ and $q^2=(q'+r)^2=-Q^2$ one can estimate $\alpha_r$ and $\beta_r$ to be:
\begin{equation}
\begin{array}{rcl} \displaystyle
\alpha_r &\simeq& r_\perp^2/s, \nonumber\\
\beta_r &\simeq& - (Q^2+r_\perp^2)/s.
\end{array}
\end{equation}

\section{Amplitude factorization}
The amplitude
of deeply virtual Compton scattering is determined by contracting the nucleon matrix element \cite{guvan} 
of the T-product of two electromagnetic currents with the photon polarization vectors
\begin{equation}
H=-i\epsilon_{\mu}(q)\epsilon'^{\ast}_{\mu} (q')
\int d^4z e^{-i q\cdot z}\langle p'| T\{j^{\mu}_{e.m.}(z)j^{\nu}_{e.m.}(0)\} |p\rangle.
\label{fla2}
\end{equation}

It is known (see for example  the review \cite{lobzor}) that in the leading order in perturbation theory
 the amplitude at high energy is purely imaginary up to the
 ${Q^2\over {\textstyle s}}$ corrections. 
At high orders in perturbation theory the amplitude will be
purely imaginary in the leading logarithmic approximation (LLA)
 and one will
restore the real part using the dispersion relations.

Hence, we will first calculate the imaginary part $\Im$ of the
amplitude $H$
\begin{equation}
V={1\over \pi} \Im H
\label{fla3}.
\end{equation}

\begin{figure}[htb]
\mbox{
   \epsfxsize=13cm
 \epsfysize=5cm
 \hspace{1cm}  \epsffile{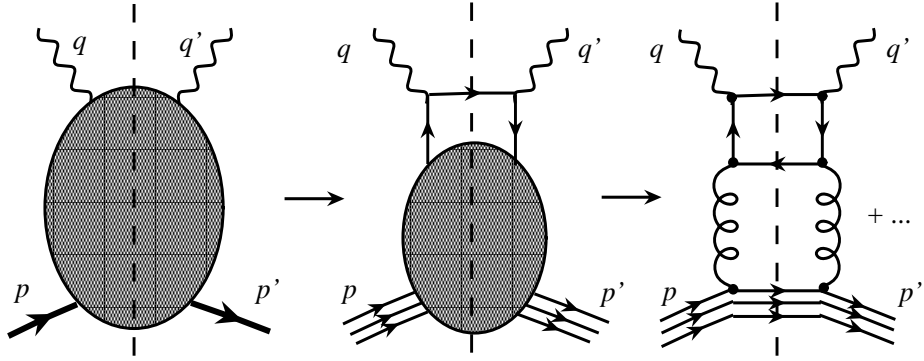}  }
{\caption{ The diagram for the virtual Compton scattering amplitude in the lowest order 
of the Leading Logarithmic Approximation (LLA).\label{ladder}
   }}
\end{figure}

\begin{figure}[htb]
\mbox{
   \epsfxsize=6cm
 \epsfysize=6cm
 \hspace{4cm}  \epsffile{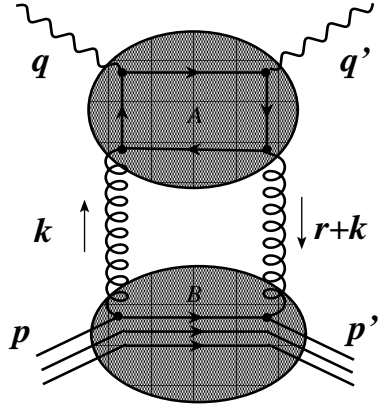}  }
{\caption{\label{dvcs_factor_2} Factorization of LLA scattering amplitude into two impact factors.
   }}
\end{figure}

The typical diagram
for the DVCS amplitude 
in the lowest order in perturbation theory is shown in Fig. \ref{ladder}
(recall that the diagrams with gluon exchanges dominate at high energies).
We are primarily looking for the imaginary part of the amplitude.
At high energy the amplitude for the colorless particle
scattering is described by the Feynman diagrams containing only two
intermediate gluons with momenta $k$ and $r+k$ in the $t$-channel (Fig. \ref{dvcs_factor_2}). 
Simple estimations show that with
good accuracy we can neglect the longitudinal momenta in their propagators:

\begin{equation}
k^2\simeq k_{\perp }^2\,,\,\,\,(r+k)^2\simeq (r+k)_{\perp }^2\,.
\end{equation}

Let us consider the integral over gluon momentum $d^4k=d^2k_{\perp}\,
\frac{d\,\alpha_k d\,\beta_k}{\textstyle 2\,s}$:
\begin{equation}\label{fla6}
V={2\over\pi}\int {d^4k\over 16\pi^4} {1\over k^2}
{1\over (r+k)^2} \Im \Phi^{ab}_{\mu\nu}(k+r, -k) 
\Im \Phi^{\mu\nu ab}_N(-k-r,k),
\end{equation}
where $\Phi^{ab}_{\mu\nu}(k+r, -k)$ and $(\Phi_N)^{ab}_{\mu\nu}(-k-r,k)$ are 
the upper and the lower blocks
of the diagram with two gluon exchanges in Fig. \ref{dvcs_factor_2} ($a,b$ and $\mu,\nu$ are the color and
Lorentz indices, respectively) corresponding to the impact-factor representation.

Due to our definition of the light-cone basis, 
all of the $\alpha$-s in the upper block could be neglected 
as well as all of the $\beta$-s in the lower block.
At high energies, the metric tensor in the numerator of the Feynman-gauge gluon 
propagator reduces to $g^{\mu\nu}\rightarrow {2\over {\textstyle s}}p_2^{\mu}p_1^{\nu}$
so the integral (\ref{fla6}) for 
the imaginary part factorizes into a product of two ``impact factors''
integrated with two-dimensional propagators
\begin{equation}
V={2s\over 4\pi}g^4\left(\sum e_q^2\right)
\int {d^2k_{\perp}\over 4\pi^2} {1\over k^2_{\perp}}
{1\over (r+k)^2_{\perp}}I(k_{\perp},r_{\perp}) I_N(k_{\perp},r_{\perp}),
\label{fla7}
\end{equation}
where
\begin{eqnarray}
I(k_{\perp},r_{\perp})&=&{1\over 2s}p_2^{\mu}p_2^{\nu}
\left.\int {d\beta_k\over 2\pi}
\Im \Phi^{aa}_{\mu\nu}(k+r, -k)\right|_{\alpha_k=0}
\label{fla8},\\
I_N(k_{\perp},r_{\perp})&=&{1\over 2s}p_1^{\mu}p_1^{\nu}
\left.\int {d\alpha_k\over 2\pi}
\Im \Phi^{aa}_{N\mu\nu}(-k-r,k)\right|_{\beta_k=0},
\label{fla9}
\end{eqnarray}
and $\left(\sum e_q^2\right)$ is the sum of squared charges of active flavors 
($u,d,s$, and possibly $c$). 

\section{Photon impact factor}

\begin{figure}[htb]
\mbox{
\epsfxsize=14cm
\epsfysize=5cm
\epsffile{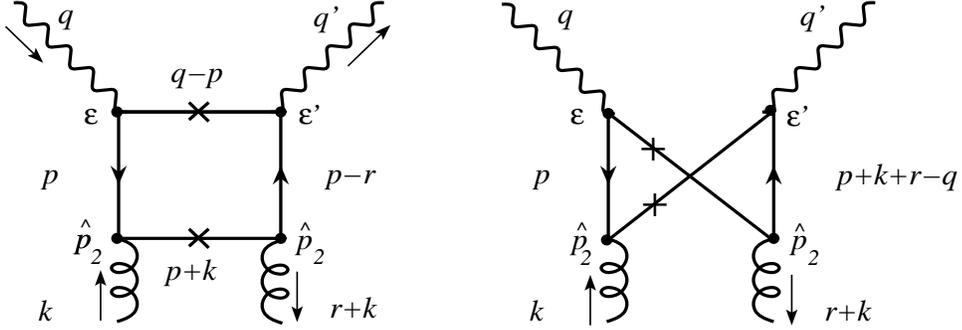}}
{\caption{\label{fig3} Photon impact factor.}}
\end{figure}

The photon impact factor is given by the two one-loop diagrams shown in 
Fig. \ref{fig3}. 

The first diagram yields:
\begin{eqnarray}
-\frac{2}{s}\int\frac{d\beta_k}{2\pi i}\int\frac{d^4p}{i(2\pi)^4}&&\frac{
Tr[\hat{\epsilon}\hat{p}\hat{p_2}(\hat{p}+\hat{k})\hat{p_2}(\hat{p}-\hat{r})
\hat{\epsilon'}(\hat{p}-\hat{q})]}{(p^2+i\varepsilon)((p-r)^2+i\varepsilon)} \times \nonumber \\
&& 2\pi i \delta((q-p)^2)\theta(q_0-p_0) 2\pi i \delta((p+k)^2)\theta(p_0+k_0).
\label{trace1}
\end{eqnarray}

%=================================
%==================================  
The second diagram differs only by trace and quark propagator: 
\begin{eqnarray}
-\frac{2}{s}\int\frac{d\beta_k}{2\pi i}\int\frac{d^4p}{i(2\pi)^4}&&\frac{Tr[\hat{\epsilon}\hat{p}\hat{p_2}(\hat{p}+\hat{k})\hat{\epsilon'}(\hat{p}+\hat{k}+\hat{r}-\hat{q})\hat{p_2}(\hat{p}-\hat{q})]}{(p^2+i\varepsilon)((p+k+r-q)^2+i\varepsilon)} \times \nonumber \\
&& 2\pi i \delta((q-p)^2)\theta(q_0-p_0) 2\pi i \delta((p+k)^2)\theta(p_0+k_0).
\label{trace2}
\end{eqnarray}

In these formulae, the following notation was used: $\hat{p}=\gamma^\mu p_\mu$, $\delta((q-p)^2)$ is a Dirac delta-function and $\theta(q_0-p_0)$ is a Heaviside step function. The singularities in the physical region were replaced using the Cutkosky rule \cite{cutk}.

In a physical region the singularity is related to external 4-momenta. It is true for the normal thresholds. 
Anomalous thresholds do not show up as main singularities in the physical region if 
the external particles are stable, which means that the 4-momenta squared for each vertex is 
less than the smallest value of the normal threshold in the given channel.

S. Coleman and R.E. Norton \cite{colnor} found the following simple interpretation. 
In the physical region the main singularity of the diagram occurs only if all the 
vertices could be considered as some space-time points while the internal lines could be 
regarded as the trajectories of the real relativistic particles on mass shell.

Due to the $\delta$-function all the particles are on the mass shell 
\[ \beta \rightarrow \frac{q^2}{s}+\frac{p^2_\perp}{\bar{\alpha}s},\ \ \ \
 \beta_k \rightarrow -\frac{q^2}{s}-\frac{p^2_\perp+\bar{\alpha}(k^2_\perp+2p_\perp k_\perp)}{\alpha \bar{\alpha}s},\]
and their energy is positive 
($\theta(q-p)_0 \rightarrow \theta(q-p,p_2) \rightarrow \theta(\bar{\alpha} \frac{s}{2})$ and $\theta(p+k)_0 \rightarrow \theta(p+k,p_2) \rightarrow \theta(\alpha \frac{s}{2})$). Here we introduced the notation $\bar{\alpha} \equiv 1-\alpha$.

The trace in the integral (\ref{trace2}) contains $k$-dependence:
\begin{eqnarray}
\lefteqn{\frac{-1}{s^2}Tr[\hat{\epsilon}\hat{p}\hat{p_2}(\hat{p}+\hat{k})\hat{\epsilon'}(\hat{p}+\hat{k}+\hat{r}-\hat{q})\hat{p_2}(\hat{p}-\hat{q}]=}\nonumber \\
&&-(\epsilon,\epsilon') 2 (p_{\perp}^2+R_{\perp} p_{\perp})\nonumber \\
&&-(\epsilon,r_{\perp})(\epsilon',\frac{p_2}{s})4 p_{\perp}^2\nonumber \\
&&+(\epsilon,R_{\perp})(\epsilon',p_{\perp})2 \nonumber \\
&&+(\epsilon,\frac{p_2}{s})(\epsilon',p_1)4((1-2\alpha\bar{\alpha})p_{\perp}^2+R_{\perp} p_{\perp})\nonumber \\
&&-(\epsilon,\frac{p_2}{s})(\epsilon',\frac{p_2}{s})8 p_{\perp}^2 (p_{\perp}^2-r_{\perp} p_{\perp})\nonumber \\
&&-(\epsilon,\frac{p_2}{s})(\epsilon',p_{\perp})4 p_{\perp}^2(1-2\alpha)\nonumber \\
&&-(\epsilon,\frac{p_2}{s})(\epsilon',R_{\perp})4 p_{\perp}^2 \nonumber \\
&&-(\epsilon,p_{\perp})(\epsilon',p_1)4 \alpha\bar{\alpha}(1-2\alpha)\nonumber \\
&&-(\epsilon,p_{\perp})(\epsilon',\frac{p_2}{s})4 ((1-2\alpha) p_{\perp}^2+2 R_{\perp} p_{\perp})\nonumber \\
&&+(\epsilon,p_{\perp})(\epsilon',p_{\perp})8\alpha\bar{\alpha}\nonumber \\
&&-(\epsilon,p_{\perp})(\epsilon',R_{\perp})2 (1-2\alpha)\nonumber \\
&&-(\epsilon,p_1)(\epsilon',p_1)8\alpha^2\bar{\alpha}^2\nonumber \\
&&+(\epsilon,p_1)(\epsilon',\frac{p_2}{s})4((1-2\alpha\bar{\alpha})p_{\perp}^2+(1-2\alpha)R_{\perp} p_{\perp})\nonumber \\
&&-(\epsilon,p_1)(\epsilon',p_{\perp})4\alpha\bar{\alpha}(1-2\alpha)\nonumber \\
&&-(\epsilon,p_1)(\epsilon',R_{\perp})4\alpha\bar{\alpha}\nonumber \\
 &&+(\epsilon,R_{\perp}+\bar{\alpha}r_{\perp})(\epsilon',\frac{p_2}{s})4 R_{\perp} p_{\perp} \nonumber \\
 &&+(\epsilon,p_{\perp})(\epsilon',R_{\perp}+\bar{\alpha}r_{\perp})4\alpha(1-2\alpha)\nonumber \\
 &&+(\epsilon,p_1)(\epsilon',R_{\perp}+\bar{\alpha}r_{\perp})8 \alpha^2\bar{\alpha}\nonumber \\
 &&+(\epsilon,\frac{p_2}{s})(\epsilon',R_{\perp}+\bar{\alpha}r_{\perp})8\alpha p_{\perp}^2 \nonumber \\
 &&-(\epsilon,r_{\perp})(\epsilon',\frac{p_2}{s})4(R_{\perp}r_{\perp}+\bar{\alpha}r_{\perp}^2)\nonumber \\
 &&-(\epsilon,p_{\perp})(\epsilon',\frac{p_2}{s})4(R_{\perp}+\bar{\alpha}r_{\perp})((1-2\alpha)R_{\perp}-\alpha(4p_{\perp}+(1-2\alpha)r_{\perp}))\nonumber \\
 &&-(\epsilon,p_1)(\epsilon',\frac{p_2}{s})8\alpha(R_{\perp}+\bar{\alpha}r_{\perp})((1-2\alpha)p_{\perp}+\bar{\alpha}R_{\perp}-\alpha\bar{\alpha}r_{\perp})\nonumber \\
 &&-(\epsilon,\frac{p_2}{s})(\epsilon',\frac{p_2}{s})8(R_{\perp}+\bar{\alpha}r_{\perp})(2p_{\perp}+R_{\perp}-\alpha r_{\perp})p_{\perp}^2,
\end{eqnarray}
hidden in the definition of $R_{\perp}$: $R_{\perp} \equiv k_\perp+\alpha r_{\perp}$.

This formula is rather long, but its beauty is in its generality.
Here one has no restrictions on photon polarizations or virtualities. 
It simplifies a lot for every specific case. 
For example, when both photons are transverse $(\epsilon,p_1)=(\epsilon,p_2)=(\epsilon',p_1)=(\epsilon',p_2)=0$, 
this integral reduces to:

\begin{eqnarray}
&&\int_0^1\frac{d\alpha}{2\pi}\int\frac{d^2p_\perp}{(2\pi)^2}\frac{1}{\alpha\bar{\alpha}q^2+p_\perp^2}\frac{1}{\alpha\bar{\alpha}q'^2+(p_\perp+R_\perp)^2}\nonumber\\
&&\times (-2)\Bigg[
(\epsilon,\epsilon') (p_{\perp}^2+R_{\perp} p_{\perp})\nonumber\\
&&~~~~~~~-(\epsilon,p_{\perp})(\epsilon',r_{\perp})2\alpha\bar{\alpha}(1-2\alpha)\nonumber\\
&&~~~~~~~-(\epsilon,p_{\perp}+R_{\perp})(\epsilon',p_{\perp})\nonumber\\
&&~~~~~~~+(\epsilon,p_{\perp})(\epsilon',R_{\perp})(1-2\alpha)^2\Bigg]. 
\label{two}
\end{eqnarray}

The integral for the first diagram differs only in overall sign and the definition of $R_{\perp} \equiv -\bar{\alpha} r_{\perp}$.
By introducing the Feynman parameter $\alpha'$ (with simultaneous shifting to $p_\perp \leftarrow p_\perp+\alpha' R_\perp$) 
and using formulae of Feynman parametrization,
one is able to get the photon impact factor for the case of arbitrary photon polarization as the following:
\begin{eqnarray}\label{general}
&&\int_0^1\frac{d\alpha}{2\pi}\int_0^1\frac{d\alpha'}{2\pi}\left\{
   \alpha'\bar{\alpha'}(R_{\perp}^{2}+\Omega^2)
   \right\}^{-1} \times \nonumber \\ &&
     \Bigg\{ (\epsilon,\epsilon') \big( \alpha\bar{\alpha}(1-2\alpha\bar{\alpha})(1-2\alpha')(q'^2-q^2)+(1-2\alpha\bar{\alpha}-2\alpha'\bar{\alpha'}+8\alpha\bar{\alpha} \alpha'\bar{\alpha'}) R_{\perp}^2 \big) \Big)  \nonumber \\ &&
    -(\epsilon,p_1)(\epsilon',p_1) 8{\alpha}^2 {\bar{\alpha}}^2   \nonumber \\ &&
    -(\epsilon,\frac{p_2}{s})(\epsilon',p_1) 2\Big( \alpha\bar{\alpha}(1\!-\!2\alpha\bar{\alpha})(1\!-\!2\alpha')(q'^2-q^2)+\nonumber \\ &&
~~~~(1\!-\!2\alpha\bar{\alpha}\!-\!2\alpha'\bar{\alpha'}\!+\!4\alpha\bar{\alpha} \alpha'(2\!-\!\alpha')) R_{\perp}^2\Big)\nonumber \\ &&
    +(\epsilon,R_{\perp})(\epsilon',p_1) 4\alpha\bar{\alpha}(1-2\alpha)\alpha'  \nonumber \\ &&
    -(\epsilon,p_1)(\epsilon',\frac{p_2}{s}) 2\Big(\alpha\bar{\alpha}(1-2\alpha\bar{\alpha})(1-2\alpha')(q'^2-q^2)-4{\alpha}^2{\bar{\alpha}}^2 r_{\perp}^2+ \nonumber \\ &&
     ~~~~ 4\alpha\bar{\alpha}(1-2\alpha)\bar{\alpha'}(R_{\perp},r_{\perp})+(1+2\alpha\bar{\alpha}-2\alpha'\bar{\alpha'}-4\alpha\bar{\alpha} \alpha'^2) R_{\perp}^2 \Big)  \nonumber \\ &&
    -(\epsilon,\frac{p_2}{s})(\epsilon',\frac{p_2}{s}) 4\Big( \alpha\bar{\alpha}(q'^2+r_{\perp}^2)(\alpha\bar{\alpha}(1-2\alpha')(q'^2-q^2)+(1-4\alpha'+2\alpha'^2)R_{\perp}^2)-\nonumber \\ &&
     ~~~~(1-2\alpha)(R_{\perp},r_{\perp})(\alpha\bar{\alpha}(1-4\alpha'\bar{\alpha'})(q'^2-q^2)+(1-6\alpha'+10\alpha'^2-6\alpha'^3)R_{\perp}^2) \Big)\nonumber \\ &&
    -(\epsilon,R_{\perp})(\epsilon',\frac{p_2}{s}) 4\alpha\bar{\alpha} \alpha' \big((1-2\alpha)(q'^2+r_{\perp}^2)+4\bar{\alpha'}(R_{\perp},r_{\perp}) \big) \nonumber \\ && 
    +(\epsilon,r_{\perp} )(\epsilon',\frac{p_2}{s}) 2\Big(\alpha\bar{\alpha}(1-2\alpha\bar{\alpha})(1-2\alpha')(q'^2-q^2)+\nonumber \\ &&
~~~~(1-2\alpha\bar{\alpha}-2\alpha'\bar{\alpha'}+8\alpha\bar{\alpha} \alpha'\bar{\alpha'}) R_{\perp}^2 \Big) \nonumber \\ &&
    -(\epsilon,p_1)(\epsilon',R_{\perp}) 4\alpha\bar{\alpha} (1-2\alpha) \bar{\alpha'} \nonumber \\ &&
    +(\epsilon,p_1)(\epsilon',r_{\perp}) 8{\alpha}^2{\bar{\alpha}}^2 \nonumber \\ &&
    +(\epsilon,\frac{p_2}{s})(\epsilon',R_{\perp}) 2(1-2\alpha) \Big( \alpha\bar{\alpha} (1\!\!-\!\!2\alpha')^2 (q'^2\!\!-\!\!q^2)+(1\!\!-\!\!6\alpha'\!\!+\!\!10\alpha'^2\!\!-\!\!6\alpha'^3) R_{\perp}^2 \big)  \nonumber \\ &&
    -(\epsilon,\frac{p_2}{s})(\epsilon',r_{\perp}) 4\alpha\bar{\alpha} \big(\alpha\bar{\alpha}(1-2\alpha')(q'^2-q^2)+(1-4\alpha'+2\alpha'^2)R_{\perp}^2 \big) \nonumber \\ &&
    -(\epsilon,R_{\perp})(\epsilon',R_{\perp}) 8\alpha\bar{\alpha} \alpha'\bar{\alpha'}   \nonumber \\ &&
    -(\epsilon,R_{\perp})(\epsilon',r_{\perp}) 4\alpha\bar{\alpha} (1-2\alpha)\alpha' \Big]\Bigg\} \\ &&
       ~~ -\Big[ R_{\perp}=k_{\perp} +\alpha r_{\perp} \to R_{\perp}'=-\bar{\alpha}r_{\perp}\Big], \nonumber
\end{eqnarray}
\begin{equation}\label{omega}
\Omega^2\equiv\frac{\alpha\bar{\alpha}}{\alpha'\bar{\alpha'}}(\alpha'q'^2+\bar{\alpha'}q^2).
\end{equation}

One can see from Eq.~(\ref{general}), that even in the most general case, the difference between the two diagrams of Fig. \ref{fig3} is in the presence of $k_{\perp}$ in one and the absence ($k_{\perp} \rightarrow 0$) in the other.
It means one can write \cite{ifak}:
\begin{equation}
I(k_{\perp},r_{\perp})=\bar{I}(k_{\perp},r_{\perp})-
\bar{I}(0,r_{\perp}).
\label{fla10}
\end{equation}

For the particular case of two transverse photons one gets (cf. \cite{ing}):
 \begin{eqnarray}
\bar{I}^{TT}(k_{\perp},r_{\perp})&=&
   {1\over 2}\int _{0}^{1} \frac {d\alpha}{2\pi }
   \int _{0}^{1} \frac {d\alpha'}{2\pi }
   \left\{
  \alpha'\bar{\alpha'}(R_{\perp}^{2}+\Omega^2) 
   \right\}^{-1} \times
\label{fla11}\nonumber\\
&&
   \Bigg\{
      (1-2\alpha\bar{\alpha})
      R_{\perp}^{2} (\epsilon, \epsilon')_{\perp}\nonumber\\
&&+4\alpha\bar{\alpha}\bar{\alpha'}[R_{\perp}^{2} (\epsilon, \epsilon')-
2(\epsilon, R)_{\perp}(\epsilon', R)_{\perp}] \nonumber\\
&&-4\alpha\bar{\alpha}(1-2\alpha)(r,\epsilon)_{\perp} (R,\epsilon')_{\perp}
\Bigg\}, 
\end{eqnarray}
and for the longitudinal polarization
 \begin{equation}
 \epsilon^3(q)={1\over Q}(p_1+xp_2)
 \label{fla13}
 \end{equation}
 of the incoming photon the formula is also simple:
\begin{eqnarray}
\bar{I}^{LT}(k_{\perp},r_{\perp}) &=&
   {1\over 2Q}\int _{0}^{1} \frac {d\alpha}{2\pi }
   \int _{0}^{1} \frac {d\alpha'}{2\pi }
   \left\{
  \alpha'\bar{\alpha'}(R_{\perp}^{2}+\Omega^2)
   \right\}^{-1} \times
\label{fla12}\nonumber\\
&&   \Bigg\{
      (1-2\alpha\bar{\alpha})
      R_{\perp}^{2} (r, \epsilon')_{\perp}\nonumber\\
&&+4\alpha\bar{\alpha}\bar{\alpha'}[R_{\perp}^{2} (r, \epsilon')_{\perp}-
2(r, R)_{\perp}(\epsilon', R)_{\perp}] \\
&&      -4\alpha\bar{\alpha}(1-2\alpha)Q^2(R,\epsilon')_{\perp}
   \Bigg\}. \nonumber
\end{eqnarray}

Here $(a,b)_{\perp}$
denotes the (positive) scalar
product of transverse components of vectors $a$ and $b$. At
large transverse momenta $k_{\perp}^2\gg r_{\perp}^2$, the impact factor
(\ref{fla10})
reduces to:
\begin{equation}
 I(k_{\perp},r_{\perp})\rightarrow
 {(\epsilon, \epsilon')_{\perp}\over 4\pi^2}{k_{\perp}^2\over Q^2}
 \ln{Q^2\over r_{\perp}^2}
\label{fla14}.
\end{equation}

The impact factor for the proton, which describes the pomeron-nucleon
coupling, cannot be calculated in
perturbation theory. However, in the next section it is demonstrated that at high
momenta $k_{\perp}^2\gg M^2$ this impact factor reduces to
\begin{equation}
I_N(k_{\perp},r_{\perp})\stackrel{k_{\perp}^2\gg M^2}{=}
F_1^{p+n}(t),
\label{fla15}
\end{equation}
where $F_1^{p+n}(t)$ is the sum of the proton and neutron
Dirac form factors.

\section{Nucleon impact factor}
\begin{figure}[htb]
%\vspace{-1cm}
\mbox{
\epsfxsize=13cm
\epsfysize=5cm
\hspace{1cm}
\epsffile{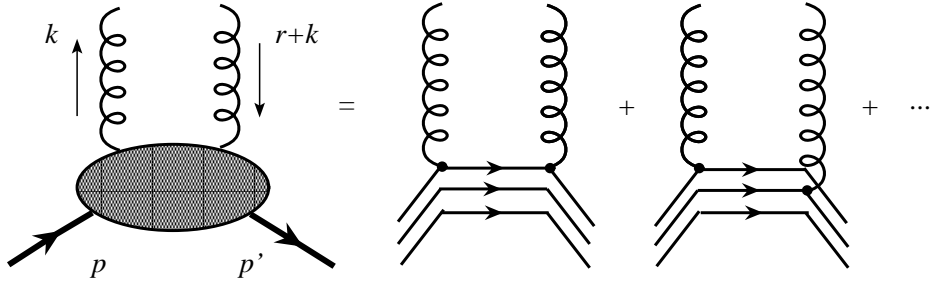}}
%\vspace{-2cm}
{\caption{\label{nucleon_if} Nucleon impact factor.}}
\end{figure}

In the lowest order in perturbation theory there is no difference
between the diagrams for the nucleon impact factor
shown in Fig. \ref{nucleon_if}
%=================================
%==================================
and similar diagrams with two gluons
replaced by two photons. One has to add the trivial numerical factor
$c_F={4\over 3}$ and to 
replace  $e \leftrightarrow g$. In this case the lower part of
the diagram can be formally written as follows:
\begin{equation}
\Phi_N(-k-r, k)\stackrel{\rm def}
=
{2\over 3}i {p_1^{\mu} p_1^{\nu}\over s}\int dz e^{ikz}
\langle p'|T^*\{J_{\mu}(z)J_{\nu}(0)\}| p\rangle ,
\label{fla19}
\end{equation}
where $J_{\mu}=\bar{u}\gamma_{\mu}u+\bar{d}\gamma_{\mu}d $.
The  $T^*$ means that we take into account only the T-product of the diagrams with
pure gluon exchanges in the $t$-channel excluded. By definition, such diagrams 
contribute to subsequent ranks of the BFKL ladder rather than to the impact
factor. This is also the reason why we have not included strange quarks in the definition of electromagnetic current $J$.
Since $k^2$ in
our case is large and negative (-$k^2=k_{\perp}^2\gg M^2$) we can expand
the T-product of two currents near the light cone (see e.g. \cite{bal83})
\begin{equation}
\Phi_N(-k-r, k)=\!\!
{2\over 3s}\!\!\int\!\! dz e^{ikz}{zp_1\over \pi^2 z^4}
\langle p'|\!-\!\bar{\psi}(z)[z,0]\hat{p_1}\psi(0)\!+\!\bar{\psi}(0)[0,z]\hat{p_1}\psi(z)| p\rangle^*_{z^2=0},
\label{fla20}
\end{equation}
where again $\langle...\rangle^*$ stands for the matrix element with pure gluon
exchanges excluded. Here $[x,y]$ denotes
the gauge link connecting the points $x$ and $y$
($[x,y]\equiv {\cal P}exp\left( ig\int_0^1 du(x-y)^{\mu}A_{\mu}(ux+(1-u)y)\right)$).
The matrix element (\ref{fla20}) can be parametrized
in terms of skewed parton distributions \cite{rad2} as follows:
\begin{eqnarray*}
\langle p',\lambda'|\bar{q}(z)[z,0]\hat{p_1}q(0)| p,\lambda\rangle^*_{z^2=0}
=\bar{u}(p',\lambda')\hat{p_1}u(p,\lambda)
\int_0^1\!\! dX e^{i(X-x)pz}{\cal V}^q_x(X,t)&&\nonumber\\
+{1\over 2M}\bar{u}(p',\lambda')\hat{p_1}\!\hat{r}_{\perp} u(p,\lambda)
\int_0^1\!\! dX e^{i(X-x)pz}{\cal W}^q_x(X,t),&&
\end{eqnarray*}
\begin{eqnarray}\label{fla21}
\langle p',\lambda'|\bar{q}(0)[0,z]\hat{p_1}q(z)| p,
\lambda\rangle^*_{z^2=0}=
\bar{u}(p',\lambda')\hat{p_1}u(p,\lambda)
\int_0^1\!\! dX e^{-iXpz}{\cal V}^q_x(X,t)&&\\
+{1\over 2M}\bar{u}(p',\lambda')\hat{p_1}\!\hat{r}_{\perp} u(p,\lambda)
\int_0^1\!\! dX e^{-iXpz}{\cal W}^q_x(X,t),&&
\nonumber
\end{eqnarray}
 where ${\cal V}^u_x(X,t)$ and ${\cal W}^u_x(X,t)$ are the non-flip and 
spin-flip
 skewed parton distributions for
the {\it valence} $u$ quark 
(recall that one must take into account only valence quarks since 
diagrams with pure gluon exchanges are forbidden). Similarly, 
${\cal V}^d_x(X,t)$ and ${\cal W}^d_x(X,t)$ refer to the valence $d$-quark
distributions. 
At large energies
$\bar{u}(p',\lambda')\hat{p_1}u(p,\lambda)=s\delta_{\lambda\lambda'}$,  so
\begin{eqnarray}
\lefteqn{\langle p',\lambda'|\bar{q}(0)[0,z]\hat{p_1}q(z)-
\bar{q}(z)[z,0]\hat{p_1}q(0)| p,\lambda\rangle^*_{z^2=0}=}\label{fla22}\\
&&
\int_0^1\!\! dX \left(e^{-iXpz}-
e^{i(X-x)pz}\right)\left[s\delta_{\lambda\lambda'}{\cal V}^q_x(X,t)+
\bar{u}(p',\lambda'){\hat{p_1}\!\hat{r}_{\perp}\over 2M} u(p,\lambda)
{\cal W}^q_x(X,t)\right]
\nonumber.
\end{eqnarray}
After integration over $z$ the lower block (\ref{fla19}) reduces to
\begin{eqnarray}
\lefteqn{\Phi_N(-k-r, k)=}\label{fla23}\\
&&
{2\over 3s}
\int_0^1 dX 
\left[{(X-x)s+2p_1\cdot k\over -k^2-2p\cdot k (X-x)-i\varepsilon}
-{-Xs+2p_1\cdot k\over -k^2+2p\cdot k X-i\varepsilon}\right]
\nonumber\\
&&
\left(\delta_{\lambda\lambda'}({\cal V}^u_{x}(X,t)+{\cal V}^d_x(X,t))+
\bar{u}(p',\lambda'){\hat{p_1}\!\hat{r}_{\perp}\over 2Ms} u(p,\lambda)
({\cal W}^u_{x}(X,t)+{\cal W}^d_x(X,t)) \right).
\nonumber
\end{eqnarray}
The nucleon impact factor (\ref{fla9}) is the integral of
the imaginary part of
right hand side of Eq.~(\ref{fla23}) over energy:
\begin{eqnarray}\label{fla24}
&&I_N(k_{\perp},r_{\perp})=\int_0^1 {d\alpha_k\over 2\pi}
\Im \Phi_N(-(\alpha_k-{r_{\perp}^2\over s}) p_1-k_{\perp}-
r_{\perp},\alpha_k p_1+k_{\perp} )\\
&&~~~={1\over 3}
\int_0^1 d\alpha_k\int_x^1 dX \left[s(X-x)
\delta (k_{\perp}^2-\alpha_k s(X-x))-sX\delta (k_{\perp}^2+ \alpha_k sX)\right]
\nonumber\\
&&\left(\delta_{\lambda\lambda'}({\cal V}^u_x(X,t)+{\cal V}^d_x(X,t))+
\bar{u}(p',\lambda'){\hat{p_1}\!\hat{r}_{\perp}\over 2Ms} u(p,\lambda)
({\cal W}^u_{x}(X,t)+{\cal W}^d_x(X,t))\right).\nonumber
\end{eqnarray}
And finally,
\begin{eqnarray}
&&I_N(k_{\perp},r_{\perp})={1\over 3}\int_x^1 dX \Bigg(\delta_{\lambda\lambda'}({\cal V}^u_x(X,t)+{\cal V}^d_x(X,t)) \\
&&~~~~~~~~~~~~~~~+{1\over 2Ms}\bar{u}(p',\lambda')\hat{p_1}\!\hat{r}_{\perp} u(p,\lambda)
({\cal W}^u_{x}(X,t)+{\cal W}^d_x(X,t))\Bigg).\nonumber
\end{eqnarray}
Since valence quark distributions decrease at $x\rightarrow 0$
one can extend the lower limit
of integration in the r.h.s. of Eq.~(\ref{fla24}) to 0 and obtain:
\begin{eqnarray}\label{fla25}
&I_N(k_{\perp}, r_{\perp})&\stackrel{k_{\perp}^2\gg M^2}{=}{1\over 3}
\int_0^1 dX
\Bigg( \delta_{\lambda\lambda'}({\cal V}^u_x(X,t)+{\cal V}^d_x(X,t))\nonumber \\ 
&&~~~+
{1\over 2Ms}\bar{u}(p',\lambda')\hat{p_1}\!\hat{r}_{\perp} u(p,\lambda)
({\cal W}^u_{x}(X,t)+{\cal W}^d_x(X,t))\Bigg).
\end{eqnarray}
Let us recall the sum rules \cite{ji,rad2}
\begin{eqnarray}
\int_0^1 dX
\left({\cal F}^q_x(X,t)-{\cal F}^{\bar q}_x(X,t)\right)&=&F^q_1(t),
\nonumber\\
\int_0^1 dX
\left({\cal K}^q_x(X,t)-{\cal K}^{\bar q}_x(X,t)\right)&=&F^q_2(t),
\label{fla26}
\end{eqnarray}
where ${\cal F}^q_x(X,t)$ and ${\cal K}^q_x(X,t)$ are the total
(valence $+$ sea)
non-flip and spin-flip skewed quark
distributions while  ${\cal F}^{\bar q}_x(X,t)$ and
${\cal K}^{\bar q}_x(X,t)$ are
the antiquark ones. In these equations, $F^q_1(t)$ and $F^q_2(t)$ stand for
the $q$-quark
components of the Dirac and Pauli
form factors of the proton.
Since the contribution of sea quarks drops from the difference
${\cal F}^q-{\cal F}^{\bar q}$ one can rewrite Eqs.~(\ref{fla26}) as the sum
rules for valence quark distributions:
\begin{equation}
\int_0^1 dX {\cal V}^q_x(X,t)= F^q_1(t),~~~~~~~~~
\int_0^1 dX {\cal W}^q_x(X,t)= F^q_2(t).
\label{fla27}
\end{equation}
Substituting this estimate to Eq.~(\ref{fla25}) and
using the isospin invariance, one can get the final result
for the nucleon impact factor at large transverse momenta
\begin{equation}
I_N(k_{\perp}, r_{\perp})\stackrel{k_{\perp}^2\gg M^2}{=}
\delta_{\lambda\lambda'}F_1^{p+n}(t)+
{1\over 2Ms}\bar{u}(p',\lambda')\hat{p_1}\!\hat{r}_{\perp}
u(p,\lambda)F_2^{p+n}(t),
\label{fla28}
\end{equation}
where
$F_1^{p+n}(t)\equiv F_1^p(t)+F_1^n(t)$ and
$F_2^{p+n}(t)\equiv F_2^p(t)+F_2^n(t)$.
As usual, $F_1^{p(n)}$ and
$F_2^{p(n)}$ are the Dirac and Pauli form factors of
the proton (neutron), respectively.
With our accuracy they can be approximated
by the dipole formulas:
\begin{equation}
\begin{array}{lll}
F_1^p+F_2^p t/4M^2&=G_E^p&=1/(1+|t|/\alpha_t)^2,\nonumber\\
F_1^p+F_2^p&=G_M^p&=2.79/(1+|t|/\alpha_t)^2,\nonumber\\
\\
F_1^n+F_2^n t/4M^2&=G_E^n&=0,\nonumber\\
F_1^n+F_2^n&=G_M^n&=-1.91/(1+|t|/\alpha_t)^2,
\end{array}
\label{fla29}
\end{equation}
which leads to
\begin{eqnarray}\label{fla31}
F_1^{p+n}(t)=& {1\over \left(1+
|t|/\alpha_t \right)^2}{1+0.88 |t|/ 4M^2\over
1+|t|/ 4M^2}&\rightarrow{1\over \left(1+
|t|/\alpha_t \right)^2},\nonumber\\
F_2^{p+n}(t)=&{0.12\over \left(1+
|t|/\alpha_t\right)^2}&\rightarrow 0.
\end{eqnarray}
Here  the notation $\alpha_t=0.71$ GeV$^2$ is introduced. 
In further calculations we will use the estimate (\ref{fla31}).
It is obvious that the spin-flip term turned out to be negligible
for our values of $t$.
Besides, it vanishes at $t=0$, which suggests that it
is numerically small at all $t$.

 Our final estimate of the nucleon
impact factor is:
\begin{equation}
I_N(k_{\perp}, r_{\perp})\stackrel{k_{\perp}^2\gg M^2}{=}
\delta_{\lambda\lambda'}F_1^{p+n}(t),
\label{fla32}
\end{equation}
where $F_1^{p+n}$ is given by the dipole formula (\ref{fla31}).

The dipole formula for the neutron form factor does not seem to work
as well as the dipole formula for the proton form factor. As a measure
of the uncertainty one can compare the results obtained from Eq.~(\ref{fla31})
to those obtained using the model from Ref. \cite{prof} (which
was fit only to the proton form factor):
\begin{eqnarray}
F^{p+n}_1(t)&=&
{1\over 3}\int_0^1dX\left({\cal V}^u_x(X,t)+{\cal V}^d_x(X,t)\right),\nonumber\\
{\cal V}^u_x(X,t)&=&1.89 X^{-0.4}\bar{X}^{3.5}(1+6X)
\exp\left(-{\bar{X}\over X}{|t|\over 2.8{\rm GeV}^2}\right),
\nonumber\\
{\cal V}^d_x(X,t)&=&0.54 X^{-0.6}\bar{X}^{4.2}(1+8X)
\exp\left(-{\bar{X}\over X}{|t|\over 2.8{\rm GeV}^2}\right).
\label{fla33}
\end{eqnarray}
The results for the DVCS cross section in this model are about
$1.5$ times bigger than the results obtained from
the dipole formula (\ref{fla31}).

In what follows the factor $\delta_{\lambda\lambda'}$ is omitted
(as it was done
in Eq.~(\ref{fla15})) since all the
amplitudes are diagonal in the proton's spin.

\section{The BFKL ladder}
As we shall see below, 
the characteristic transverse momenta in our
gluon loop are large, so the estimate (\ref{fla15}) is sufficient for our
purposes. Substituting the
nucleon impact factor (\ref{fla15}) into Eq.~(\ref{fla7}) one obtains:
\begin{equation}
V={2s\over \pi}g^4(\sum e_q^2)
F_1^{p+n}(t)
\int {d^2k_{\perp}\over 4\pi^2}
{I(k_{\perp},r_{\perp})\over k_{\perp}^2(r+k)_{\perp}^2}
\label{fla16}.
\end{equation}
The final integration over $k_{\perp}$ reveals the logarithmic dependence of the photon impact factor $\ln{{r_{\perp}^2}/{\Omega^2}} \to \ln{|t|/Q^2}$:
\begin{eqnarray} 
I_0\equiv\int\frac{d^2 k_{\perp} }{k_{\perp} ^2(k_{\perp} +r_{\perp})^2}\Bigg(\frac{1}{(k_{\perp} +\alpha r_{\perp})^2+\Omega^2}-\frac{1}{\bar{\alpha}^2 r_{\perp}^2+\Omega^2}\Bigg)&&\nonumber \\
=\frac{\pi}{\Omega^2-\alpha\bar{\alpha}r_{\perp}^2} 
\Big\{\frac{-\alpha}{\Omega^2+\alpha^2r_{\perp}^2}(2\ln{\frac{\Omega^2+\alpha^2r_{\perp}^2}{\Omega^2}}+\eta) \nonumber \\
+\frac{-\bar{\alpha}}{\Omega^2+\bar{\alpha}^2r_{\perp}^2}(2\ln{\frac{\Omega^2+\bar{\alpha}^2r_{\perp}^2}{\Omega^2}}+\eta) 
+\frac{\bar{\alpha}}{\Omega^2+\bar{\alpha}^2r_{\perp}^2}(2\ln{\frac{r_{\perp}^2}{\Omega^2}}+2\eta) \Big\}, 
\end{eqnarray}
\begin{eqnarray}
I_1\equiv\int  \frac{d^2 k_{\perp} }{k_{\perp} ^2(k_{\perp} +r_{\perp})^2}\Bigg(\frac{(k_{\perp} +\alpha r_{\perp})^{\mu}}{(k_{\perp} +\alpha r_{\perp})^2+\Omega^2}-\frac{(-r_{\perp}+\alpha r_{\perp})^{\mu}}{\bar{\alpha}^2r_{\perp}^2+\Omega^2}\Bigg)&&\nonumber \\
=\frac{\pi}{\Omega^2-\alpha\bar{\alpha}r_{\perp}^2}\frac{r_{\perp}^{\mu}}{r_{\perp}^2}
\Big\{\frac{\Omega^2-\alpha^2r_{\perp}^2}{\Omega^2+\alpha^2r_{\perp}^2}\ln{\frac{\Omega^2+\alpha^2r_{\perp}^2}{\Omega^2}}-\frac{\Omega^2-\bar{\alpha}^2r_{\perp}^2}{\Omega^2+\bar{\alpha}^2r_{\perp}^2}\ln{\frac{\Omega^2+\bar{\alpha}^2r_{\perp}^2}{\Omega^2}} \nonumber \\
+\frac{\Omega^2-\bar{\alpha}^2r_{\perp}^2}{\Omega^2+\bar{\alpha}^2r_{\perp}^2}\ln{\frac{r_{\perp}^2}{\Omega^2}} 
+(1-\frac{\alpha^2r_{\perp}^2}{\Omega^2+\alpha^2r_{\perp}^2}-\frac{\bar{\alpha}^2r_{\perp}^2}{\Omega^2+\bar{\alpha}^2r_{\perp}^2})\eta \Big\}, 
\end{eqnarray}
\begin{eqnarray}
\int  \frac{d^2 k_\perp}{k_\perp^2(k_\perp+r_\perp)^2}\Bigg(\frac{(k_\perp+\alpha r_\perp)^2}{(k_\perp+\alpha r_\perp)^2+\Omega^2}-\frac{\bar{\alpha}^2 r_\perp^2}{\bar{\alpha}^2 r_\perp^2+\Omega^2}\Bigg)=-\Omega^2* I_0,
\end{eqnarray}
\begin{eqnarray}
\int  \frac{d^2 k_\perp}{k_\perp^2(k_\perp+r_\perp)^2}\Bigg(\frac{(k_\perp+\alpha r_\perp)^2(k_\perp+\alpha r_\perp)^{\mu}}{(k_\perp+\alpha r_\perp)^2+\Omega^2}-\frac{\bar{\alpha}^2 r_\perp^2(-r_\perp+\alpha r_\perp)^{\mu}}{\bar{\alpha}^2 r_\perp^2+\Omega^2}\Bigg)\nonumber \\
=-\Omega^2*I_1+\frac{\pi r_\perp^{\mu}}{r_\perp^2}(\ln{\frac{r_\perp^2}{\Omega^2}+\eta}),
\end{eqnarray}
where
\begin{equation}
\eta\equiv\frac{1}{\varepsilon}+\gamma_E+\ln{\pi}+\ln{\Omega^2},
\end{equation}
\begin{equation}
R_{\mu}R_{\nu} \leftrightarrow \frac{1}{2}g_{\mu\nu}R^2,
\end{equation}
and $\Omega^2$ is defined in Eq.~(\ref{omega}). 

For the case of transverse photon polarizations one gets
\begin{eqnarray} 
V^{TT}&=&
{2\over x}\left({\alpha_s\over \pi}\right)^2(\sum_q e_q^2)
F_1^{p+n}(t)\\
&&
\Bigg ( (\epsilon,\epsilon')_{\perp}\left({1\over 2} \ln^2{Q^2\over |t|}+2\right)-
(\epsilon,\epsilon')_{\perp}+{2\over r_{\perp}^2}(\epsilon,r)_{\perp}(\epsilon',r)_{\perp}+{\cal O}(\frac{t}{Q^2})\Bigg ),\nonumber
\label{fla17}
\end{eqnarray}
and when the incoming photon has longitudinal polarization we have
\begin{eqnarray} \label{fla18}
V^{LT}&=&
-{2\over x}\left({\alpha_s\over \pi}\right)^2(\sum_q e_q^2)
F_1^{p+n}(t) {(r,\epsilon')_{\perp}\over Q}\nonumber\\
&&
\left({1\over 2} \ln^2{Q^2\over |t|}-5\ln{Q^2\over |t|}+
{15\over 2}-{\pi^2\over 3}+{\cal O}(\frac{t}{Q^2})\right).
\end{eqnarray}
The longitudinal amplitude (\ref{fla18})
is twist-suppressed as ${\sqrt{|t|/Q^2}}$ in comparison to
the transverse amplitude (\ref{fla17}) (as it should, due to the fact that
$t\rightarrow 0$ corresponds to a real incoming photon).

Since the integral
over $k_{\perp}$ (\ref{fla16}) converges at $k_{\perp}\sim Q$, the region
$k_{\perp}\sim M$,
where we do not know the nucleon impact factor, contributes to the terms $\sim
{\cal O}(|t|/Q^2)$ which we neglect.  

%=================================
\begin{figure}[htb]
\mbox{
\epsfxsize=13cm
\epsfysize=8cm
\hspace{0cm}
\epsffile{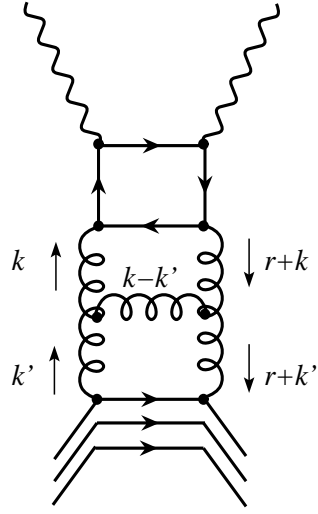}}
{\caption{\label{NLO} Typical diagram in the next-to-leading order
in perturbation theory.}}
\end{figure}
%==================================
In the next-to-leading order (NLO) in  perturbation theory the most important diagrams are
those of the type shown in Fig. \ref{NLO}.
Actually, this diagram gives the total contribution in LLA if one
replaces the three-gluon vertex in Fig. \ref{NLO} by the
effective Lipatov's vertex \cite{lobzor}.

Calculation of this diagrams in the leading log approximation yields:
\begin{eqnarray}
 \lefteqn{V={2sg^4\over\pi}(\sum e_q^2)\left(6\alpha_s\ln {1\over x}\right)}
 \nonumber\\
 &&
 \int {d^2k_{\perp}\over 4\pi^2}{d^2k'_{\perp}\over 4\pi^2}
  {I(k_{\perp},r_{\perp})\over k_{\perp}^2(r+k)_{\perp}^2}
  K(k_{\perp},k'_{\perp},r_{\perp})
{I_N(k'_{\perp},r_{\perp})\over (k'_{\perp})^{2} (r+k')_{\perp}^2},
\label{fla34}
\end{eqnarray}
where $K(k_{\perp},k'_{\perp},r_{\perp})$ is the BFKL kernel \cite{bfkl}
\begin{eqnarray}
K(k',k,r)=
2\Bigg[\frac{k'\cdot (k'+r)}{k'^2(k'+r)^2}-\frac{k'\cdot (k'-k)}{k'^2(k'-k)^2}-\frac{(k'-k)\cdot (k'+r)}{(k'-k)^2(k'+r)^2}\nonumber \\
+\frac{k'\cdot (k'-k)}{(k'-k)^2[k'^2+(k'-k)^2]}+\frac{(k'-k)\cdot (k'+r)}{(k'-k)^2[(k'-k)^2+(k'+r)^2]}\Bigg].
\label{fla35}
\end{eqnarray}
As we shall see below, the integral over $k'_{\perp}$ converges at
$|k'_{\perp}|\gg M$ so we can again use the approximation (\ref{fla15})
for the nucleon impact factor. One obtains
\begin{equation}
 \int d^2k'_{\perp}
 K(k_{\perp},k'_{\perp},r_{\perp})
{I_N(k'_{\perp},r_{\perp})\over (k'_{\perp})^{2} (r+k')_{\perp}^2}=
\pi F_1^{p+n}(t)
\left(\ln {k_{\perp}^2\over r_{\perp}^2}+
\ln {(k+r)_{\perp}^2\over r_{\perp}^2}\right),
\label{fla36}
\end{equation}
and therefore the amplitude (\ref{fla34}) takes the form
\begin{equation}
 V={g^4 s\over \pi}F_1^{p+n}(t)
 \left({3\alpha_s\over\pi}\ln {1\over x}\right)
 \int {d^2k_{\perp}\over 4\pi^2}
 {I(k_{\perp},r_{\perp}) \over k_{\perp}^2(r+k)_{\perp}^2}
 \left(\ln {k_{\perp}^2\over r_{\perp}^2}+
\ln {(k+r)_{\perp}^2\over r_{\perp}^2}\right)
\label{fla37}.
\end{equation}
Finally, the integration over $k$ yields:
\begin{eqnarray}
\lefteqn{ V=}\nonumber\\
&&{2\over x}\left({\alpha_s\over\pi}\right)^2(\sum_q e_q^2) 
F_1^{p+n}(t)\left({3\alpha_s\over\pi}\ln {1\over x}\right)\\
&&
\Bigg ( (\epsilon,\epsilon')_{\perp}\Big({1\over 6} \ln^3{Q^2\over |t|}+
2\ln{Q^2\over |t|}-2+\zeta(3)\Big)+
\Big({2\over r_{\perp}^2}(\epsilon,r)_{\perp}(\epsilon',r)_{\perp} -
(\epsilon,\epsilon')_{\perp}\Big)
\Bigg ), \nonumber
\label{fla38}
\end{eqnarray}
where the accuracy is $O({1/ \ln x})$ and $\zeta(3)$ is a Riemann zeta function.

\begin{figure}[htb]
%\hspace{4cm}
\mbox{
\epsfxsize=13cm
\epsfysize=8cm
%\hspace{3cm}
\epsffile{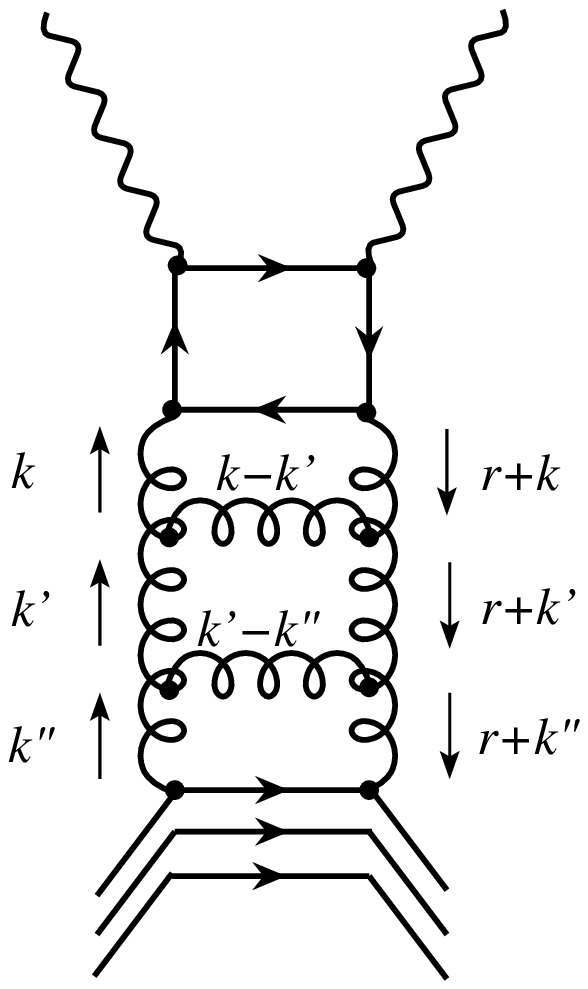}}
%\vspace{-1cm}
{\caption{\label{fig6} Typical diagram in the next-to-next-to-leading order 
in perturbation theory.}}
\end{figure}

In the next order in the BFKL approximation (see Fig. \ref{fig6}) it is
still possible  to obtain the DVCS amplitude (\ref{fla3}) in 
the explicit form.
%=================================
%==================================
It is possible to write down the result of the summation of the BFKL
ladder in the form of a Mellin integral over complex momenta using the Lipatov's
conformal eigenfunctions of the BFKL equation in the coordinate space.
Unfortunately, we were not able to perform explicitly the integration of the
Lipatov's eigenfunctions with impact factors and without it the Mellin
representation  of the DVCS amplitude is useless for practical
applications.

The amplitude in the next-to-next-to-leading order is:
\begin{eqnarray}
\lefteqn{V={g^4s\over\pi}(\sum e_q^2)
\left(6\alpha_s\ln {1\over x}\right)^2
\int {d^2k_{\perp}\over 4\pi^2}{d^2k'_{\perp}\over 4\pi^2}
{d^2k''_{\perp}\over 4\pi^2}{I(k_{\perp},r_{\perp})\over k_{\perp}^2(r+k)_{\perp}^2}\label{fla39}}\\
&&  K(k_{\perp},k'_{\perp},r_{\perp}) 
{1\over (k'_{\perp})^{2}(r+k')_{\perp}^2}K(k'_{\perp},k''_{\perp},r_{\perp})
{1\over (k''_{\perp})^2(r+k'')_{\perp}^2}I_N(k''_{\perp},r_{\perp}).
\nonumber
\end{eqnarray}
Once again, if we use the fact that the integral over $k'_{\perp}$
converges at
$|k'_{\perp}|\gg M$  we can approximate the nucleon impact factor
by Eq.~(\ref{fla32}), and obtain:
\begin{eqnarray}
&&\int {d^2k'_{\perp} \over 4\pi^2} \int {d^2k''_{\perp} \over 4\pi^2} 
K(k_{\perp},k'_{\perp},r_{\perp}) 
{1\over (k'_{\perp})^{2}(r+k')_{\perp}^2}K(k'_{\perp},k''_{\perp},r_{\perp})
{I_N(k''_{\perp},r_{\perp})\over (k''_{\perp})^2(r+k'')_{\perp}^2}
\nonumber\\
&&
={1\over 4\pi} F_1^{p+n}(t)
\int {d^2k'_{\perp} \over 4\pi^2}
{K(k_{\perp},k'_{\perp},r_{\perp})\over (k')_{\perp}^{2} (r+k')_{\perp}^2}
\left(\ln {(k'_{\perp})^2\over r_{\perp}^2}+
\ln {(k'+r)_{\perp}^2\over r_{\perp}^2}\right)\nonumber\\
&&
={1\over 16\pi^2}F_1^{p+n}(t)
\left(\ln^2 {k_{\perp}^2\over r_{\perp}^2}+
\ln^2 {(k+r)_{\perp}^2\over r_{\perp}^2}\right)
\label{fla40}.
\end{eqnarray}
The resulting integration over $k_{\perp}$ yields:
\begin{eqnarray}
\lefteqn{ V=}\nonumber\\
&&{9\over x}\left({\alpha_s\over\pi}\right)^4(\sum e_q^2)
F_1^{p+n}(t)\ln^2x
\Bigg[  (\epsilon,\epsilon')_{\perp}\Big({1\over 24} \ln^4{Q^2\over |t|}+
\ln^2{Q^2\over |t|}
-2\ln{Q^2\over |t|}+\nonumber\\
&&2(\zeta(3)-1)+1.46 \Big)+
\left({2\over r_{\perp}^2}(\epsilon,r)_{\perp}(\epsilon',r)_{\perp} -
(\epsilon,\epsilon')_{\perp}\right)
\Bigg]
\label{fla41}.
\end{eqnarray}
As it was mentioned, we are unable yet to obtain the explicit expressions
for the amplitude in higher orders in perturbation theory.
It turns out, however, that for HERA energies the achieved accuracy is
reasonably good.
The estimation of the next term gives $\sim$ 30\% of the answer at not
too low $x$ (see the discussion in next section).

In the leading logarithmic approximation it is impossible to
distinguish between $\alpha_s(Q)$ and $\alpha_s(\sqrt{|t|})$ -- to this end one
needs to
use the NLO BFKL approximation\cite{nlobfkl} (see also \cite{cia})
which is beyond the scope
of this paper.

The final result for the DVCS amplitude with transversely polarized photons is:
\begin{eqnarray}
\lefteqn{ V=}\label{fla42}\\
&&{2\over x}\left({\alpha_s(Q)\over\pi}\right)^2(\sum_q e_q^2)
F_1^{p+n}(t)\Bigg [
(\epsilon,\epsilon')_{\perp} v
+\left({2\over r_{\perp}^2}(\epsilon,r)_{\perp}(\epsilon',r)_{\perp}-
(\epsilon,\epsilon')_{\perp}\right)v'
\Bigg ],
\nonumber
\end{eqnarray}
where
\begin{eqnarray}
&&v(x,Q^2/t)=\left({1\over 2}\ln^2{Q^2\over |t|}+ 2\right)+{3\alpha_s(Q)\over \pi}
\ln{1\over x}\left({1\over 6} \ln^3{Q^2\over |t|} +2\ln{Q^2\over
|t|}-2+\zeta(3)\right)\nonumber\\
&&~~~~+{1\over 2}\left({3\alpha_s(Q)\over \pi}\ln {1\over
x}\right)^2\left( {1\over 24} \ln^4{Q^2\over |t|}+\ln^2{Q^2\over |t|}
+2(\zeta(3)\!\!-\!\!1)\ln{Q^2\over |t|}+1.46\right),\label{fla43}\\
&&v'(x,Q^2/t)=
1+{3\alpha_s(Q)\over \pi}\ln {1\over x}+
{1\over 2}\left({3\alpha_s(Q)\over \pi}\ln {1\over x}\right)^2
\label{fla44}.
\end{eqnarray}

Note that the spin-dependent part $\sim v'$ does not contain any
$\ln{Q^2/|t|}$
 and is, hence, much smaller than the spin-independent part $\sim v$. For the
longitudinal polarization (\ref{fla13}) the amplitude is twist-suppressed as
$\simeq\sqrt {{|t|/ Q^2}}$ so we
have not calculated any terms beyond Eq.~(\ref{fla18}). In the numerical
analysis carried out in the next sections only the spin-independent
part of the amplitude is kept:
\begin{eqnarray}
V_{\perp}\equiv{1\over 4}\sum \epsilon_{\perp}\epsilon'_{\perp} V=
{2\over x}\left({\alpha_s(Q)\over\pi}\right)^2(\sum_q e_q^2)
F_1^{p+n}(t)v(x,Q^2,t)
\label{fla45}.
\end{eqnarray}

The expressions above give us the imaginary part of the DVCS amplitude.
For the calculation of the
DVCS cross section one needs to know also the real
part $\Re H$ of this amplitude,
which can be estimated via the dispersion relation.
For the positive-signature amplitude $H_{\perp}$
($\equiv{1\over 4}\sum \epsilon_{\perp}\epsilon'_{\perp}H$) we get \cite{bronzan}
(see also \cite{strikfurt1})
\begin{equation}
\Re H_{\perp}(s)={\pi\over 2}\tan\left(s{d\over ds}\right)
\Im H_{\perp}(s)
\label{fla46},
\end{equation}
which amounts to the substitution
\begin{equation}
\ln s \rightarrow {1\over 2}\big(\ln (-s-i\varepsilon)+\ln s\big)
\label{fla47}
\end{equation}
in our amplitude (\ref{fla45}). Thus, the real part is:
\begin{equation}
R\equiv {1\over \pi}\Re H_{\perp}={2\over x}\left({\alpha_s\over\pi}\right)^2
(\sum_q e_q^2)
F_1^{p+n}(t)r(x,Q^2,t),
\end{equation}
\begin{eqnarray}
&&r(x,Q^2,t)={\pi\over 2}\Bigg[{3\alpha_s\over\pi}\left({1\over 6} \ln^3{Q^2\over
|t|} +2\ln{Q^2\over |t|}-2+\zeta(3)\right)+
\nonumber\\
&&\left({3\alpha_s\over \pi}\right)^2\ln {1\over
x}\left( {1\over 24} \ln^4{Q^2\over |t|}+\ln^2{Q^2\over |t|}
+2(\zeta(3)-1)\ln{Q^2\over |t|}+1.46\right)\Bigg]
\label{fla48}.
\end{eqnarray}

\section{DVCS cross section}
It is instructive to compare the DVCS amplitude $V$ given by
Eq.~(\ref{fla3}) with
the corresponding amplitude for the forward $\gamma^*$ scattering
\begin{equation}
H=-i\epsilon_{\nu}\epsilon'_{\mu}
\int dz e^{-iq\cdot z}\langle p| T\{j^{\mu}(z)j^{\nu}(0)\} |p\rangle
\label{fla49}.
\end{equation}
The imaginary part of this amplitude is the total cross section for
deep inelastic scattering (DIS)
\begin{eqnarray}
\lefteqn{{1\over \pi}\Im H =W=}\nonumber\\
&\epsilon_{\nu}\epsilon'_{\mu}
\Bigg[\left({q_{\mu}q_{\nu}\over q^2}-g_{\mu\nu}\right)F_1(x,Q^2)+
{1\over pq}\left(p_{\mu}-q_{\mu}{pq\over q^2}\right)
\left(p_{\nu}-q_{\nu}{pq\over q^2}\right)F_2(x,Q^2)\Bigg].
\label{fla50}
\end{eqnarray}

For example,
$W$ averaged over the transverse polarizations of the photons is:
\begin{equation}
W_{\perp}\stackrel{\rm def}{\equiv}
{1\over 4}\sum \epsilon_{\perp}\epsilon'_{\perp}W=F_1(x, Q^2)={1\over 2x}
F_2(x, Q^2),
\label{fla51}
\end{equation}
(at the leading twist level we have the Callan-Gross relation
$F_2=2xF_1$).
We will compare the imaginary part of the
DVCS amplitude $V_{\perp}$ given by Eq.~(\ref{fla45})
to the result for $W_{\perp}$ calculated with the same accuracy.
(The notation $W_{\perp}(x)$ is used
rather than $F_1(x)$ in order to avoid confusion with $F_1(t)$).

\begin{figure}[htb]
\mbox{
\epsfxsize=6cm
\epsfysize=5cm
\epsffile{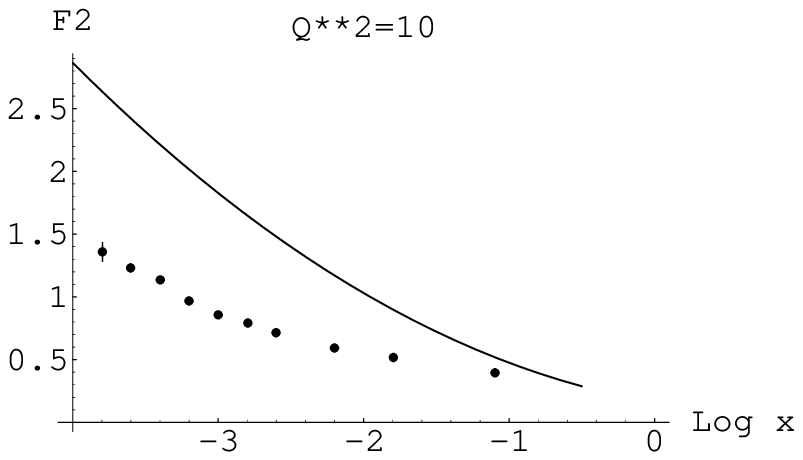}}
\hspace{1cm}
\mbox{
\epsfxsize=6cm
\epsfysize=5cm
\epsffile{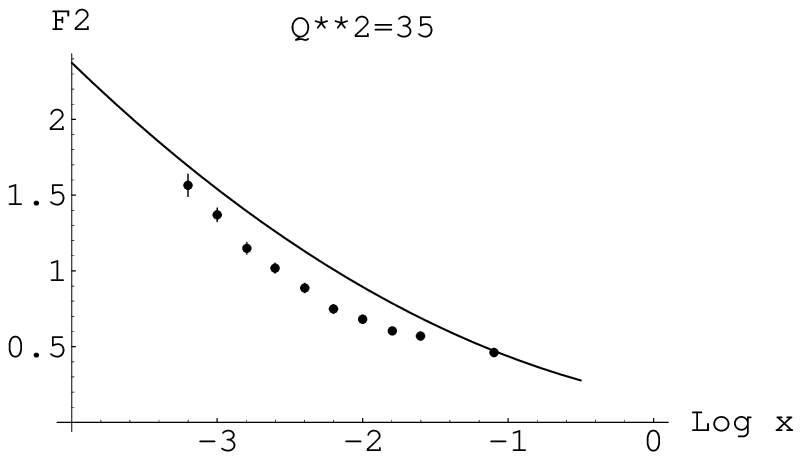}}
{\caption{\label{fig7} $F_2(x)$ from Eq.~(\ref{fla56}) versus experimental
data at $Q^2=10$GeV$^2$ and $Q^2=35$GeV$^2$. }}
\end{figure}

  Similarly to the DVCS case, the DIS amplitude has the form
(cf. Eqs.~(\ref{fla16}),(\ref{fla34}), and (\ref{fla39})):
\begin{eqnarray}\label{fla52}
W_{\perp}&=&
{2g^2s\over \pi}\left(\sum e_q^2\right)
\int {d^2k_{\perp}\over 4\pi^2} {1\over k_{\perp}^4}
I_{\perp}(k_{\perp},0)\nonumber\\ 
&&\Bigg[1+
{3g^2\over 8\pi^3}\ln {1\over x}\int d^2k'_{\perp}
 K(k_{\perp},k'_{\perp},0){1\over (k'_{\perp})^2}I_N(k'_{\perp},0)
+\nonumber\\
&&{9g^4\over 128\pi^6}\ln^2 {1\over x}\int d^2k'_{\perp}{K(k_{\perp},k'_{\perp},0)\over (k'_{\perp})^2}\int d^2k''_{\perp}
{K(k'_{\perp},k''_{\perp},0)\over (k''_{\perp})^2}
I_N(k''_{\perp},0)\Bigg],
\end{eqnarray}
where $I_{\perp}(k_{\perp},0)$ is the virtual photon impact factor
averaged over
the transverse polarizations \cite{mes}:
\begin{equation}
I_{\perp}(k_{\perp},0)=
   {1\over 2}\int _{0}^{1} \frac {d\alpha}{2\pi }
   \int _{0}^{1} \frac {d\alpha'}{2\pi }
    {k_{\perp}^2(1-2\alpha\bar{\alpha})(1-2\alpha'\bar{\alpha'})\over
      \alpha'\bar{\alpha'}(k_{\perp}^{2}+\Omega^2)}.
   \label{fla53}
\end{equation}
The nucleon impact factor $I_N(k'_{\perp},0)$ cannot be calculated
in perturbation theory since it is determined by the
large-scale nucleon dynamics. However, we know the
asymptotics at large $k_{\perp}\gg M$
\begin{equation}
I_N(k_{\perp},0)\stackrel{k_{\perp}^2\gg M^2}{=}
F_1^{p+n}(0)=1.
\label{fla54}
\end{equation}
Also, at $I_N(k_{\perp},0)\rightarrow 0$ at $k\rightarrow 0$ due to the
gauge invariance. It seems reasonable to model this impact factor by
the simple formula:
\begin{equation}\label{fla55}
I_N(k_{\perp},0)={k_{\perp}^2\over k_{\perp}^2+M^2},
\end{equation}
which has the correct behavior both at large and small $k_{\perp}$.
With this model, the DIS amplitude (\ref{fla52}) takes the form:
\begin{eqnarray}\label{fla56}
W_{\perp}={F_2\over 2x}={4\over 3x}\left(\alpha_s(Q)\over \pi\right)^2(\sum_q e_q^2)
\Bigg[\left({1\over 2}\ln^2{Q^2\over M^2}+{7\over 6}\ln{Q^2\over M^2}+
{77\over 18}\right)+&&\\
{3\alpha_s\over \pi}\ln {1\over x}\left({1\over 6}
 \ln^3{Q^2\over M^2}+{7\over 12}\ln^2{Q^2\over M^2}+
 {77\over 18}\ln{Q^2\over M^2}+{131\over 27}+2\zeta(3)\right)+&&\nonumber\\
 {1\over 2}
 \left({3\alpha_s\over \pi}\ln {1\over x}\right)^2
  \left({1\over 24}\ln^4{Q^2\over M^2}+{7\over 36}\ln^3{Q^2\over M^2}+
 {77\over 36}\ln^2{Q^2\over M^2 }+\right. &&\nonumber\\
\left. \Big({131\over 27}+4\zeta(3)\Big)
 \ln{Q^2\over M^2}+
 {1396\over 81}-{\pi^4\over 15} +{14\over 3}\zeta(3)
  \right)
\Bigg].&&\nonumber
\end{eqnarray}
Note that the coefficients in front of leading logs of ${Q^2}$, determined by
the anomalous dimensions of twist-2 operators, coincide up to the factor
$2/3$. The graph of the model (\ref{fla56}) versus the
experimental data is presented in Fig. \ref{fig7} for $Q^2=10$GeV$^2$
and  $Q^2=35$GeV$^2$ (we take
$\sum e_q^2={10\over 9}$).

%=================================
%==================================  
For DIS it is possible to write down the total BFKL sum as a
Mellin integral
and  unlike DVCS, the integrals of impact factors with the BFKL
eigenfunctions $(k_{\perp}^2)^{-{1\over 2}+i\nu}$ can be calculated explicitly.
Eqs.~(\ref{fla56}) and (\ref{fla57}) correspond to the expansion of this
explicit expression in powers of $\alpha_s\ln x$.

For example, the next term in
BFKL series (\ref{fla56}) has the form:
\begin{eqnarray}\label{fla57}
&&
{4\over 3x}\left(\alpha_s(Q)\over \pi\right)^2(\sum_{\rm flavors}e_q^2){1\over 6}\left({3\alpha_s\over \pi}\ln {1\over x}\right)^3 \\ &&
 \left({1\over 120}\ln^5{Q^2\over M^2}+
 {7\over 144}\ln^4{Q^2\over M^2}+{77\over 108}\ln^3{Q^2\over M^2}+
 ({131\over 54}+3\zeta(3))\ln^2{Q^2\over M^2 }+\right.\nonumber\\
 &&\left.+({1396\over 81}-
 {\pi^4\over 15}+7\zeta(3))
 \ln{Q^2\over M^2}+
 {4736\over 243}-{7\pi^4\over 90} +{77\over 3}\zeta(3)+6\zeta(5)
  \right). \nonumber
\end{eqnarray}

%=================================
\begin{figure}[htb]
\vspace{0cm}
\hspace{0cm}
\mbox{
\epsfxsize=6cm
\epsfysize=5cm
\hspace{0cm}
\epsffile{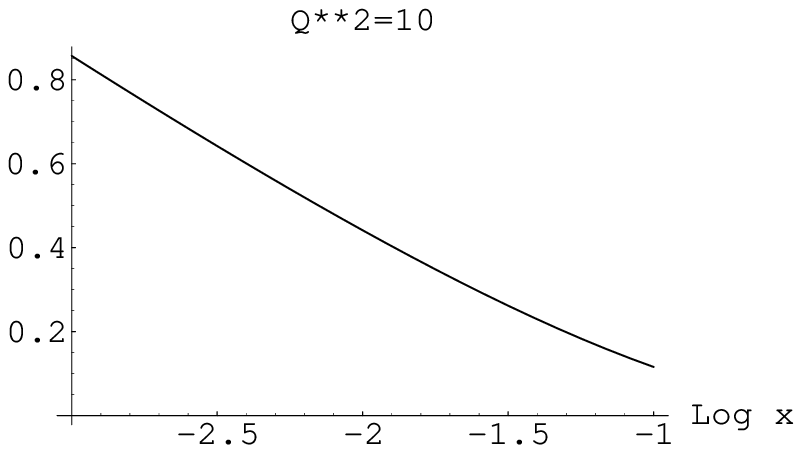}}
\hspace{1cm}
\mbox{
\epsfxsize=6cm
\epsfysize=5cm
\hspace{0cm}
\epsffile{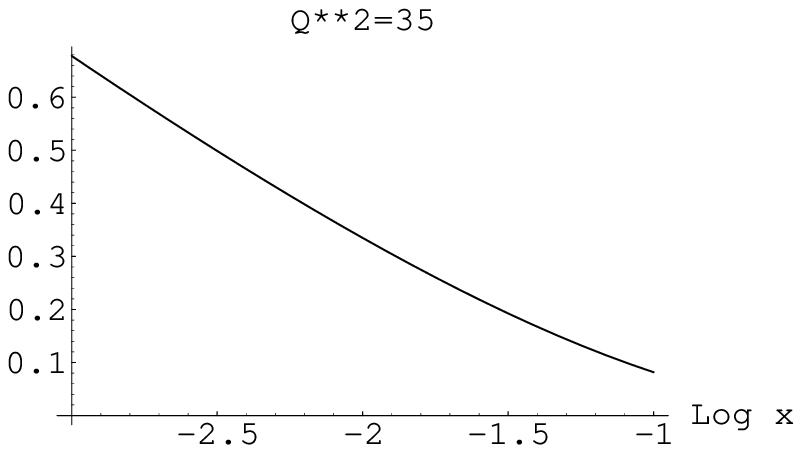}}
\vspace{0cm}
{\caption{\label{fig8} The ratio of the $\ln^3x$ term (\ref{fla57})
to Eq.~(\ref{fla56})
at $Q^2=10$GeV$^2$ and $Q^2=35$GeV$^2$.}}
\end{figure}
%==================================

The ratio of this
$\left(\alpha_s\ln x\right)^3$ term to the sum of the first
three ones (\ref{fla56}) is presented in Fig. \ref{fig8} for $Q^2=10$GeV$^2$
and $Q^2=35$GeV$^2$.
From these graphs we see that the sum of the first three terms gives 
the reliable
estimate of the DIS amplitude at not too low $x$. It is
expected that the same will also be true for the DVCS amplitude.

At very small $x\sim 10^{-3} - 10^{-5}$ the full BFKL result for
$F_2$ in our model
 is growing more rapidly than Fig. \ref{fig7}.
On the other hand if one takes into account the NLO BFKL
corrections \cite{nlobfkl, cia} the result for $F_2$ at very small $x$
 goes well under the experimental points. This indicates,
 that at such  $x$ we need to unitarize
 the BFKL pomeron, which is currently an unsolved problem.
 (The best hope is to find the effective action for
 the BFKL pomeron (see e.g. \cite{lipac, efek})).
On the contrary, at ``intermediate'' $x\sim 0.1 - 0.001$, we
see from Fig. \ref{fig7}
that, since
the corrections almost cancel each other, it makes sense to
take into account only a few first terms in BFKL series.

It is instructive to compare  the t-dependence of our DVCS amplitude
(\ref{fla43}) with the model used in the paper \cite{strikfurt1}
\begin{eqnarray}
V_1(x,t,Q^2)&=&{1\over R}F_1(x,Q^2)e^{bt/2}, \label{fla58}\\
V_2(x,t,Q^2)&=&{1\over R}F_1(x,Q^2){1\over \big(1+{|t|\over \alpha_t}\big)^2},
\label{fla59}
\end{eqnarray}
where $R\simeq 0.5$ for our energies. (Literally, the model used in 
ref. \cite{strikfurt1} corresponds to $V_1$ but it is more natural to
approximate the $t$ - dependence by the dipole formula \cite{private}).

\begin{figure}[htb]
\mbox{
\epsfxsize=6cm
\epsfysize=5cm
\epsffile{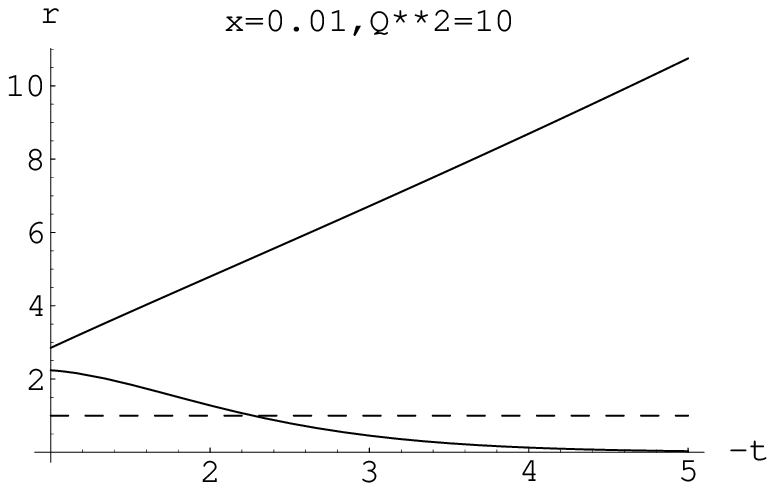}}
\hspace{1cm}
\mbox{
\epsfxsize=6cm
\epsfysize=5cm
\epsffile{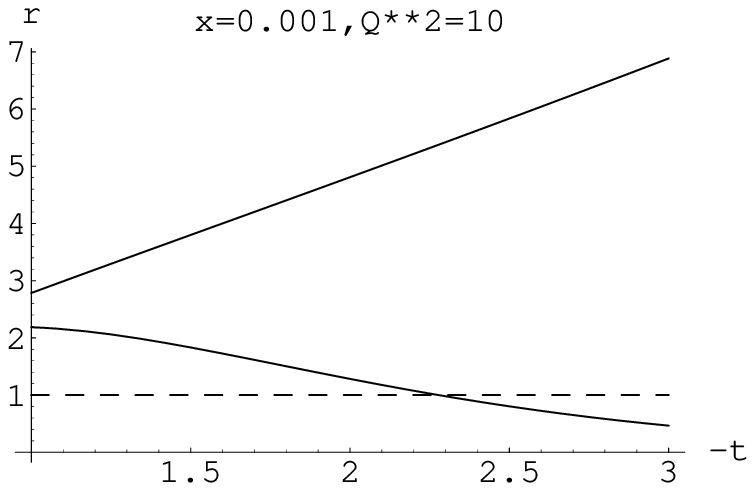}}\\
\mbox{
\epsfxsize=6cm
\epsfysize=5cm
\epsffile{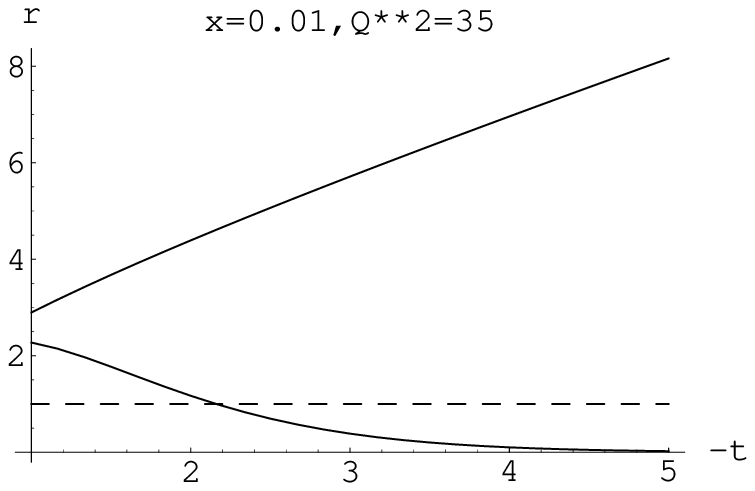}}
\hspace{1cm}
\mbox{
\epsfxsize=6cm
\epsfysize=5cm
\epsffile{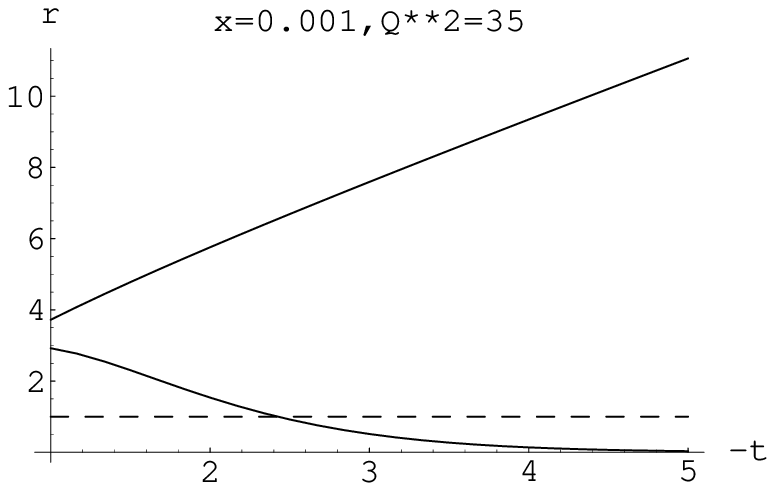}}
{\caption{\label{fig9} The ratio $V_1/V_{\perp}$ (lower curve) and
 $V_2/V_{\perp}$ (upper curve).}} 
\end{figure}

The comparison is shown in Fig.
\ref{fig9} for $Q^2=10$GeV$^2$, $Q^2=35$GeV$^2$ and $x$=0.01, $x$=0.001.

In order to estimate the cross section for DVCS at HERA kinematics
($Q^2>6$GeV$^2$ and $x<10^{-2}$)  we will use formulae from Ref.
\cite{strikfurt1} (see also Ref. \cite{dil}) with the trivial substitution
${1\over 2x}F_2(x)R^{-1}e^{bt/2}\rightarrow V_{\perp}(x,Q^2,t)$.
The expressions for
the DVCS cross section, the quantum electrodynamics (QED) Compton (Bethe-Heitler) cross section,
and the interference term have the form ($\bar{y}\equiv 1-y$):
 \begin{eqnarray}
{d\sigma^{\rm DVCS}\over dxdydtd\phi_r}&=& \pi\alpha^3x{1+\bar{y}^2\over Q^4y}
(V_{\perp}^2(x,Q^2,t)+R_{\perp}^2(x,Q^2,t)), \label{fla60} \\
{d\sigma^{\rm QEDC}\over dxdydtd\phi_r}&=& {\alpha^3\over\pi x}
{y(1+\bar{y}^2)\over |t|Q^2\bar{y}} \left((F^p_1(t))^2+
{|t|\over 4M^2}(F^p_2(t))^2\right), \label{fla61}\\
{d\sigma^{\rm INT}\over dxdydtd\phi_r}&=& \mp 2\alpha^3
{(1+\bar{y}^2)\over Q^3\sqrt{\bar{y}|t|}} R_{\perp}(x,Q^2,t)
F^p_1(t)\cos\phi_r. \label{fla62}
\end{eqnarray}
%=================================
The expression for the interference term from ref. \cite{strikfurt1}
is corrected by factor 2 \cite{private}, \cite{belitsky}.
Here $y=1-{E'\over E}$ ($E$ and $E'$ are the incident and scattered
electron energies, respectively, as defined in the proton rest frame)
and $\phi_r=\phi_e+\phi_N$ where $\phi_N$ is the azimuthal angle
between the plane defined by $\gamma^*$ and the final state proton and the
$x~-~z$ plane and $\phi_e$ is the azimuthal angle between the plane defined by
the initial and final state electron and $x~-~z$ plane 
(see Ref. \cite{strikfurt1}). As
mentioned above, we
approximate the Dirac and Pauli form factors of the proton by
the dipole formulas (\ref{fla29}).
%=================================
\begin{figure}[htb]
\mbox{
\epsfxsize=6cm
\epsfysize=5cm
\epsffile{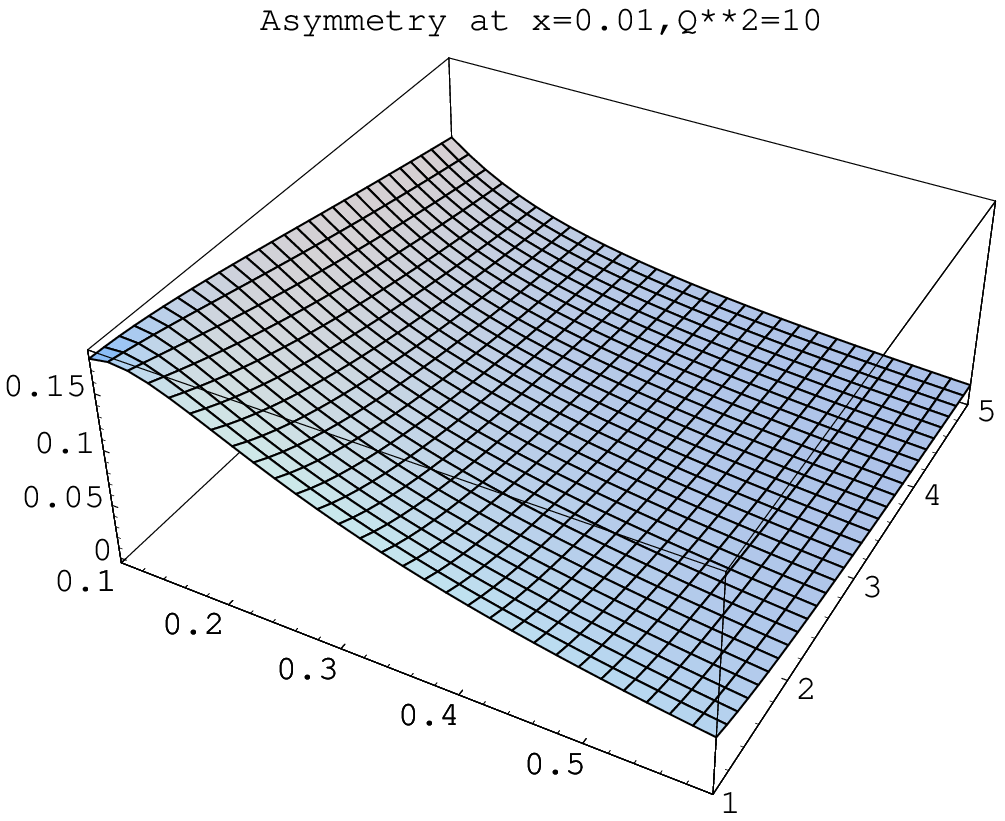}}
\hspace{1cm}
\mbox{
\epsfxsize=6cm
\epsfysize=5cm
\epsffile{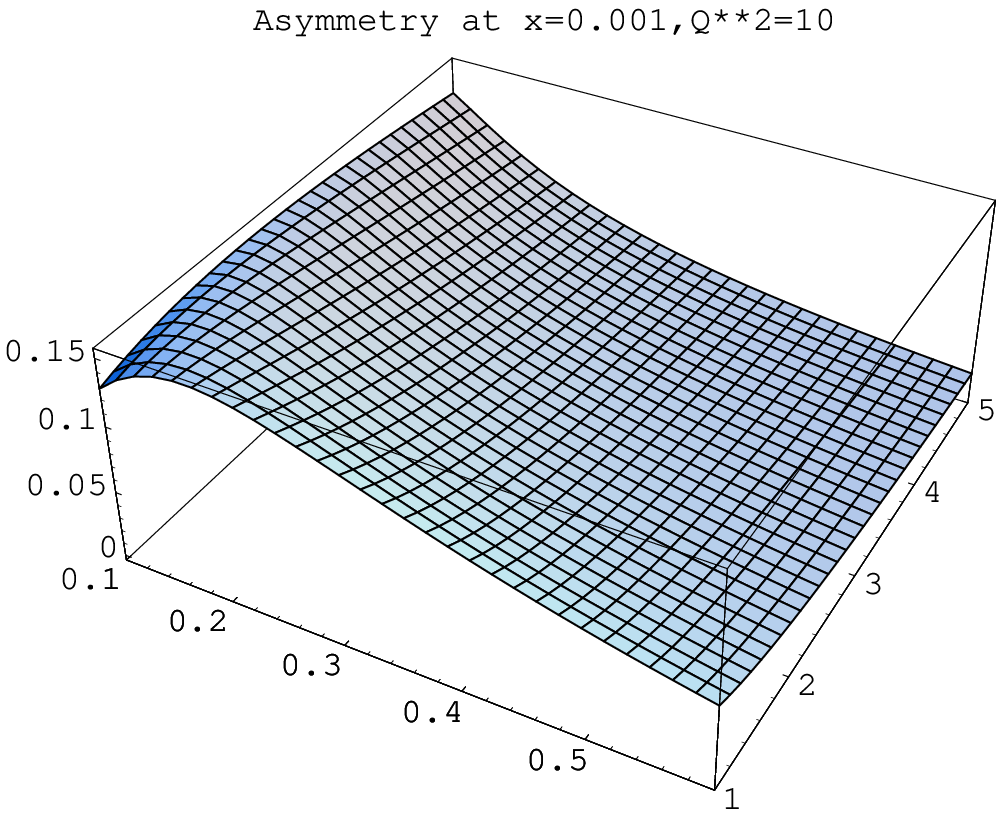}}\\
\mbox{
\epsfxsize=6cm
\epsfysize=5cm
\epsffile{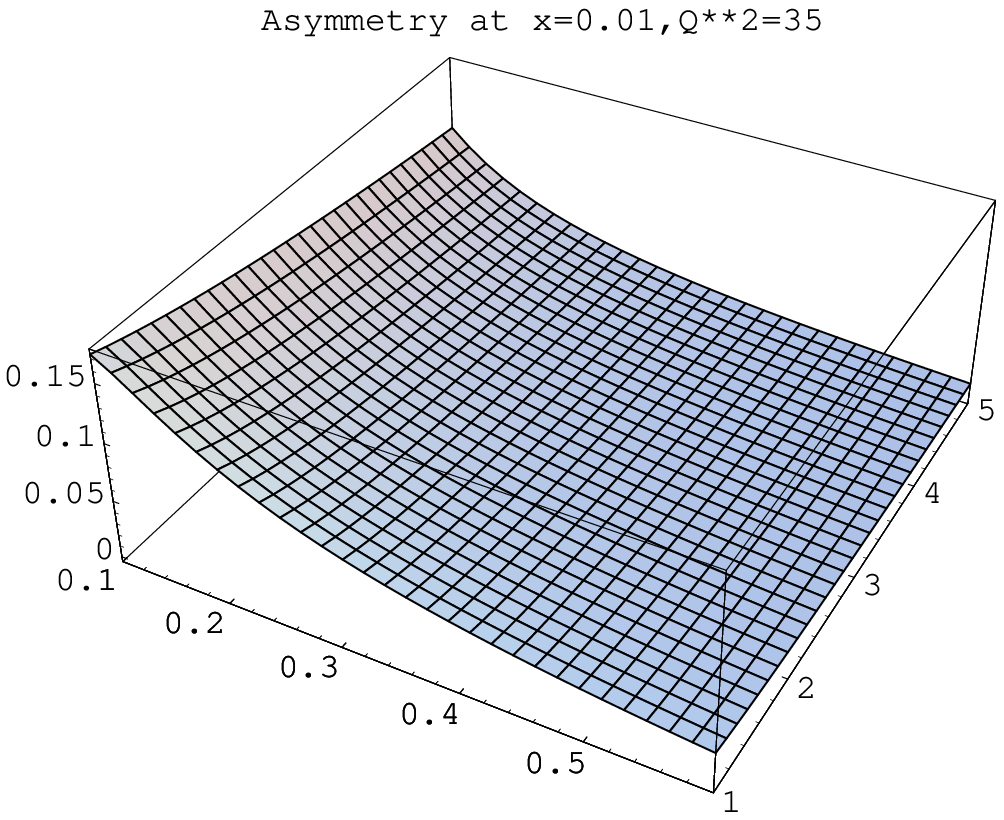}}
\hspace{1cm}
\mbox{
\epsfxsize=6cm
\epsfysize=5cm
\epsffile{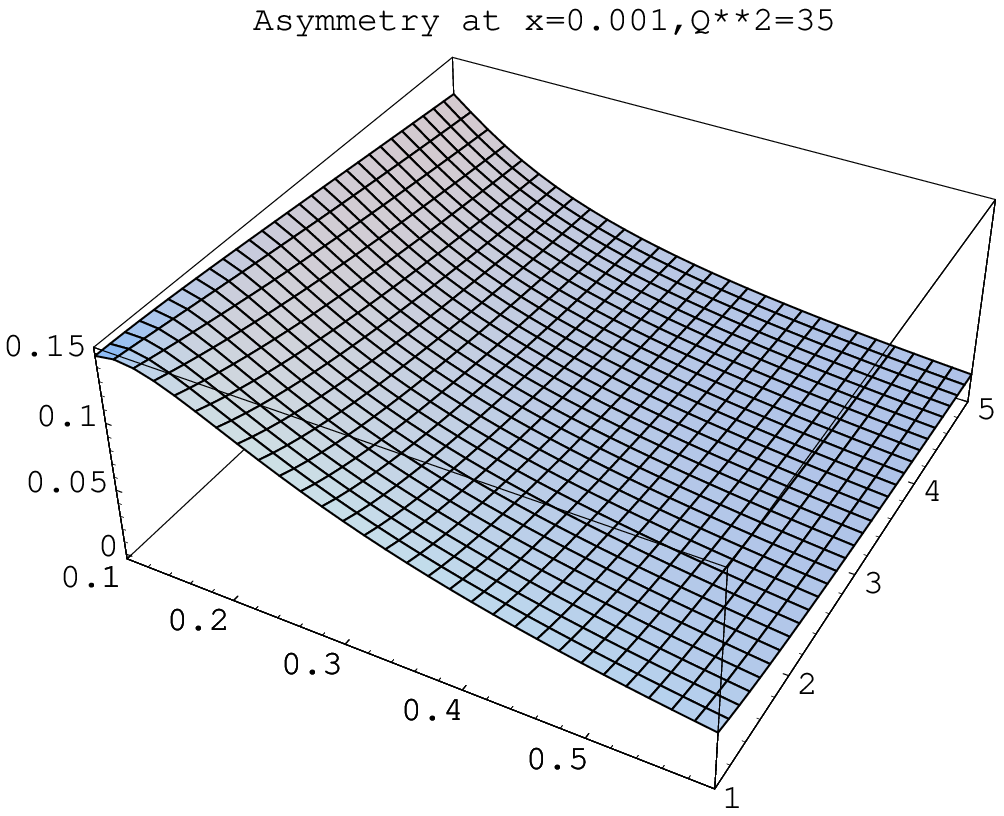}}
{\caption{\label{fig10} Asymmetry versus $y=0.1 - 0.6$ and $|t|=1 - 5$
GeV$^2$. }}
\end{figure}
%==================================

At first let us discuss the relative weight of the above cross sections.
We start with the asymmetry defined in ref. \cite{strikfurt2}
\begin{equation}
A={\int_{-\pi/2}^{\pi/2}d\phi_rd\sigma^{\rm DQI}-
\int_{\pi/2}^{3\pi/2}d\phi_rd\sigma^{\rm DQI}\over
\int_{0}^{2\pi}d\phi_rd\sigma^{\rm DQI}},
\label{fla63}
\end{equation}
where 
\begin{equation} d\sigma^{\rm DQI}\equiv d\sigma^{\rm DVCS}+
d\sigma^{\rm QEDC}+d\sigma^{\rm INT}.
\label{fla64}
\end{equation} 
The asymmetry shows the relative importance of the interference term,
which is proportional to the real part of the DVCS amplitude.
In our
approximation the asymmetry is  
\begin{equation}
A(y,t)=
{4y\sqrt{{Q^2\over |t|\bar{y}}}(\sum e_q^2)\left({\alpha_s\over\pi}\right)^2
\left(1+2.8{|t|\over4M^2}\right)r\over
4\pi^2(\sum e_q^2)^2(v^2+r^2)\left({\alpha_s\over\pi}\right)^4
\left(1+{|t|\over 4M^2}\right)+
 {y^2Q^2\over \bar{y}|t|}\left(1+7.84{|t|\over 4M^2}\right)}.
\label{fla65}
\end{equation}
The plots of asymmetry versus $y$ and $|t|$ are given by Fig. \ref{fig10}.
%=================================
\begin{figure}[htb]
\mbox{
\epsfxsize=6cm
\epsfysize=5cm
\epsffile{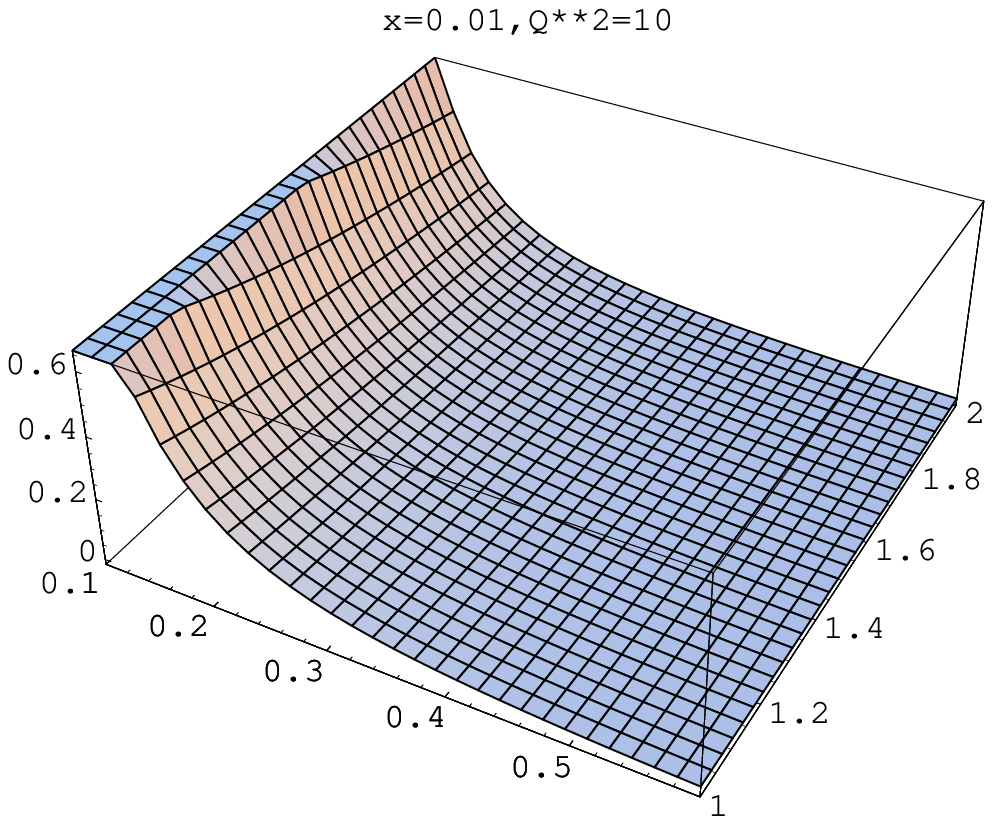}}
\hspace{1cm}
\mbox{
\epsfxsize=6cm
\epsfysize=5cm
\epsffile{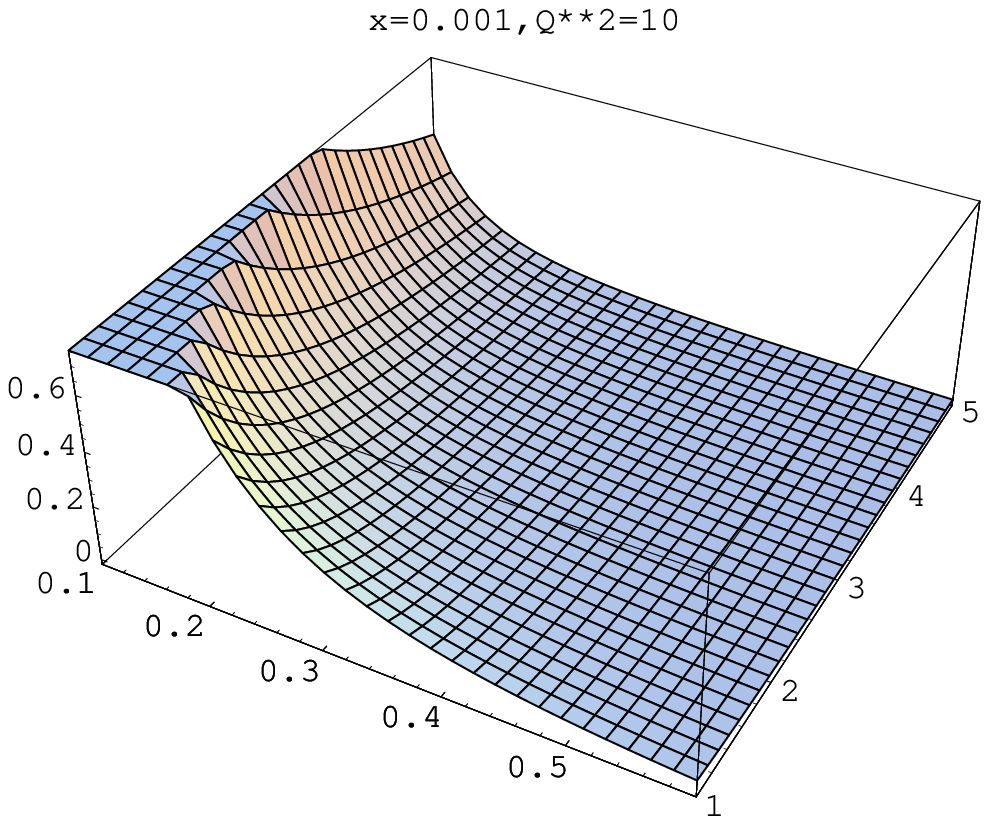}}\\
\mbox{
\epsfxsize=6cm
\epsfysize=5cm
\epsffile{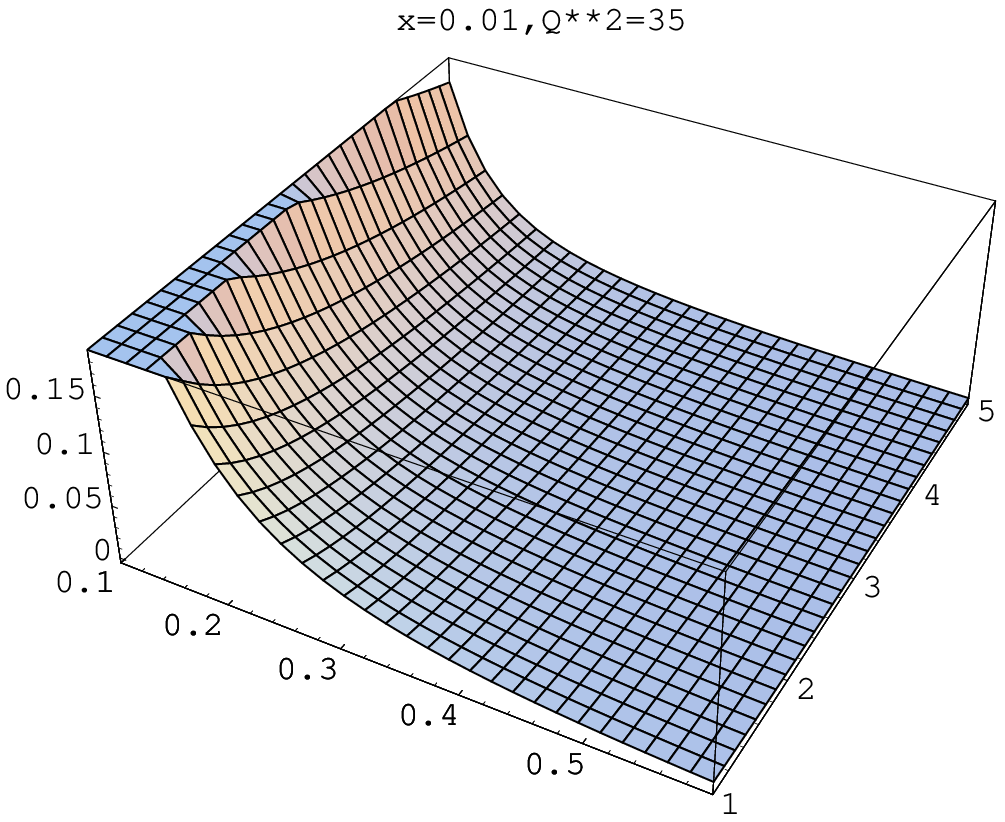}}
\hspace{1cm}
\mbox{
\epsfxsize=6cm
\epsfysize=5cm
\epsffile{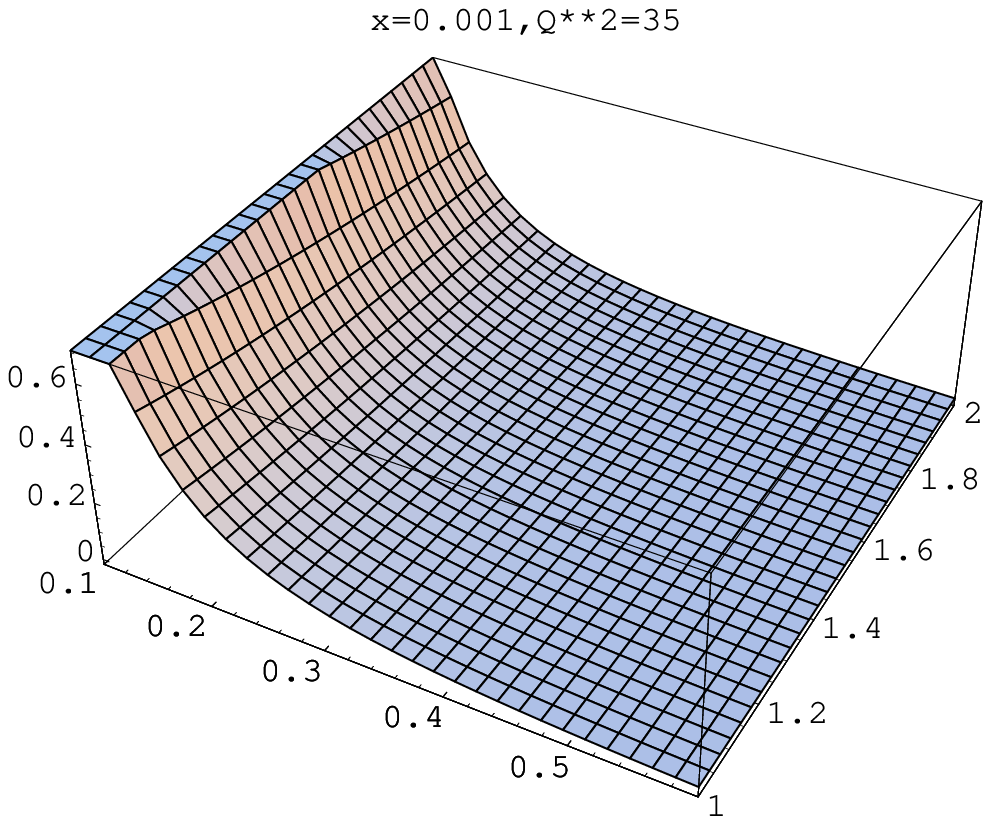}}
{\caption{\label{fig11} The ratio $D(x,Q^2/t)$ versus $y=0.1 - 0.6$
and $\mid t\mid=1 - 5$
GeV$^2$.}}
\end{figure}
%==================================

Second, we define the ratio of the DVCS and Bethe-Heitler cross
sections \cite{strikfurt1}
\begin{equation}
D(y,t)\equiv {d\sigma_{DVCS}\over d\sigma_{QEDC}}=
{4\pi^2(\sum e_q^2)^2(v^2+r^2)\left({\alpha_s\over\pi}\right)^4
\left(1+{|t|\over 4M^2}\right)\over
\frac{y^2Q^2}{\bar{y}|t|}\left(1+7.84{|t|\over 4M^2}\right)}.
\label{fla66}
\end{equation}
This ratio is presented on Fig. \ref{fig11}.

We see that there is a sharp dependence on $y$;
at $y>0.2$ the DVCS part is negligible in comparison
to Bethe-Heitler background whereas at $y<0.05$
the QEDC background is small in comparison to DVCS.
%=================================
\begin{figure}[htb]
\mbox{
\epsfxsize=6cm
\epsfysize=5cm
\epsffile{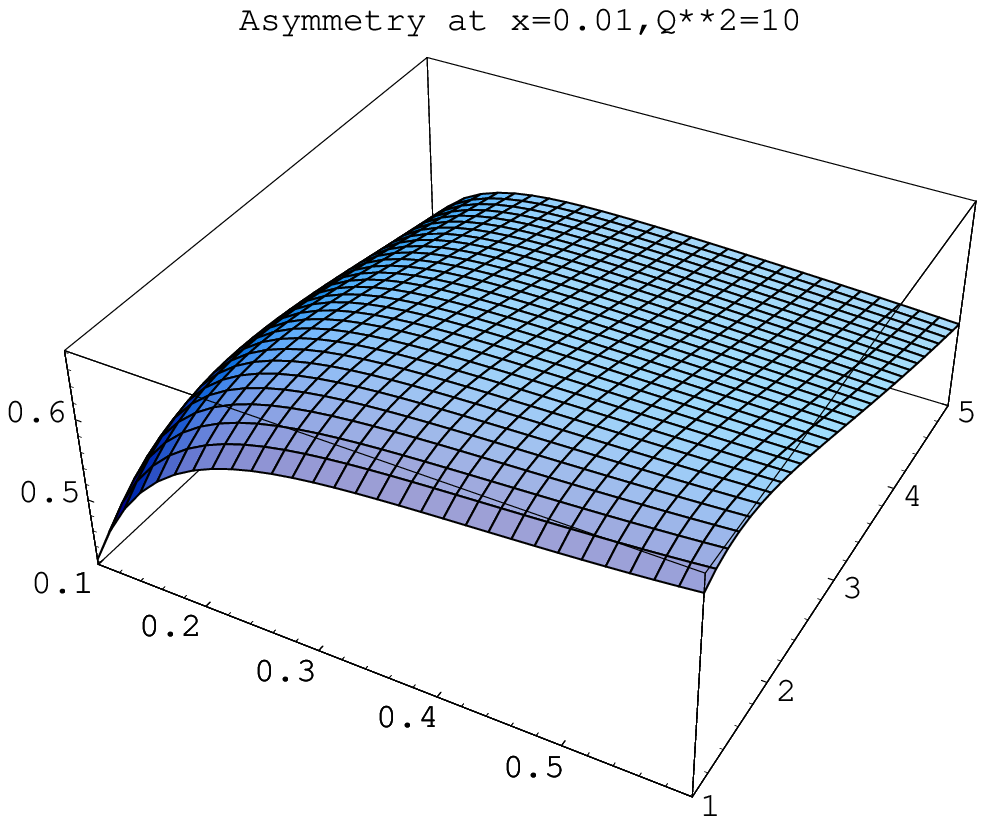}}
\hspace{1cm}
\mbox{
\epsfxsize=6cm
\epsfysize=5cm
\hspace{0cm}
\epsffile{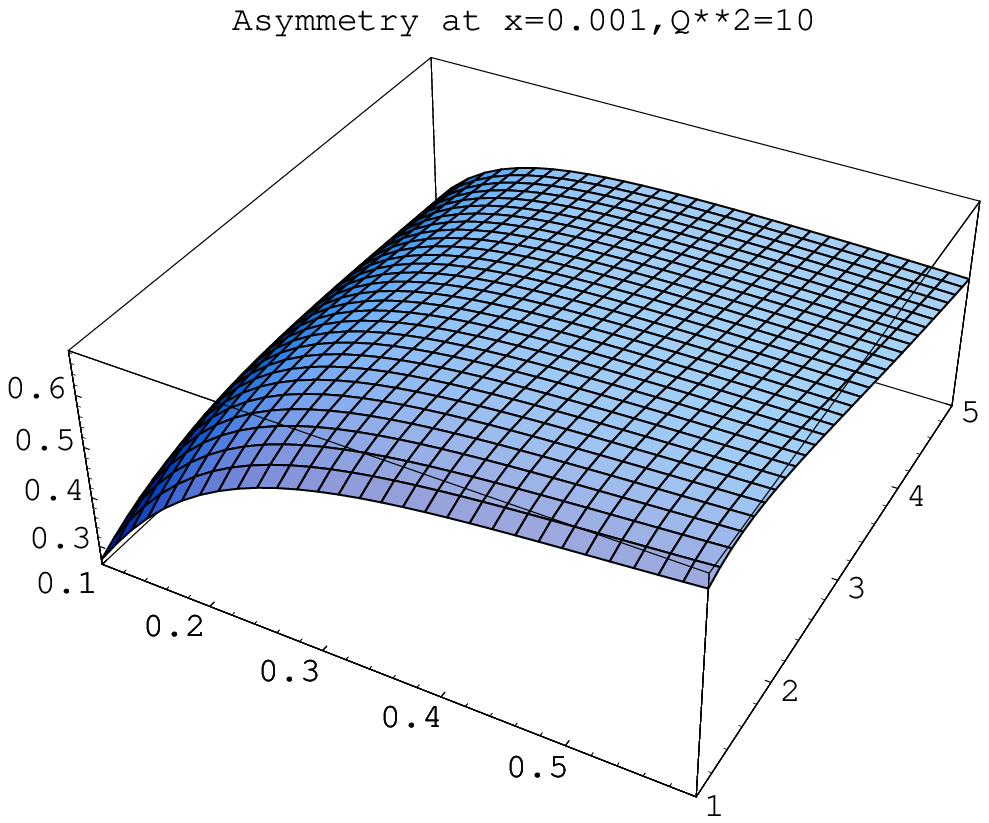}}\\
\mbox{
\epsfxsize=6cm
\epsfysize=5cm
\epsffile{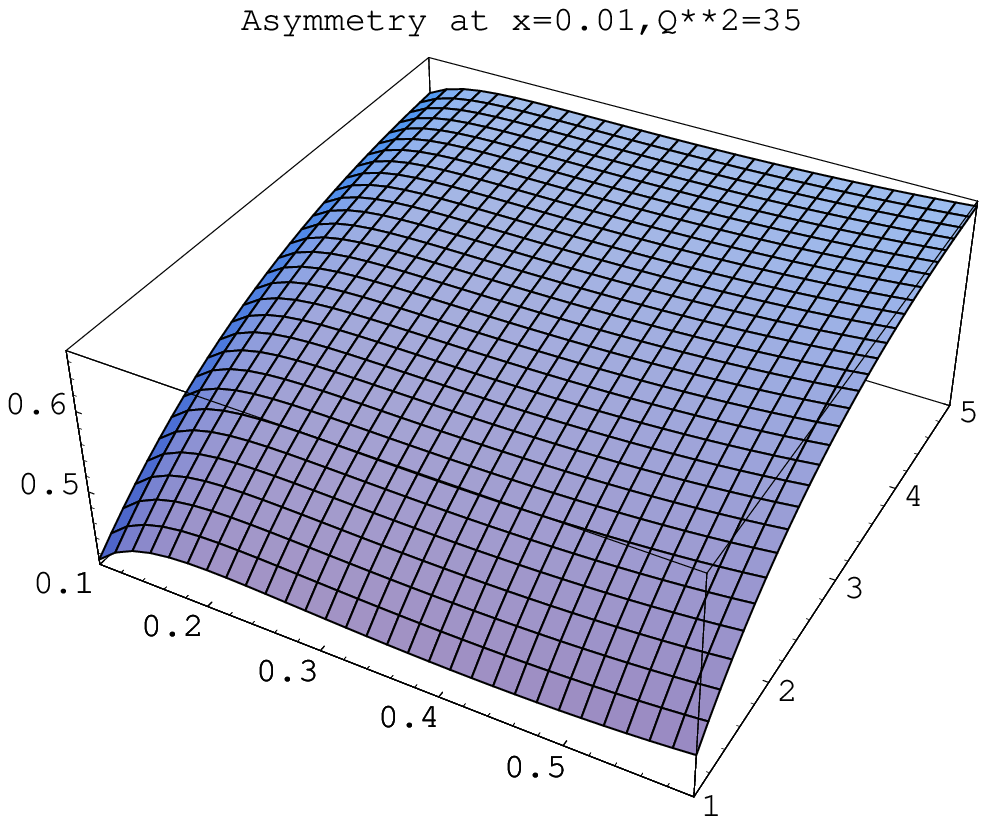}}
\hspace{1cm}
\mbox{
\epsfxsize=6cm
\epsfysize=5cm
\epsffile{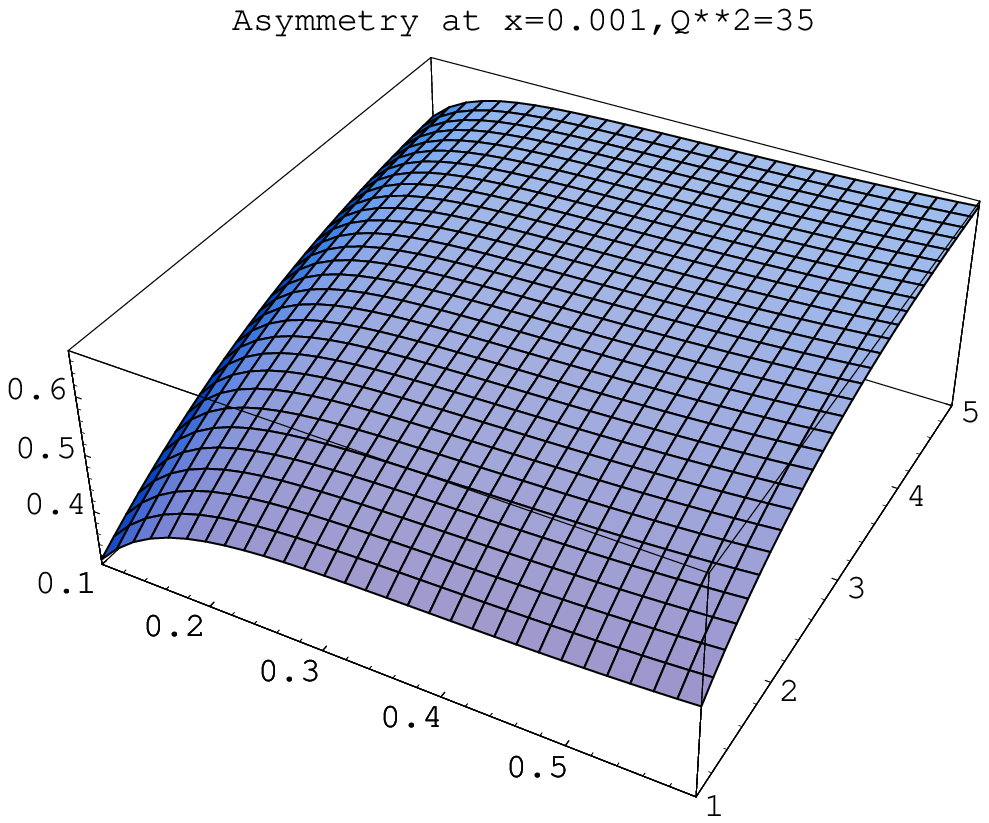}}
{\caption{\label{fig12} Asymmetry with the correction factor (\ref{fla68}). }}
\end{figure}
%==================================

 Finally let us estimate the relative weight of the DVCS signal (starting
 from $|t|=1$ GeV$^2$) as compared to the DIS
background. We define (cf. ref. \cite{strikfurt1})
\begin{eqnarray}
\lefteqn{R_{\gamma}=
{\sigma(\gamma^*+p\rightarrow \gamma+p)\over \sigma(\gamma^*+p\rightarrow
\gamma^*+p)}\simeq}\nonumber\\
&{4\pi\alpha\over Q^2F_2(x,Q^2)}\left({\alpha_s\over \pi}\right)^4
\left(\sum e_q^2\right)^2\int^{Q^2}_1 dt
\left(F_1^{p+n}(t)\right)^2(v^2(x,Q^2/t)+r^2(x,Q^2/t).)
\label{fla67}
\end{eqnarray}
At $Q^2=10$GeV$^2$ we find $R_\gamma=1.56\times 10^{-5}$ for $x=0.01$ and
$R_\gamma=2.36\times 10^{-5}$ for $x=0.001$, while for $Q^2=35$GeV$^2$
 we find $R_\gamma=0.62\times 10^{-5}$ for $x=0.01$ and
$R_\gamma=0.71\times 10^{-5}$ for $x=0.001$.

The expressions (\ref{fla60})-(\ref{fla62}) are correct if $Q^2\gg |t|$ up to
$O({|t|\over Q^2})$ accuracy with the notable exception of
the correction $O({\sqrt{{|t|\over Q}}})$ coming from the expansion of
electron propagator in the u-channel of the Bethe-Heitler
amplitude. As suggested in ref.  \cite{belitsky}, at intermediate $t$
one can keep the propagator in unexpanded form 
(and expand the rest of the amplitude, as we have done above).
This amounts to the replacement
\begin{equation}\label{fla68}
\bar{y}\rightarrow
\bar{y}\Bigg[(1+{|t|\over Q^2\bar{y}})(1+{|t|\bar{y}\over Q^2})-
2{(2-y)\over\sqrt{\bar{y}}}\sqrt{{|t|\over Q^2}}\cos\phi_r+
4{|t|\over Q^2}\cos^2\phi_r\Bigg]
\end{equation}
in the numerator in Eqs.~(\ref{fla61}) and (\ref{fla62}) (see ref. \cite{dil}).
%=================================
\begin{figure}[htb]
\mbox{
\epsfxsize=6cm
\epsfysize=5cm
\epsffile{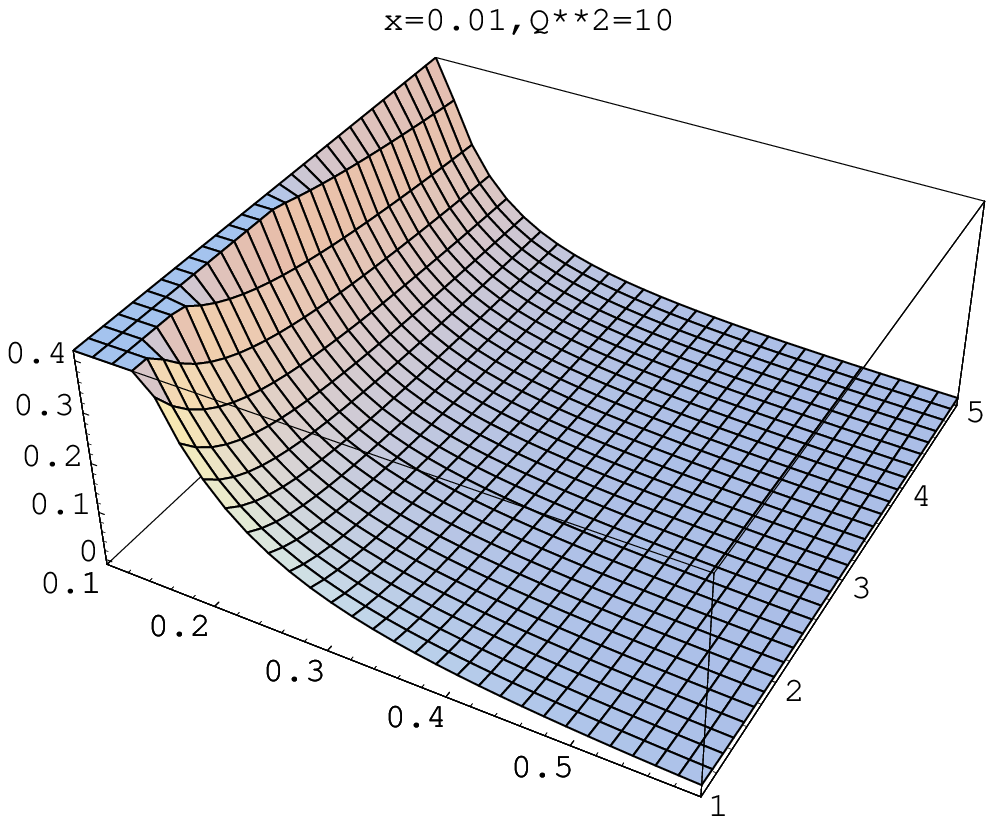}}
\hspace{1cm}
\mbox{
\epsfxsize=6cm
\epsfysize=5cm
\epsffile{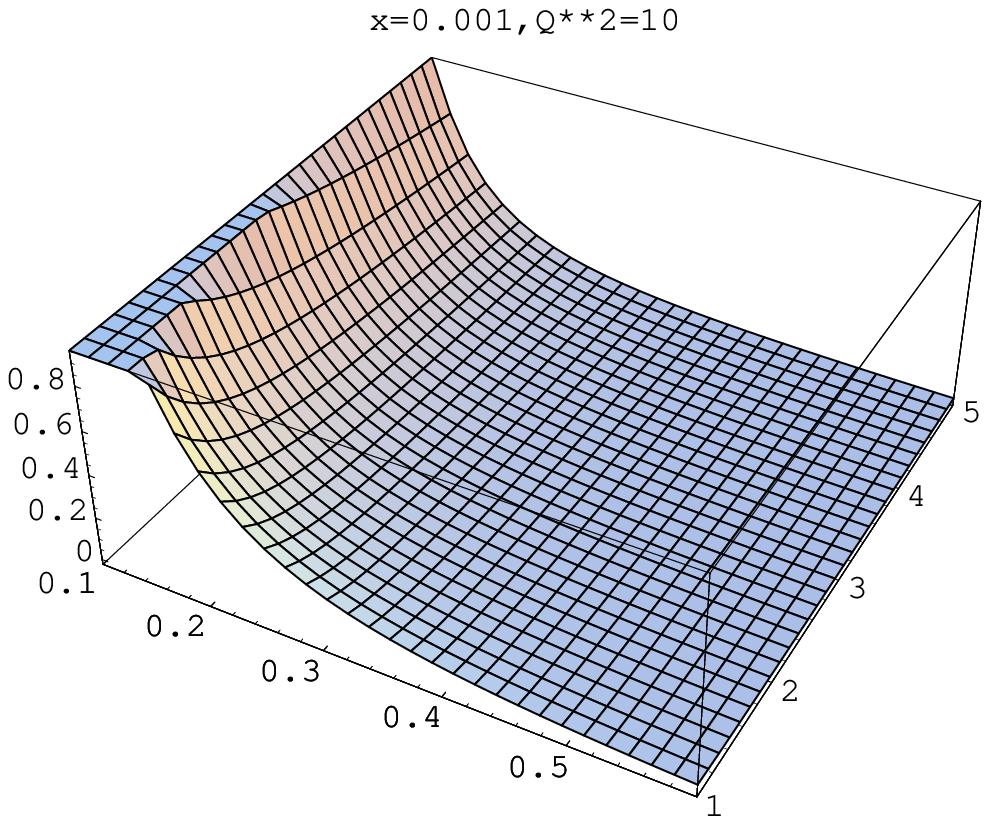}}\\
\mbox{
\epsfxsize=6cm
\epsfysize=5cm
\epsffile{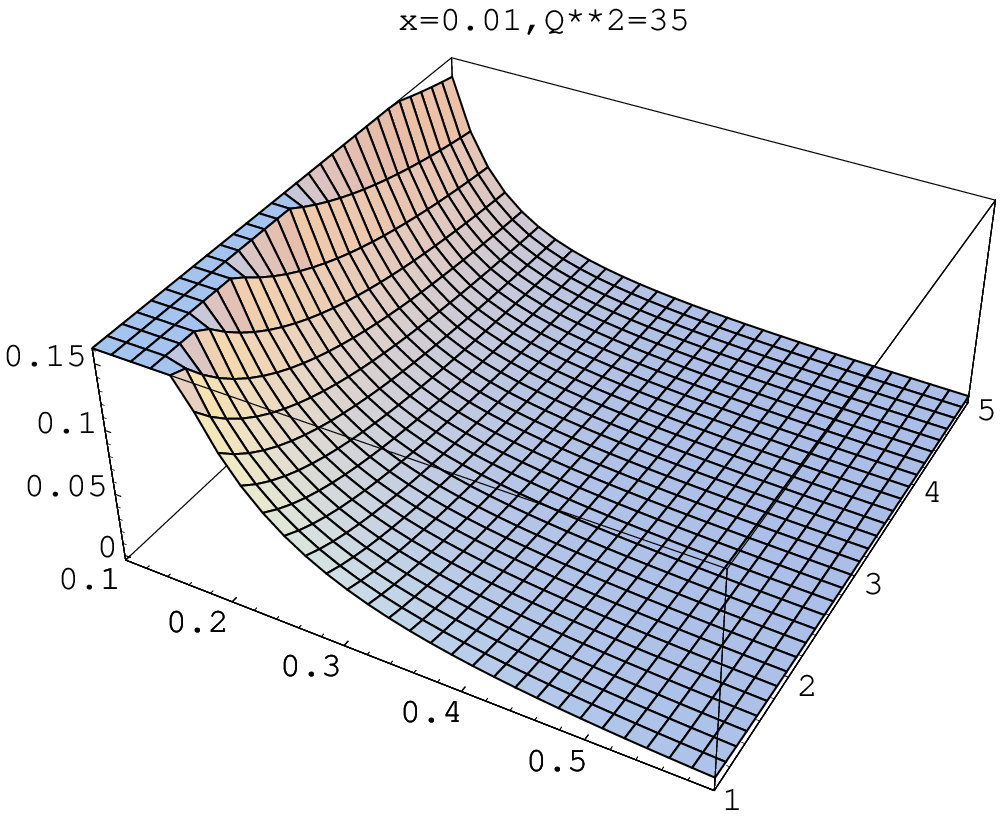}}
\hspace{1cm}
\mbox{
\epsfxsize=6cm
\epsfysize=5cm
\epsffile{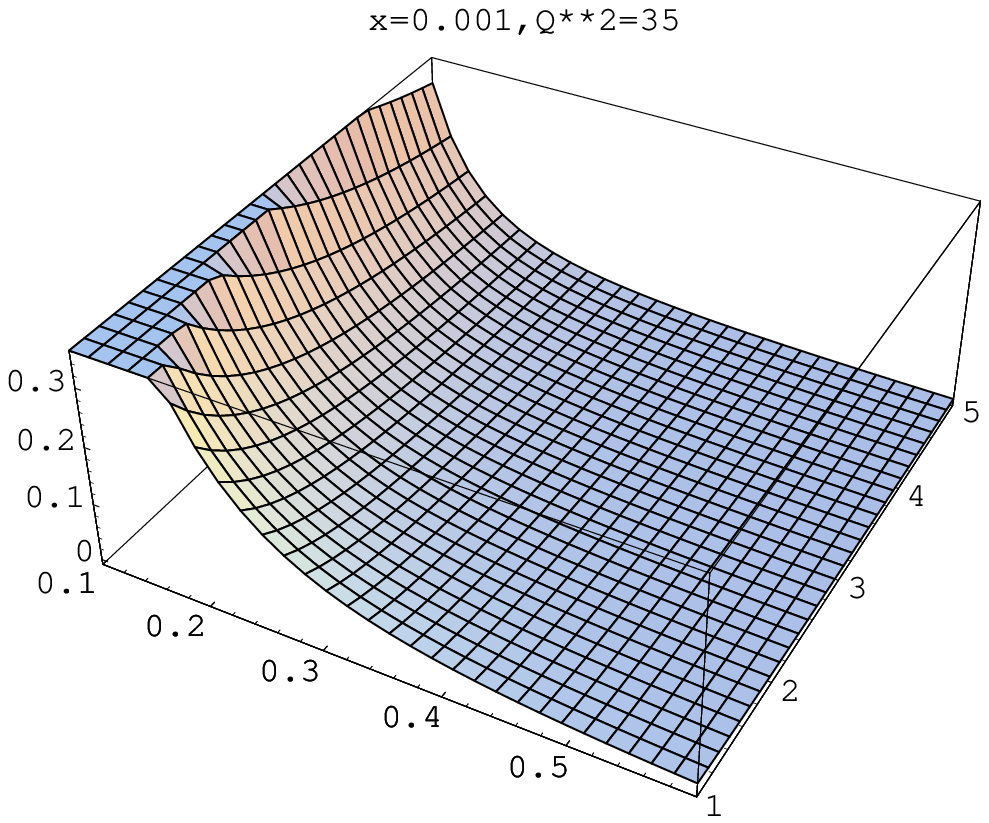}}
{\caption{\label{fig13} The ratio $D(x,Q^2/t)$ with the correction factor
(\ref{fla68}). }}  \end{figure}
%==================================

The resulting asymmetry (\ref{fla63}) is presented in Fig. \ref{fig12}.
We see that the correction factor (\ref{fla68}) crucially changes
the behavior of the asymmetry due to the fact that it restores the azimuthal
dependence of the QEDC amplitude which was not taken into account in Eqs.~(\ref{fla60}-\ref{fla62}). In order to find asymmetry at these $Q^2$
and $t$ with greater accuracy one should take into
account other twist-4 contributions as well.  On the contrary, the ratio
$D(x,Q^2/t)$ does not change much (see Fig. \ref{fig13}) so we hope that our
leading-twist results for the ratio presented in Fig. \ref{fig11} are reliable.

\section{Conclusion}
 The DVCS in the kinematical region (\ref{fla1}) is probably the best place
 to test the momentum transfer dependence of the BFKL pomeron.
 Without this dependence, the  model
 (\ref{fla59}) would be exact, hence the upper curves in Fig. \ref{fig9}
 indicate how important is the $t$-dynamics of the pomeron. We see
 that the $t$-dependence of the BFKL pomeron changes the cross section
  at $t>2$GeV$^2$ by orders of magnitude and therefore it should  be
  possible to
 detect it \cite{baku}.

 The pQCD calculation of the DVCS amplitude in the region (\ref{fla1}) is
 in a sense
 even more reliable than the calculation of usual DIS amplitudes since
 it does not rely on the models for nucleon parton distribution. Indeed,
 all the non-perturbative
 nucleon input is contained in the Dirac form factor of the nucleon,
  which is known to a reasonable accuracy. There are, of course, the non-perturbative corrections to the
 BFKL pomeron itself.
 It is not clear how to take them into account at this moment.
Of course,
 any reasonable models of nucleon parton distributions such as (\ref{fla24})
 should reproduce the form factor after integration over $X$.

 Finally, let us discuss uncertainties in our approximation and
 possible ways to improve it. One obvious improvement would be to
 calculate (at least numerically) the next
 $\sim\left(\alpha_s\ln x\right)^3$ term in the BFKL series for
 the DVCS amplitude. Hopefully, it will be as small as the corresponding
 calculation of the DIS amplitude suggests.
 Besides, there are non-perturbative corrections to the BFKL pomeron
 which are mentioned above. These non-perturbative corrections
 correspond to the situation like the ``aligned jet model'' when one
 of the two gluons in Fig. \ref{ladder} is soft and all the momentum
 transfers through the other gluon. It is not
 clear how to take these corrections into account, but one should expect
 them to be smaller than the corresponding corrections to $F_2(x)$ which
 come from two  non-perturbative gluons in Fig. \ref{ladder} (in other words, from
 the ``soft pomeron''
 contribution to $F_2(x)$).

 The biggest uncertainty in our calculation is the argument
 of coupling constant $\alpha_s$ which we take to be $Q^2$. As 
 mentioned above, it is not possible to fix the argument of
 $\alpha_s$ in the LLA, so one could have used $\alpha_s(|t|)$ instead.
 We hope to overcome
 this difficulty by using the results of NLO BFKL in our future work.

%% file: GPD/GPD.tex
\section{Introduction}
Let me start with the historical overview of the problem.
The first information about quark distributions inside the nucleon
was received from the experiments on deep inelastic scattering:
\begin{eqnarray}
e(k)+N(p)\to e(k')+X(p_n) \, ,
\label{eq_dis}
\end{eqnarray}
where $X(p_n)$ are possible final hadronic states with the 4-momentum $p_n$. 

The scattering amplitude for the $n$-channel is given by:
\begin{equation}
T_n=e^2\bar{u}(k',\lambda')\gamma_{\mu}u(k,\lambda)\frac{1}{q^2}\langle n\mid J^{\mu}_{e.m}(0)\mid p,\sigma\rangle,
\end{equation} 
where $e$ is electron's charge, $k$($k'$) is the initial(final) momentum of the electron, $\lambda$($\lambda'$) is electron's initial(final) polarization, 
$q$ is the momentum of the virtual photon, that hits the nucleon ($q^2=(k-k')^2=-4EE' \sin^2\frac{\theta}{2}\leq 0$, $Q^2=-q^2$), $\theta$ is electron's scattering angle and 
$J^{\mu}_{e.m}$ is hadronic electromagnetic current. 

The differential cross section for this scattering for unpolarized particles has the form:
\begin{equation}
d\sigma_n=\frac{1}{\mid v\mid}\frac{1}{2M}\frac{1}{2E}\frac{d^3k'}{(2\pi)^32k'_0}\prod^n_{i=1}\Bigg[\frac{d^3p_i}{(2\pi)^32p_{i0}}\Bigg]
\times\frac{1}{4}\sum_{\sigma,\lambda,\lambda'}\mid T_n\mid^2(2\pi)^4\delta^4(p+k-k'-p_n),
\end{equation}
where $p_n=\sum^n_{i=1}p_i$. 

If we sum over all possible final hadronic states, we will get the inclusive cross section
\begin{equation}
\frac{d^2\sigma}{d\Omega dE'}=\frac{\alpha^2}{q^4}\left(\frac{E'}{E}\right) l_{\mu\nu}W^{\mu\nu},
\end{equation}
where $\alpha=e^2/4\pi$ - fine structure constant. Lepton tensor $l_{\mu\nu}$ has the following form:
\begin{equation}
l_{\mu\nu}=\frac{1}{2}Tr[\hat{k'}\gamma_\mu\hat{k}\gamma_\nu]=2\left(k_\mu k'_\nu+k'_\mu k_\nu+\frac{q^2}{2}g_{\mu\nu}\right),
\end{equation}
and the hadronic one $W^{\mu\nu}$ has the form:
\begin{eqnarray}
W^{\mu\nu}=\frac{1}{4M}\sum_\sigma \int\frac{d^4x}{2\pi}e^{iq\cdot x}\langle p,\sigma\mid [J^{\mu}_{e.m}(x)J^{\nu}_{e.m}(0)] \mid p,\sigma\rangle.
\end{eqnarray}
From the current conservation $\partial_\mu J^{\mu}_{e.m}=0$ one gets the following equations: 
\begin{equation}
q_\mu W^{\mu\nu}=q_\nu W^{\mu\nu}=0, 
\end{equation} 
which brings us to the requirements that hadronic tensor $W^{\mu\nu}$ should satisfy: it should be a Lorentz invariant tensor of rank two and it should depend on momenta $p_\mu$ and $q_\mu$ (the only momenta we have in this case), which brings us to the
following representation of the hadronic tensor:
\begin{equation}
W_{\mu\nu}(p,q)=\left[-W_1\left(g_{\mu\nu}-\frac{q_\mu q_\nu}{q^2}\right)+W_2\frac{1}{M^2}\left(p_\mu-\frac{p \cdot q}{q^2}q_\mu\right)\left(p_\nu-\frac{p \cdot q}{q^2}
q_\mu\right)\right],
\end{equation}
where $W_{1,2}$ are the first Lorentz invariant structure functions of the nucleon target that had ever been introduced. Those structure functions depend on two variables
$q^2$ and $\nu=E-E'$.

In the case, when the final hadronic state $X(p_n)$ is also a nucleon (the case of the elastic electron-nucleon scattering), the matrix element for the electromagnetic current has the form:
\begin{equation}
\langle N(p')\mid J^{\mu}_{e.m}(0)\mid N(p)\rangle=\bar{u}(p')\left[\gamma_\mu F_1(q^2)+i\sigma_{\mu\nu}q^\nu \frac{1}{2M}F_2(q^2)\right]u(p),
\end{equation}
with $q=p-p'$. For the proton $F^p_1(0)=1$ is equal to the electric charge and $F^p_2(0)=1.79$ is equal to the anomalous magnetic moment. For the neutron those number are
$F^n_1(0)=0$ and $F^n_2(0)=-1.91$, respectively.

Therefore, the differential cross section measurements for the elastic electron-nucleon scattering give us information about electrical and magnetic form factors.
If the nucleon had no internal structure, those form factors were independent on $q^2$. The experimental fact that they have momentum transfer dependence opened 
a new era in nucleon structure
investigations. 

The new parton model was created by Feynman, which described a nucleon  as a particle consisting of three partons (quarks) interacting by gluon exchange. Lepton-nucleon scattering has been 
 transformed to the scattering of virtual photon off one of the partons. Accordingly, the structure functions $W_{1,2}(p,q)$ describing the response of the nucleon with momentum $p$ to the momentum transfer $q$ were replaced by structure functions $F_{1,2}(x)$:
\begin{equation}
MW_1(p,q) \to F_1(x)=\frac{1}{2} n(x),
\end{equation}
\begin{equation}
\nu W_2(p,q) \to F_2(x)=x n(x),
\end{equation}
where $n(x)$ represents the probability to find the parton (quark) inside the nucleon carrying the fraction $x$ of initial longitudinal nucleon momenta $xp$ within the range from $x$ to $x+dx$.
The corresponding function for the antiquark is denoted as $\bar{n}(x)$. 

%%%%%%%%%%%%%%%%%%%%%%%%%%%%%%%%%%%%%%%%%%%%%%%%%%%%%%
\begin{figure}
\hspace{.5cm}
\mbox{
\epsfxsize=6cm
\epsfysize=5cm
\hspace{0cm}
\epsffile{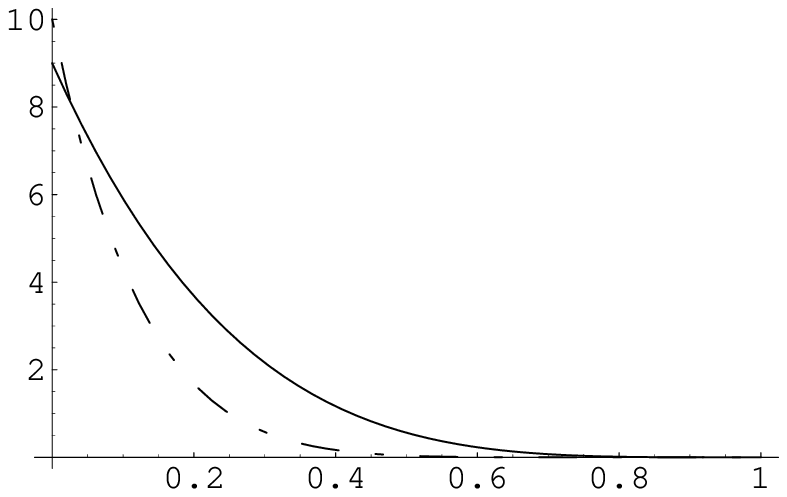}}
\hspace{1cm}
\mbox{
\epsfxsize=6cm
\epsfysize=5cm
\epsffile{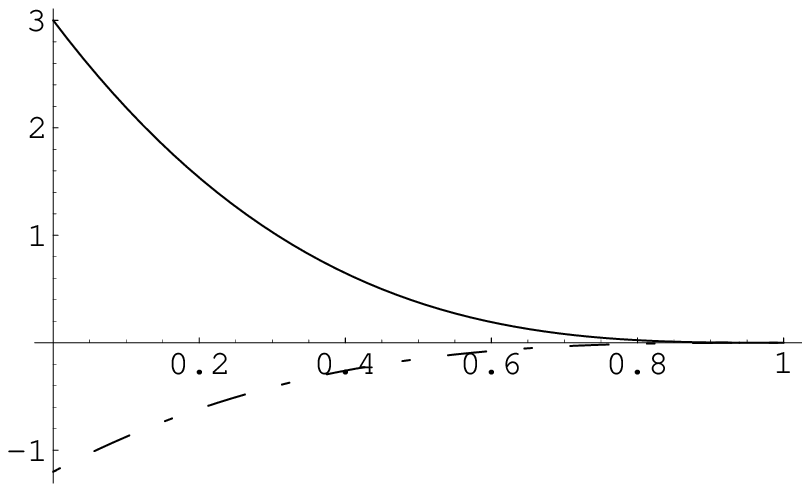}}
{\caption{\label{quark} Forward quark-parton density (left) and helicity (right) distributions used in our calculations for u (solid) and d (dashed) quark.}}
\end{figure}
%%%%%%%%%%%%%%%%%%%%%%%%%%%%%%%%%%%%%%%%%%%%%%%%%%%%%

Thus, through the experiments on deep inelastic lepton-hadron scattering one can extract information on forward quark distribution functions (Fig. \ref{quark}).

In the following, we introduce the kinematical variables
entering the DVCS process. We then discuss the DVCS amplitude 
 in the leading power in $Q$ (twist-2 accuracy), and
show its dependence upon the general parton distribution (GPD) functions.

\section{DVCS in twist-2 approximation}
As was already mentioned in the first chapter,  the deeply virtual Compton
scattering process
\begin{eqnarray}
\gamma^{\ast}(q)+N(p)\to \gamma(q') +N(p') \, ,
\label{eq:proc}
\end{eqnarray}
in the limit of vanishing momentum transfer $t=(p-p')^2$ and large virtuality $Q^2 \to \infty$ offers us much more complex insight on parton dynamics in the nucleon.
In the lowest approximation (twist-2) the structure functions $F(x,y)$, called double distributions ~\cite{ji,ji2}, carry information on the quark distributions 
with different fraction $x$ of the initial nucleon momentum and different fraction  $y$ of the momentum transfer $r=p'-p$ carried by the quark. 
This nucleon structure information can be parametrized in terms of four structure functions.
Those functions look like  usual distribution functions with respect to $x$ and like distribution amplitudes with respect to $y$. 

\subsection{Kinematical variables and distribution functions}
\label{variables-2}

Since $q'^2=0$ ($q'$ - is the final real photon four-momenta), it is natural to use it as one of the basic Sudakov light-cone 4-vectors. 
Another basic light-cone vector we use is $p$, which is also understandable, since $p^2=0$ can be neglected compared to the virtuality $Q^2=-q^2$ of the initial photon
and the energy invariant $p\cdot q=M \nu$. In this limit ($p^2=0$ and $r^2=0$), the simple equation $p'^2=(p+r)^2=p^2$ requires $p\cdot r=0$ which can be 
satisfied in the light-cone 
basis only if they are proportional to each other $r=\zeta p$.

Despite the proportionality between $p$ and $r$, they specify the momentum flow in two different channels, $s$ and $t$ respectively. The coefficient of 
their proportionality $\zeta$ is the parameter characterizing the ``asymmetry'' or the ``skewedness'' of the matrix elements. 

%%%%%%%%%%%%%%%%%%%%%%%%%%%%%%%%%%%%%%%%%%%%%%%%%%%%%%
\begin{figure}\label{double}
\hspace{.5cm}
\mbox{
\epsfxsize=6cm
\epsfysize=5cm
\hspace{0cm}
\epsffile{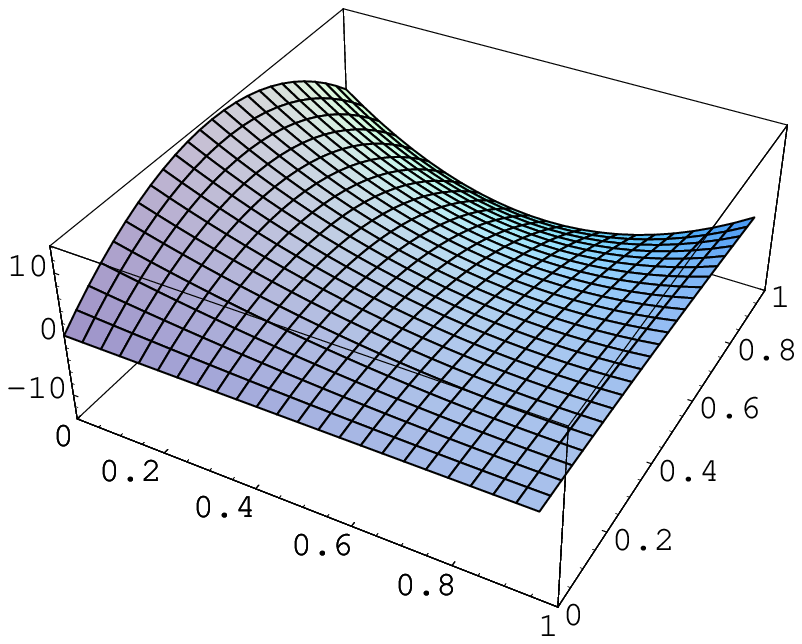}}
\hspace{1cm}
\mbox{ 
\epsfxsize=6cm
\epsfysize=5cm
\epsffile{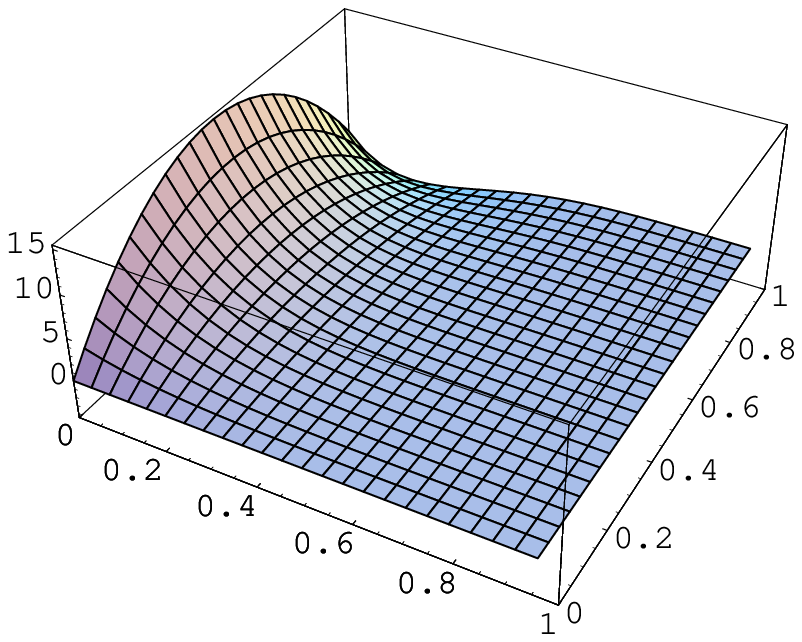}}
{\caption{ Double distributions $F(x,y;t=0)$ for u (left) and d (right) quark. }}  
\end{figure}
%%%%%%%%%%%%%%%%%%%%%%%%%%%%%%%%%%%%%%%%%%%%%%%%%%%%%
The total momentum fraction of the quark can either be positive or negative. Since positive momentum fraction corresponds to quark and the negative one - to antiquark, 
there are three major regions that could be found. Two of them, namely when we have two active quarks or antiquarks, are the regions accessible in  inclusive processes like DIS and giving us the usual forward parton distributions. The third region represents the one-quark-one-antiquark case, which cannot be measured through DIS, 
since it vanishes in the forward limit $\Delta \to 0$. 
In the limit $r=0$ we come up 
with the following reduction formulas for the double distributions $F(x,y)$:
\begin{equation}
\int^{1-x}_0 F(x,y;t=0)|_{x>0} dy=n(x);~~~~~
\int^1_{-x} F(x,y;t=0)|_{x<0} dy=-\bar{n}(-x).
\end{equation}
\begin{equation}
\int^{1-x}_0 \tilde{F}(x,y;t=0)|_{x>0} dy=n(x);~~~~~
\int^1_{-x} \tilde{F}(x,y;t=0)|_{x<0} dy=\bar{n}(-x).
\end{equation}
Positive-$x$ and negative-$x$ components of the double distributions can be treated as non-forward generalizations of quark and antiquark densities, respectively:
$F(x,y;t)|_{x>0}$ ($-F(-x,1-y;t)|_{x<0}$).

The reduction formulas and interpretation of the $x$-variable suggests, that the profile of $F(x,y)$ in $x$-direction is basically determined by the shape of $n(x)$.
\begin{figure}[h]
\mbox{
   \epsfxsize=13cm
 \epsfysize=6cm
 \hspace{1cm}  \epsffile{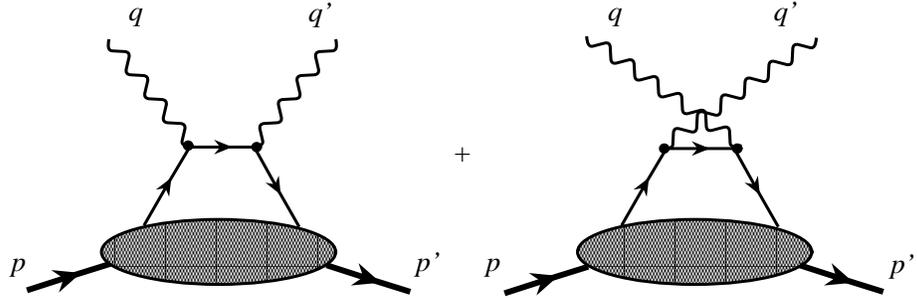}  }
{\caption{\label{handbags}''Handbag'' diagrams for DVCS.
   }}
\end{figure}
The profile in the $y$-direction at the same time characterizes the spread of the parton momentum induced by the momentum transfer $r$. This brings us to the following 
relation:
\begin{equation}\label{dd}
F(x,y)=h(x,y)n(x),
\end{equation}
where $h(x,y)$ is an even function of $y$ and
\begin{equation}
\int^{1-|x|}_{-1+|x|}h(x,y) dy=1.
\end{equation}

 Integrating each particular double distribution over $y$ gives the non-forward parton distributions:
\begin{eqnarray}\label{evo-71}
&&{\cal F}_\zeta(X)=\int^1_0dx\int_0^{1-x}\delta(x+\zeta y-X) F(x,y)dy= \nonumber \\
&&\theta(X\geq \zeta)\int^{\bar{X}/\bar{\zeta}}_0 F(X-\zeta y,y)dy+\theta(X\leq \zeta)\int^{X/\zeta}_0 F(X-\zeta y,y)dy,
\end{eqnarray}
where $X\equiv x+\zeta y$ is the total momentum fraction and the notation $\bar{\zeta}\equiv 1-\zeta$ is used.

\subsection{Twist-2 DVCS amplitude}
\label{amplitude-2}
The leading \cite{ji,compton,Mul94} handbag contribution (Fig.~\ref{handbags}) 
to the DVCS amplitude can be represented as \cite{npd}
\begin{equation}
\begin{array}{rcl} \displaystyle
T^{\mu\nu}(p,q,q') &=& \displaystyle
{1\over 2(pq')} \left \{ \sum_a e_a^2
\left ( {1\over 2(pq')}(p^\mu q'^\nu+p^\nu q'^\mu)-g^{\mu\nu}\right )
 \right . \\[4mm]
&&  \times \displaystyle
\left [ \bar u(p'){\hat q}' u(p) T^a_F(\zeta) + 
{1\over 2M}\bar u(p')
 ({\hat q}'\hat r - \hat r{\hat q}') u(p) T^a_K(\zeta) \right ] \\[4mm]
&+&\displaystyle
 {i \epsilon^{\mu\nu\alpha\beta} p_\alpha q'_\beta \over (pq')} \\[4mm]
&& \left . \times \displaystyle
\left [ \bar u(p'){\hat q}'\gamma_5 u(p) T^a_G(\zeta) +
{(q'r)\over 2M}\bar u(p')\gamma_5  u(p) T^a_P(\zeta) \right ] \right \} \ .
\end{array}\label{tvcs_munu}
\end{equation}

The invariant amplitudes
$T^a(\zeta)$ can be calculated using model \cite{musatov} for the non-forward
quark parton distributions with $t=0$:
\begin{equation}
T^a_{F}(\zeta)=-\int_0^1\left[ {1\over X-\zeta+i\epsilon}
+ {1\over X-i\epsilon} \right ]
\left ({\cal F}^a_\zeta(X) + {\cal F}^{\bar a}_\zeta(X)\right ) dX \ .
\label{dvcs-T}
\end{equation}
%=================================
\begin{figure}[h]
\mbox{
\epsfxsize=6cm
\epsfysize=5cm
\hspace{0cm}
\epsffile{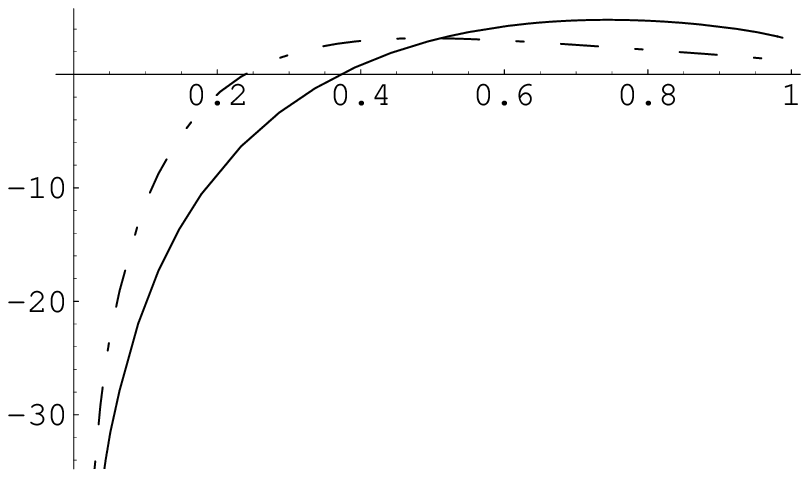}}
\hspace{1cm}
\mbox{
\epsfxsize=6cm
\epsfysize=5cm
\hspace{0cm}
\epsffile{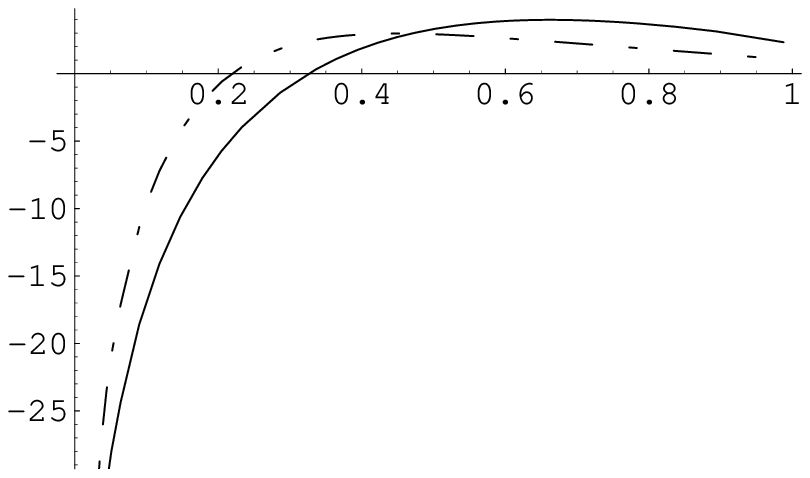}}
\vspace{1cm}
\hspace{0.3cm}
\mbox{
\epsfxsize=6cm
\epsfysize=5cm
\hspace{0cm}
\epsffile{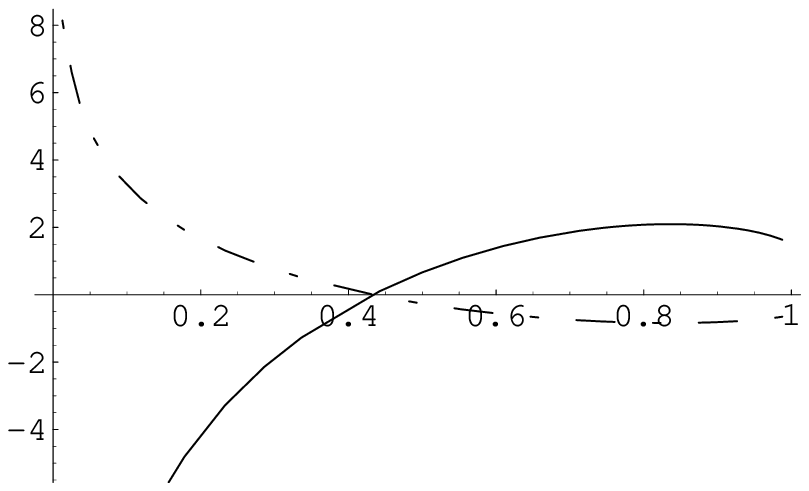}}
\hspace{1.3cm}
\mbox{
\epsfxsize=6cm
\epsfysize=5cm
\hspace{0cm}
\epsffile{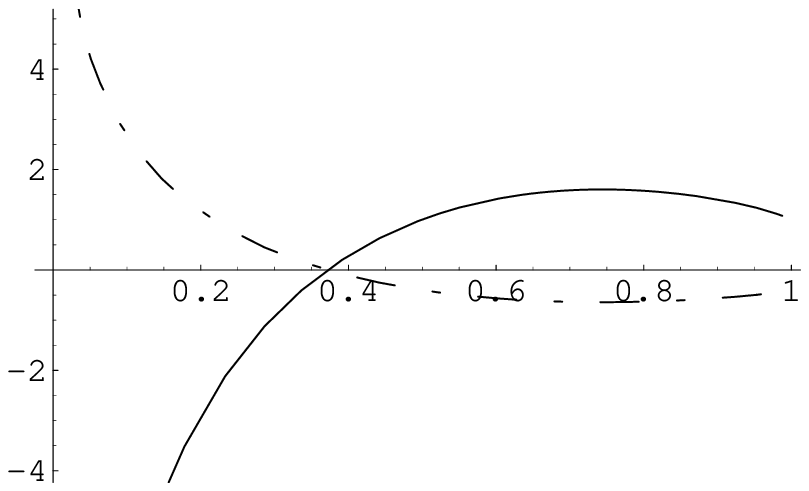}}
{\caption{\label{inv-ampl} Real parts of the invariant amplitudes $T^a_F$ (left top), $T^a_G$ (left bottom),$T^a_K$ (right top), $T^a_P$ (right bottom) for $u$ (solid) and
$d$ (dashed) quarks (Eq.~(\ref{dvcs-ReT})). }}  
\end{figure}
%%%%%%%%%%%%%%%%%%%%%%%%%%%%%%%
Note, that because the non-forward distributions are real,
the imaginary part of $T^a(\zeta)$ comes only from the singularities
of the expression in the square brackets. Since all non-forward distributions
vanish at $X=0$, only the first term in the square brackets generates
the imaginary part:
\begin{equation}
\Im T^a_F(\zeta) = \pi\left ({\cal F}^a_\zeta(\zeta) +
{\cal F}^{\bar a}_\zeta(\zeta)\right ) \ .
\label{dvcs-ImT}\end{equation}
%=================================
\begin{figure}[h]
\mbox{
\epsfxsize=6cm
\epsfysize=5cm
\hspace{0cm}
\epsffile{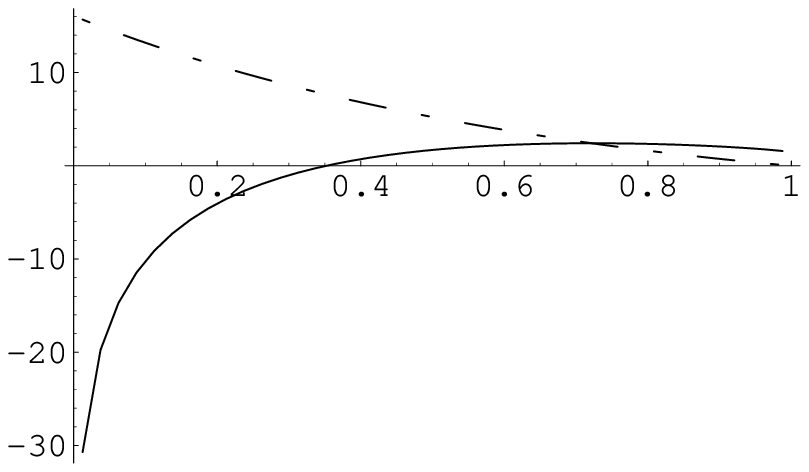}}
\hspace{1cm}
\mbox{
\epsfxsize=6cm
\epsfysize=5cm
\hspace{0cm}
\epsffile{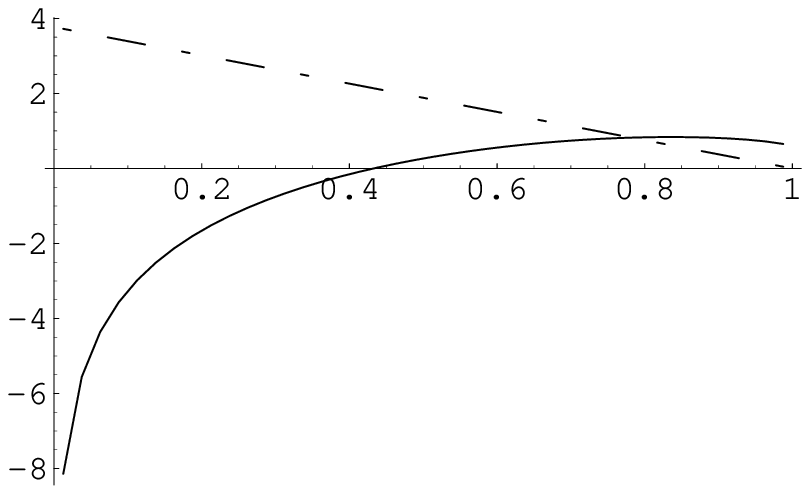}}
\vspace{1cm}
\hspace{0.3cm}
\mbox{
\epsfxsize=6cm
\epsfysize=5cm
\hspace{0cm}
\epsffile{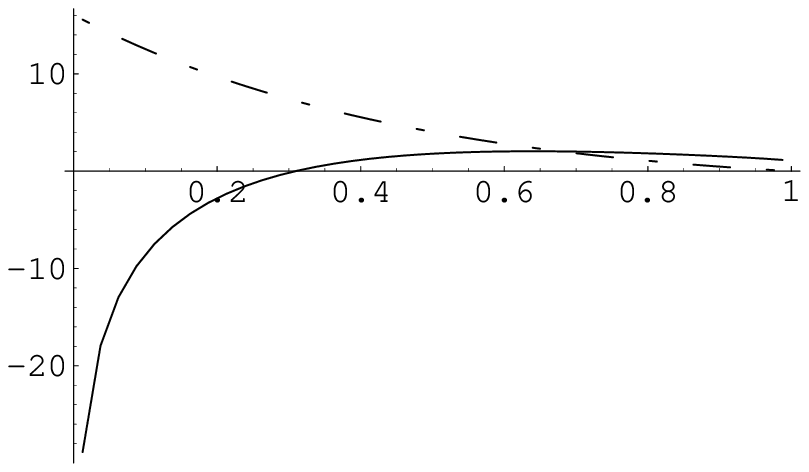}}
\hspace{1.3cm}
\mbox{
\epsfxsize=6cm
\epsfysize=5cm
\hspace{0cm}
\epsffile{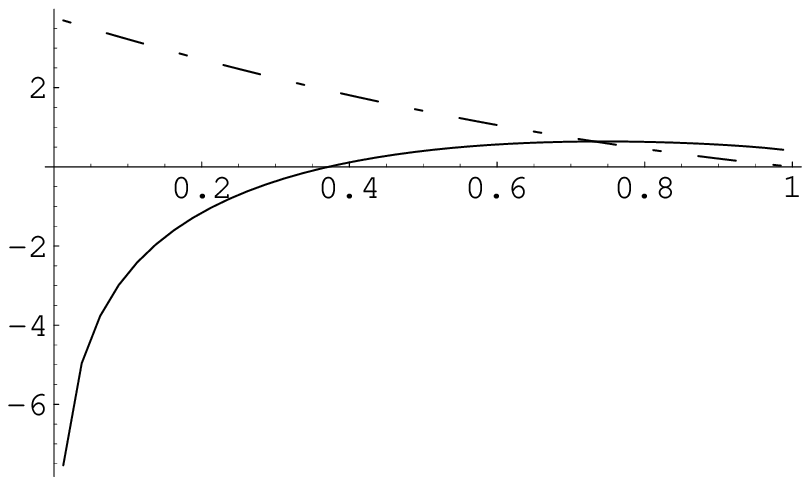}}
{\caption{\label{dvcs-Re-Im} Real (solid) and imaginary (dashed) parts of the DVCS invariant amplitudes:
$T_F$ (left top), $T_K$ (right top), $T_G$ (left bottom), and $T_P$ (right bottom).
}}  \end{figure}
%%%%%%%%%%%%%%%%%%%%%%%%%%%%%%%
The real part of the invariant amplitudes $T^a(\zeta)$ (Fig. \ref{inv-ampl}) is given by
the ``principle value integration'' in Eq.(\ref{dvcs-T})
\begin{equation}
\Re T^a_F(\zeta) = -{\cal P} \int_0^1
\left ({\cal F}^a_\zeta(X) + {\cal F}^{\bar a}_\zeta(X)\right )
 {dX\over X-\zeta}\ .
\label{dvcs-ReT}
\end{equation}
%=================================
The non-forward distributions ${\cal F}^a_\zeta(X)$
are expressed through the double distributions by Eq.(\ref{evo-71}).

In our calculations we use the  model \cite{musatov} for the double distribution given by the factorized ansatz
\begin{equation}
{\cal F}_a(x,y) = f_a (x) {h(x,y) \over h(x)}, \ \ \ h(x)=\int_0^1 h(x,y) dy,
\label{double-anzatz}\end{equation}
where $f_a (x)$ is  the forward quark -- parton distribution and
$h(x,y)$ is the profile function. The realistic
profile for the double quark distributions, both valence and sea,
is close to the ``asymptotic'' form
\begin{equation}
 h_{as}(x,y)= 6y(1-x-y).
\label{as-profile}\end{equation}

The resulting curves for the real and imaginary parts of the invariant
amplitude $T(\zeta)=\sum_a e^2_a T^a(\zeta)$ are shown in Fig. \ref{dvcs-Re-Im}.

\subsection{Twist-2 DVCS cross section}
\label{cross_section-2}

\begin{figure}[htb]
\mbox{
   \epsfxsize=13cm
 \epsfysize=5cm
 \hspace{0.5cm}  \epsffile{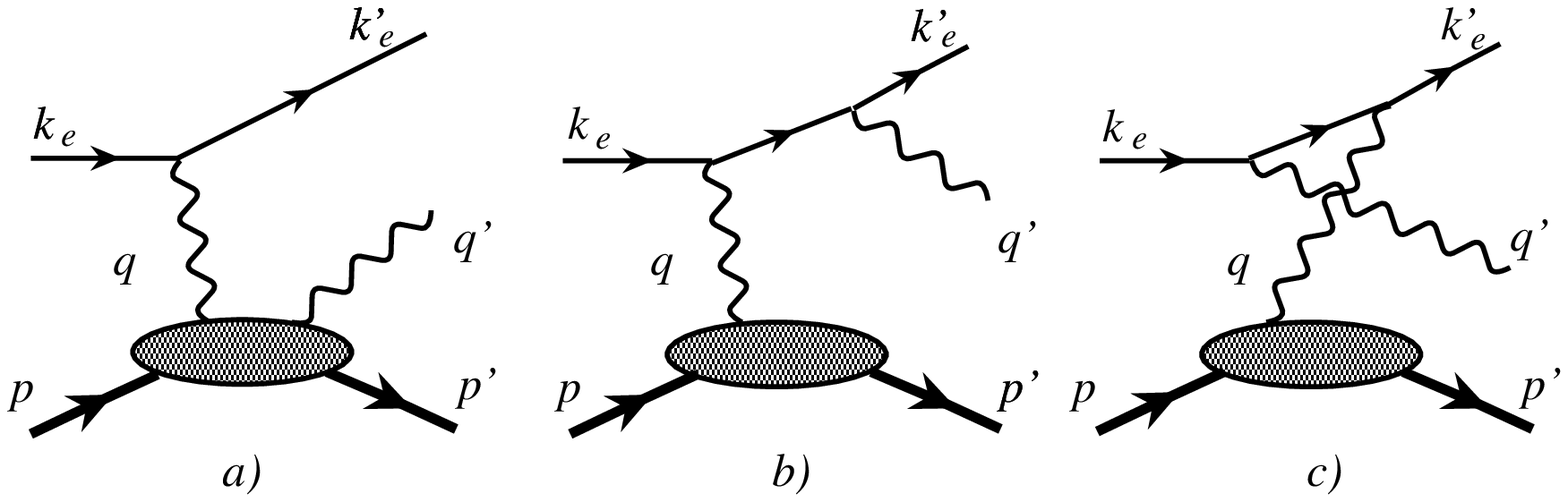}  }
{\caption{\label{dvcs-BH} $a)$ the DVCS part of the amplitude;
$b),c)$ the Bethe-Heitler part.
   }}
\end{figure}

The DVCS amplitude, as was mentioned before, may be observed in the exclusive lepton-nucleon
scattering. In the $\gamma^{\ast}(q)+N(p)\to \gamma(q') +N(p')$ reaction, the final photon can be emitted
either by the proton or by the lepton.
The three relevant diagrams  are shown in Fig.\ref{dvcs-BH}.
The blob with two photon legs in diagram Fig.\ref{dvcs-BH}$a$
stands for the DVCS nucleon amplitude
(i.e. the amplitude $T_{DVCS}$ (in Eq.~(\ref{eq:invcs})) of scattering of the virtual photon on the
nucleon, with the real photon in the final state, whereas the one-photon
blob  in diagrams Fig.\ref{dvcs-BH}$b,\ c$ represents the nucleon electromagnetic
form factor. 

We will refer to diagram Fig.\ref{dvcs-BH}$a$
as the DVCS part of the amplitude of the electron-nucleon scattering.
Figs.\ref{dvcs-BH}$b,\ c$ give together the Bethe-Heitler part.

The invariant cross section of the
$\gamma^{\ast}(q)+N(p)\to \gamma(q') +N(p')$ reaction differential with respect to
$Q^2$, $x_B$, $t$, and out-of-plane angle $\phi$ ($\phi =
0^o$ corresponds to the situation where the real photon is emitted in
the same half plane as the leptons) is given by~:
\begin{eqnarray}
{{d \sigma} \over {d Q^2 \, d x_B \, d t \, d \Phi}}  = 
{1 \over {(2 \pi)^4 \, 32}}  {{x_B \, y^2} \over {Q^4}}
 \left( 1 + {{4 M^2 x_B^2} \over {Q^2}}\right)^{-1/2}
 \biggl| T_{BH} + T_{DVCS} \biggr|^2 \, ,
\label{eq:invcs}
\end{eqnarray}
where $M$ is the nucleon mass, $y \equiv (p \cdot q) / (p \cdot k)$,
and $k$ is the initial lepton four-momentum.

The process where the photon is emitted from the
initial or final lepton is referred to as the
Bethe-Heitler (BH) process (amplitude $T_{BH}$ in
Eq.~(\ref{eq:invcs})), and can be calculated exactly:
\begin{eqnarray}
\biggl| T_{BH}\biggr|^2=-\frac{e^6}{M^2x_B^2y^2\tau^2 \beta(1+2\beta+\tau)}~~~~~~~~~~~~~~~~~~~~~~~~~~~~~~~~~~~~~~~~~~~~~&&\\
\left(
x_B^2y^2\mu\tau(1+8\beta^2+\tau^2+4\beta(1+\tau))((2F_1(t)+F_2(t))^2+2(2\mu/\tau-1)F_1^2(t)) \right. \nonumber \\
\left.+\tau(1+\bar{y}^2-x_By(1+2\beta(1+\bar{y})+\bar{y}\tau))(4\mu F_1^2(t)-\tau F_2^2(t)) \nonumber
\right),
\end{eqnarray}
where we introduced the dimensionless variables $\beta=(k\cdot q')/Q^2$, $\tau=t/Q^2$, and $\mu=M^2/Q^2$. 
The $\phi$-dependence is hidden in $\beta$ and it alone determines the
$\phi$-dependence of the Bethe-Heitler amplitude $T_{BH}$.

Light particles such as electrons radiate much more than the heavy proton.
Therefore the BH process generally dominates the DVCS amplitude, especially at small $t$. 
The best way to extract the information on general parton distribution functions and still be
in an accessible region for the experimentalists is to measure the interference term.
We could benefit from the situation if we consider the difference in cross section for electrons with opposite helicities. 
This way, in the difference of
cross sections $\sigma_{e^\rightarrow} - \sigma_{e^\leftarrow}$, the BH (whose amplitude
is purely real) drops out. The DVCS contribution is strongly suppressed and the term dominating the cross section would be the 
 BH-DVCS interference \cite{Bro72} term
\begin{equation}\label{fla:asym}
\sigma_{e^\rightarrow} - \sigma_{e^\leftarrow} \sim  \Im \left[ T^{BH} {T^{DVCS}}^* \right] \;,
\end{equation}
 where the Bethe-Heitler process will project out the imaginary part of DVCS amplitude magnifying it with its own full magnitude. 

\begin{figure}[ampl]
\mbox{
   \epsfxsize=14cm
 \epsfysize=8.4cm
 \epsffile{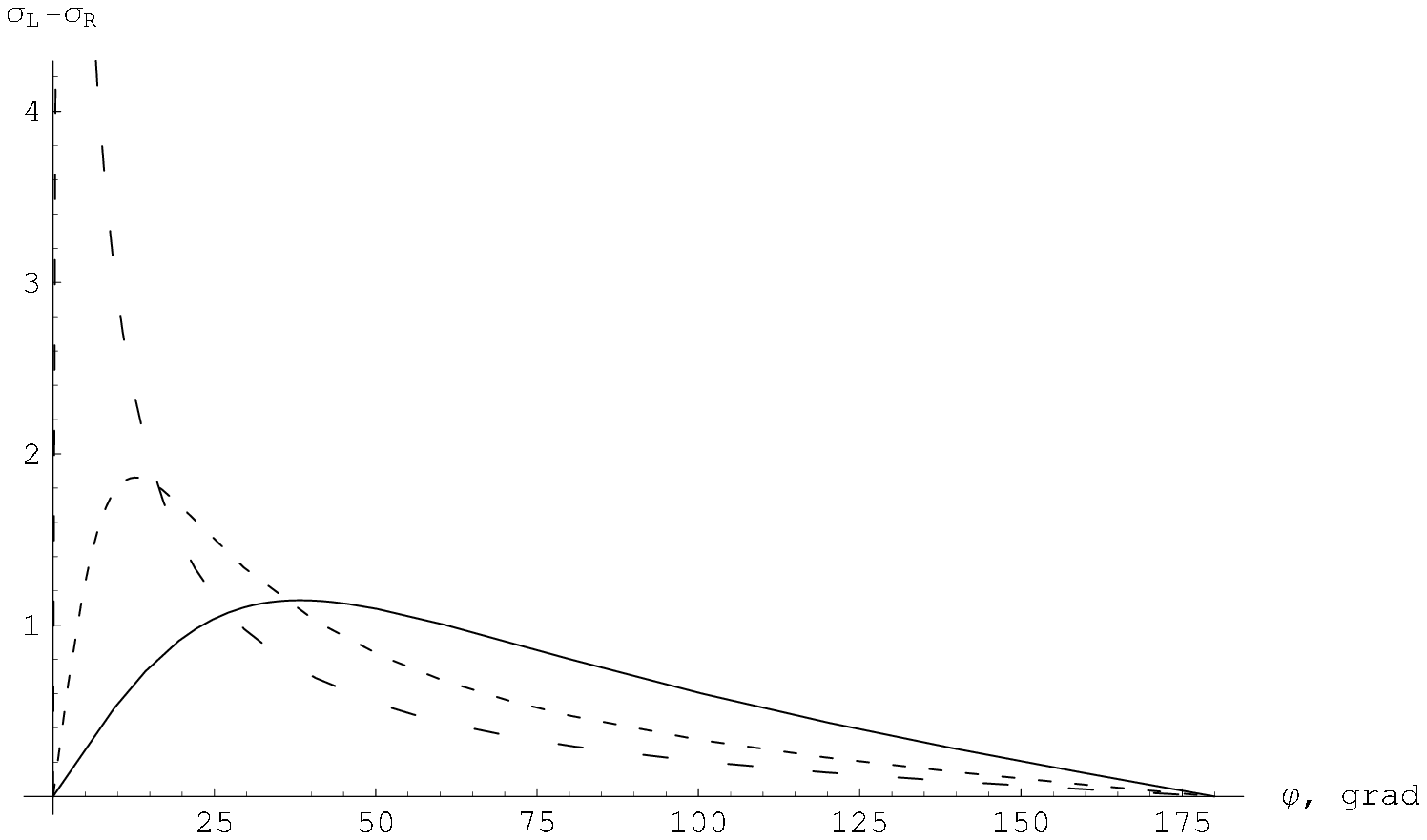}  }
\mbox{
   \epsfxsize=14cm
 \epsfysize=8.4cm
 \epsffile{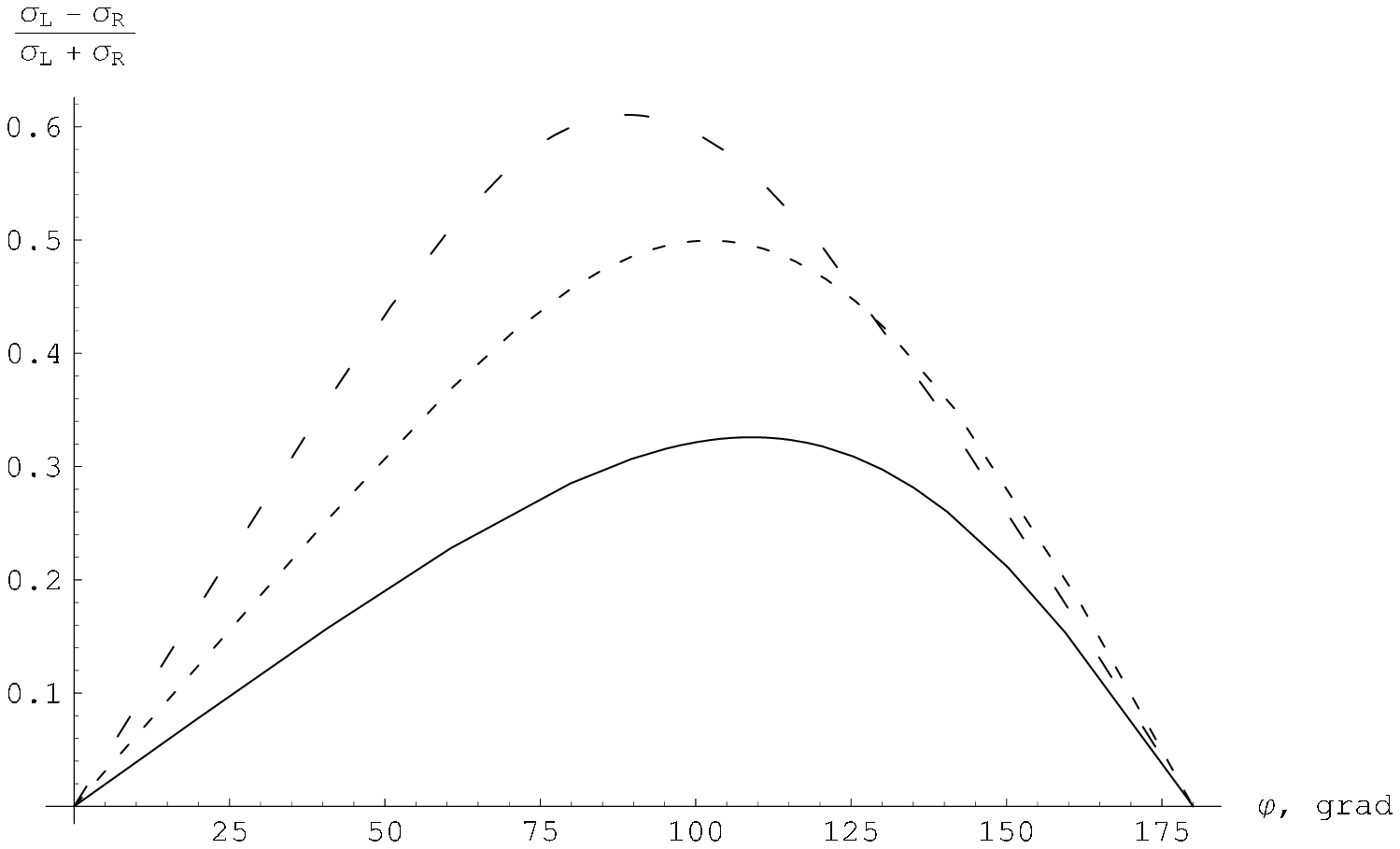}  }
{\caption{\label{res_1} Cross section calculated at twist-2 level for $E_e=4.25$GeV, $x_B=.2$, and $Q^2=1.3$GeV$^2$ for the different values of
momentum transfer:
$t=-.1$ -- solid curve, $t=-.2$ -- dotted curve, $t=-.3$ -- dashed curve, and $t=-.5$ -- dash-dotted curve.
   }}
\end{figure}
\begin{figure}[ampl]
\mbox{
   \epsfxsize=14cm
\epsfysize=8.4cm
\epsffile{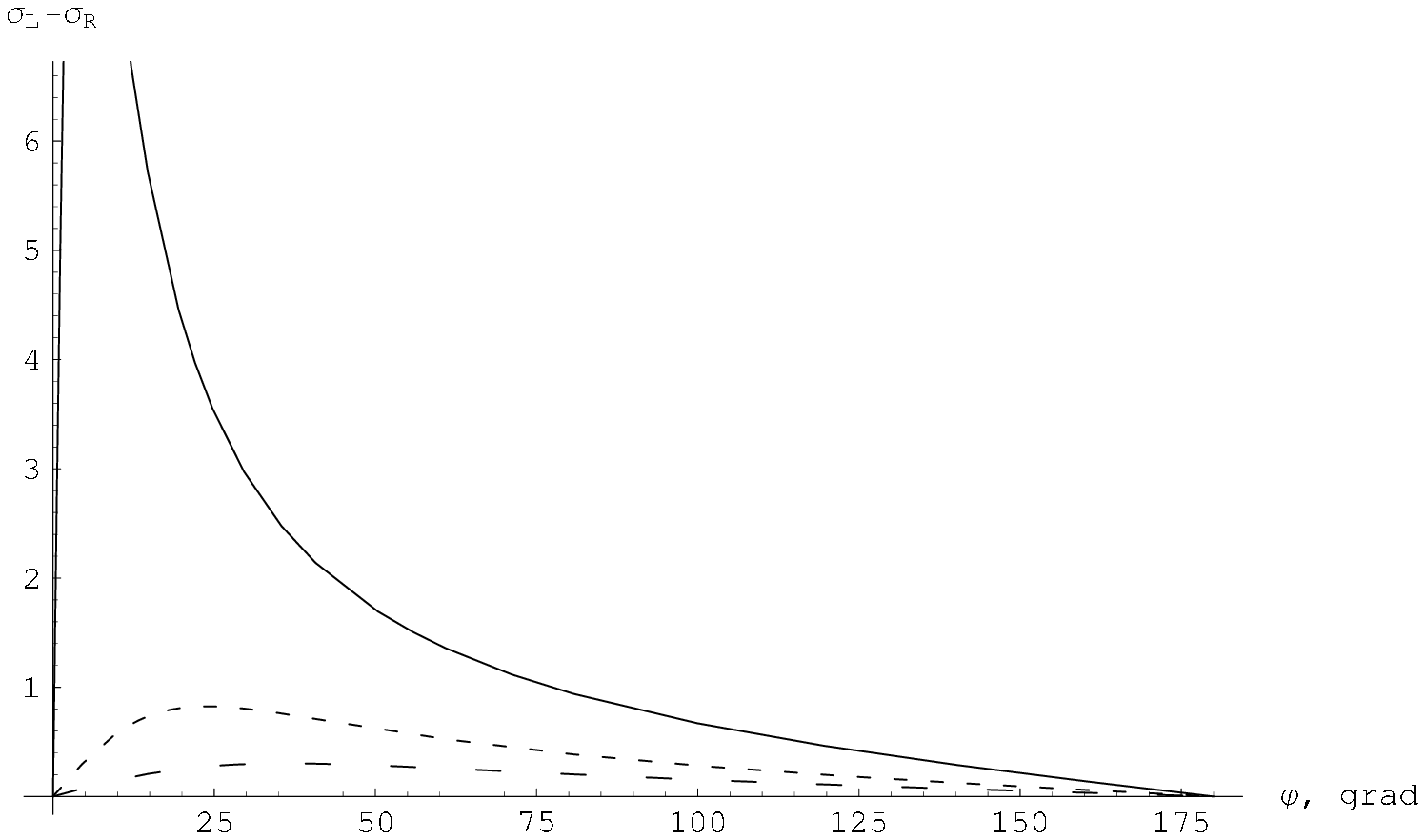}  }
\mbox{
   \epsfxsize=14cm
\epsfysize=8.4cm
 \epsffile{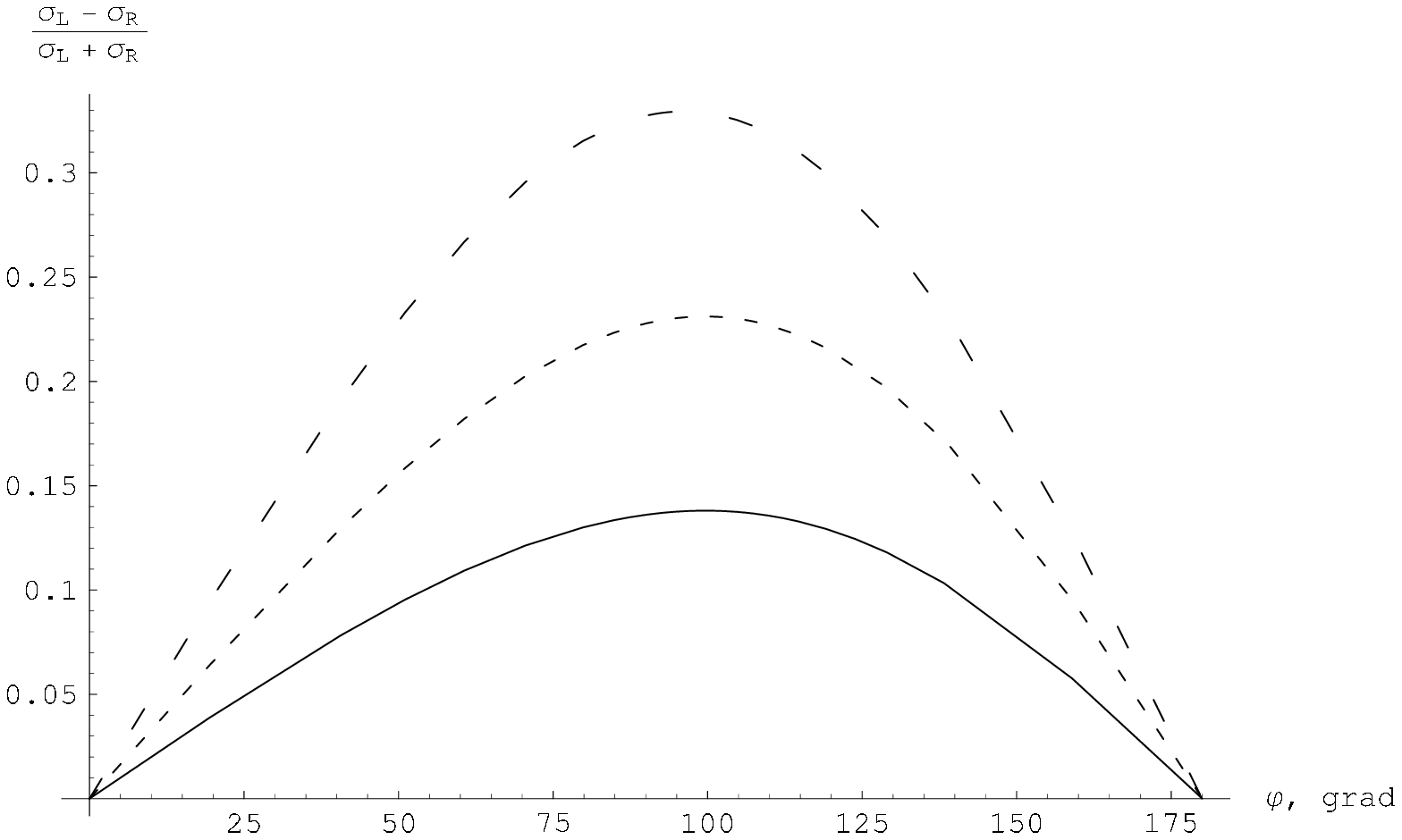}  }
{\caption{\label{res_2} The same for $E_e=4.25$GeV, $x_B=.2$, and $Q^2=1.5$GeV$^2$.
   }}
\end{figure}
\begin{figure}[ampl]
\mbox{
   \epsfxsize=14cm
\epsfysize=8.4cm
 \epsffile{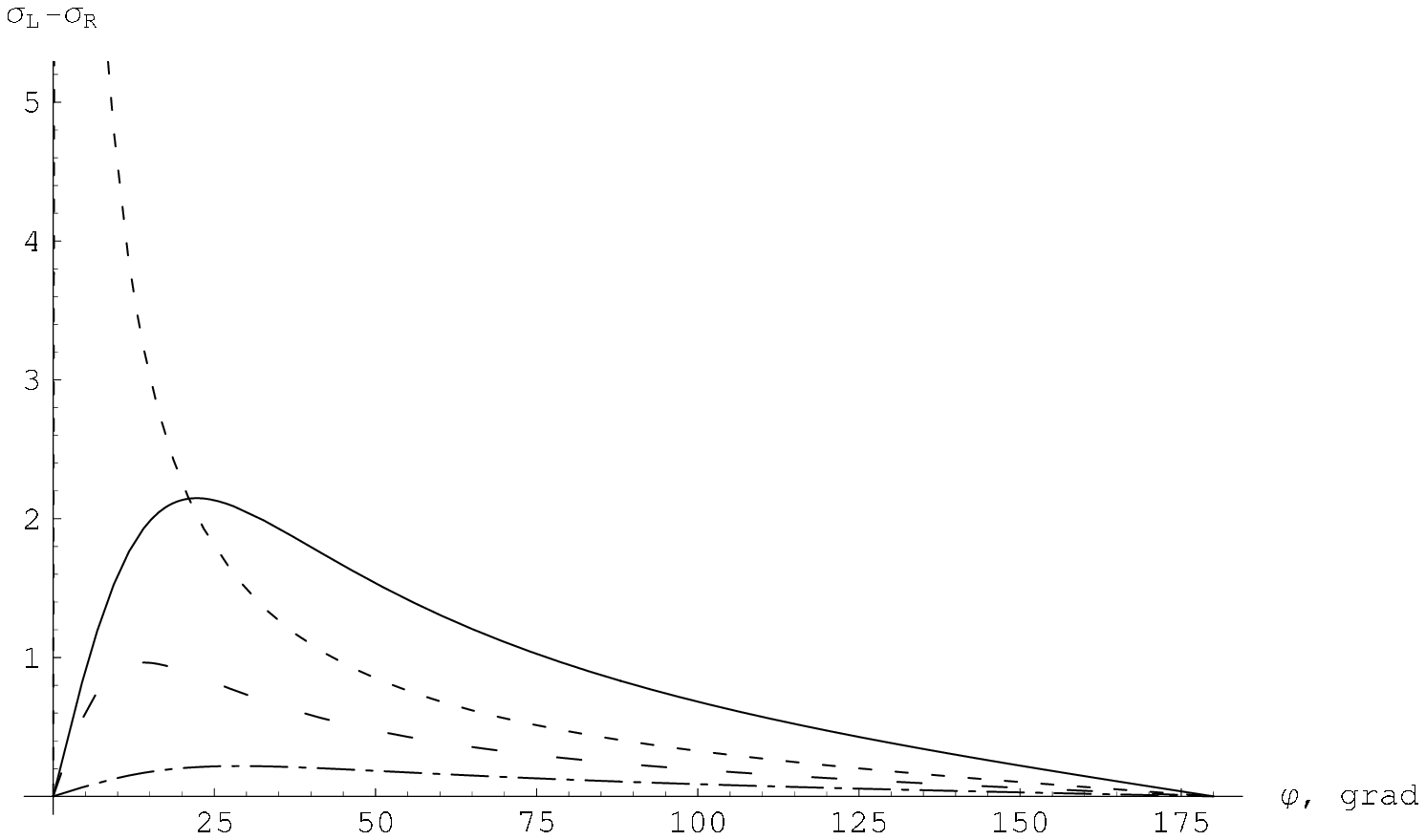}  }
\mbox{
   \epsfxsize=14cm
\epsfysize=8.4cm
 \epsffile{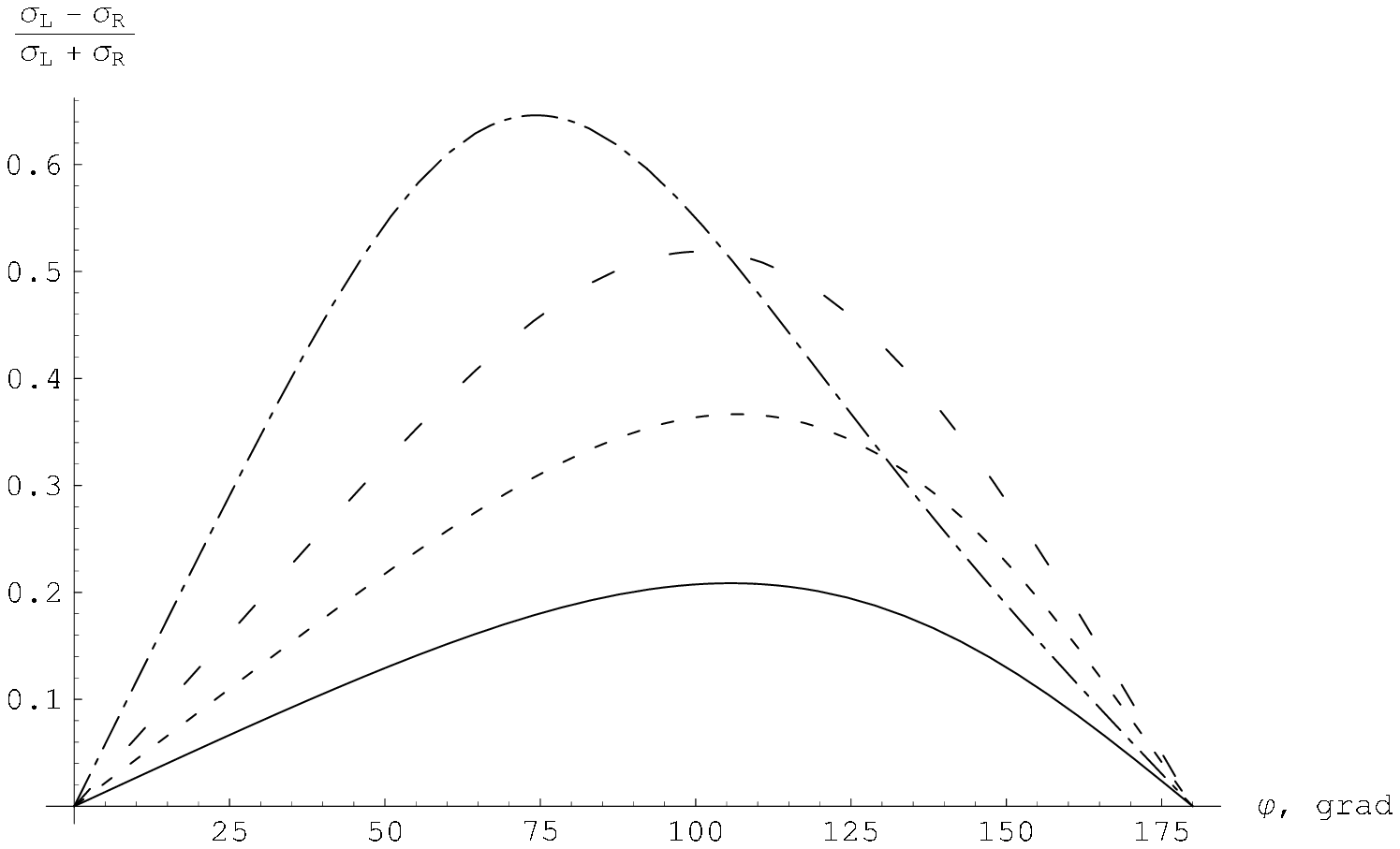}  }
{\caption{\label{res_3} The same for $E_e=6$GeV, $x_B=.15$, and $Q^2=1.5$GeV$^2$. 
   }}
\end{figure}
\begin{figure}[ampl]
\mbox{
   \epsfxsize=14cm
\epsfysize=8.4cm
  \epsffile{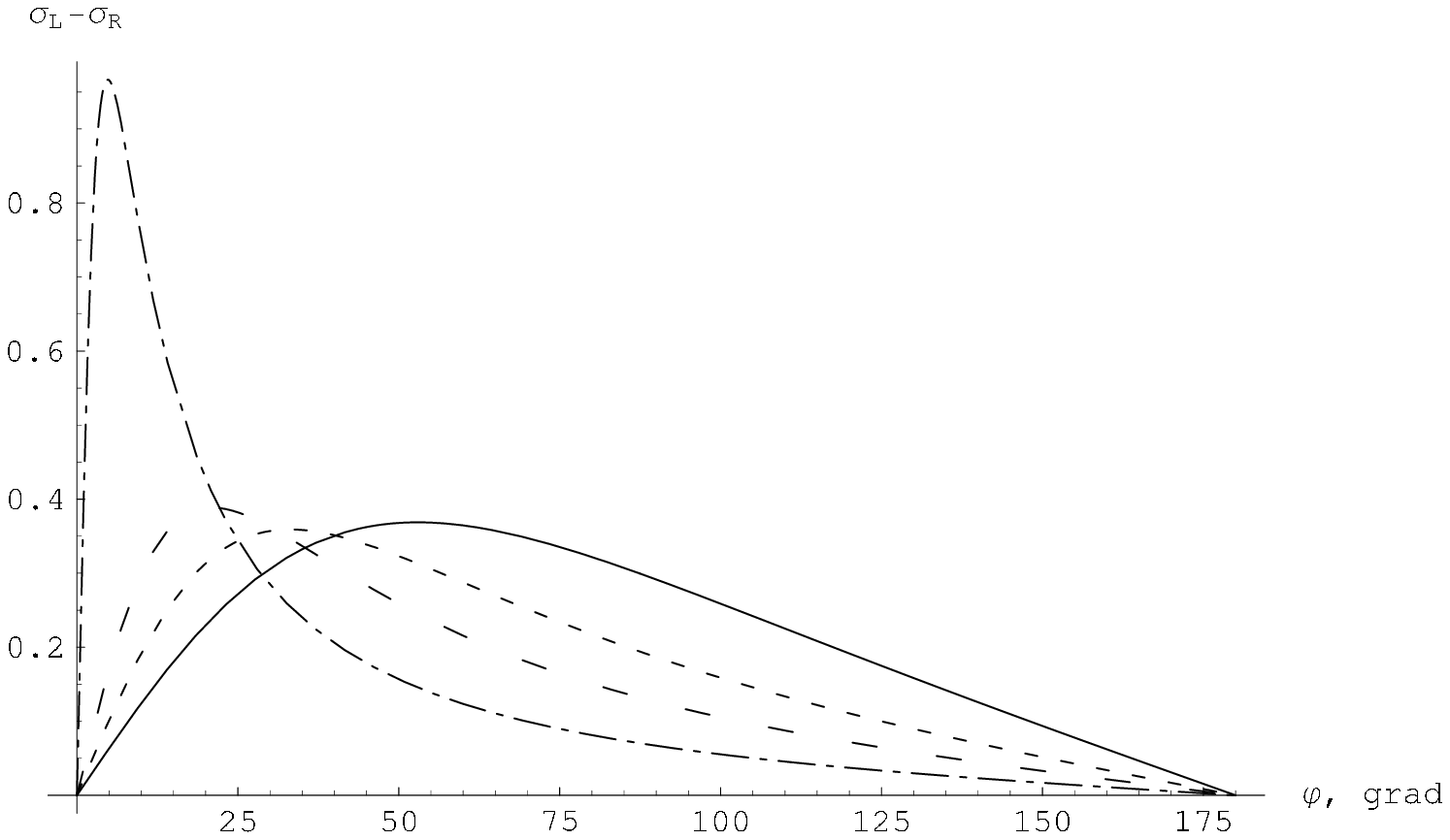}  }
\mbox{
   \epsfxsize=14cm
\epsfysize=8.4cm
 \epsffile{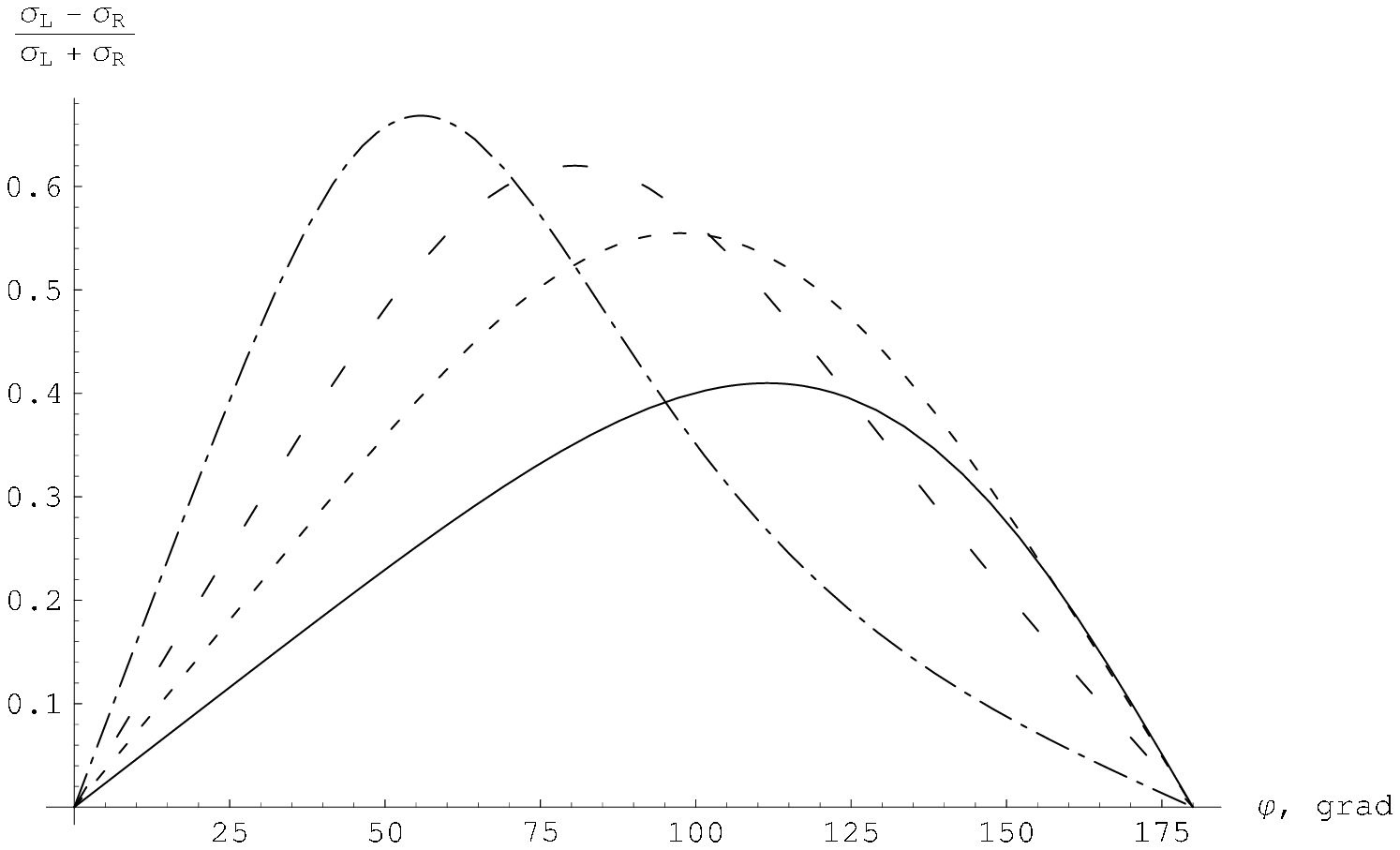}  }
{\caption{\label{res_4}The same for $E_e=6$GeV, $x_B=.2$, and $Q^2=1.5$GeV$^2$. 
   }}
\end{figure}
\begin{figure}[ampl]
\mbox{
   \epsfxsize=14cm
\epsfysize=8.4cm
 \epsffile{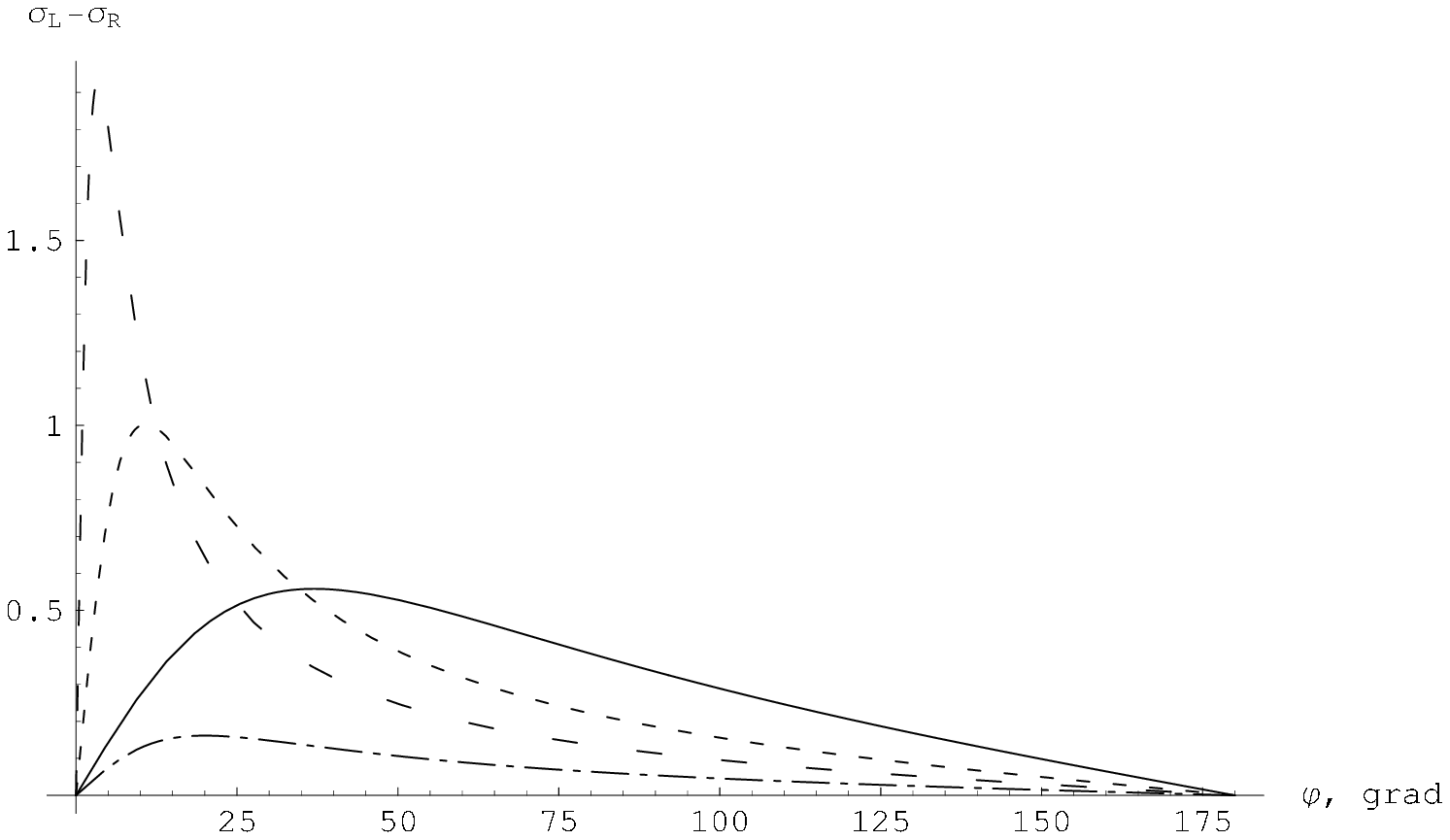}  }
\mbox{
   \epsfxsize=14cm
\epsfysize=8.4cm
 \epsffile{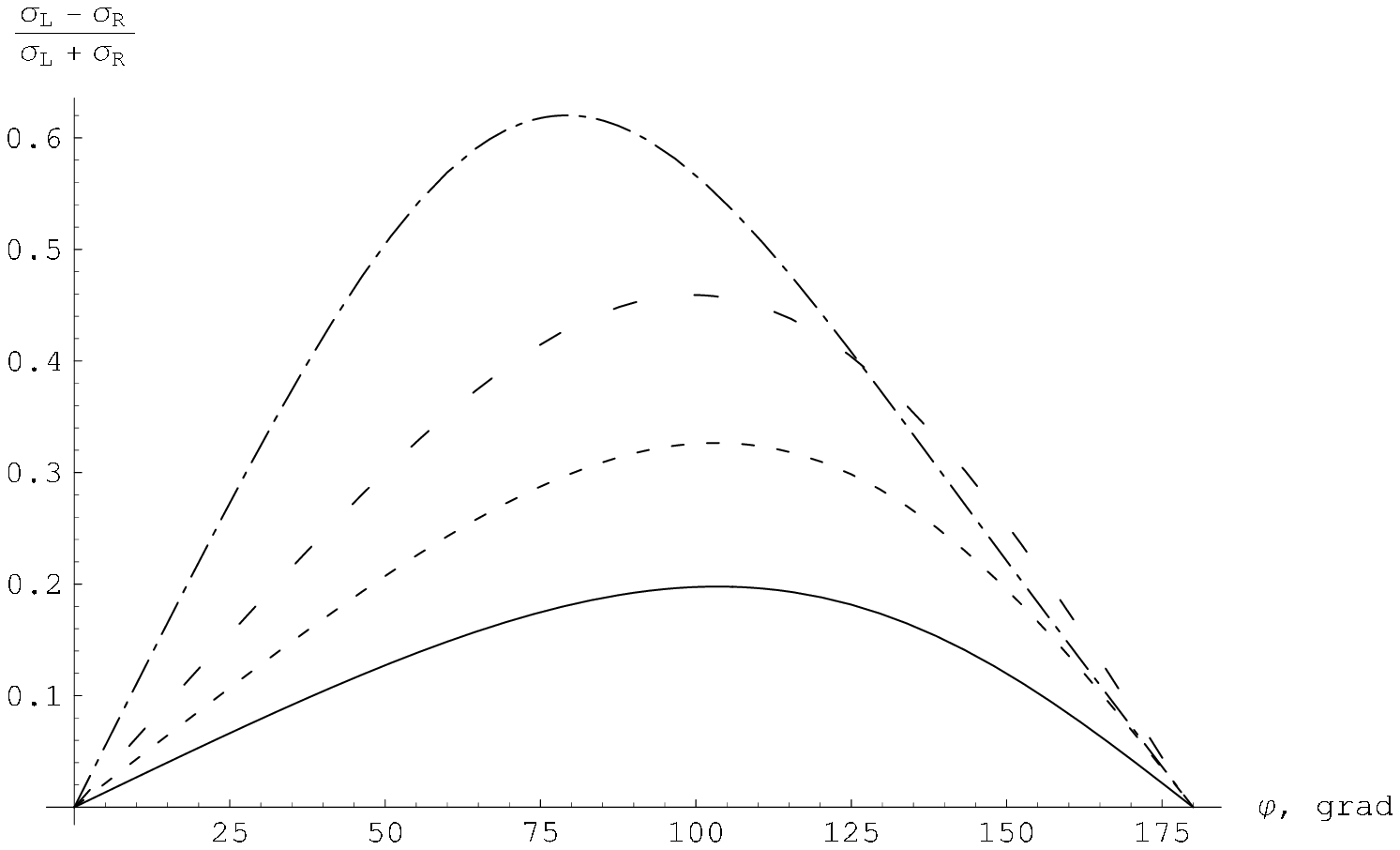}  }
{\caption{\label{res_5} The same for $E_e=6$GeV, $x_B=.2$, and $Q^2=2$GeV$^2$.
   }}
\end{figure}
\begin{figure}[ampl]
\mbox{
   \epsfxsize=14cm
\epsfysize=8.4cm
 \epsffile{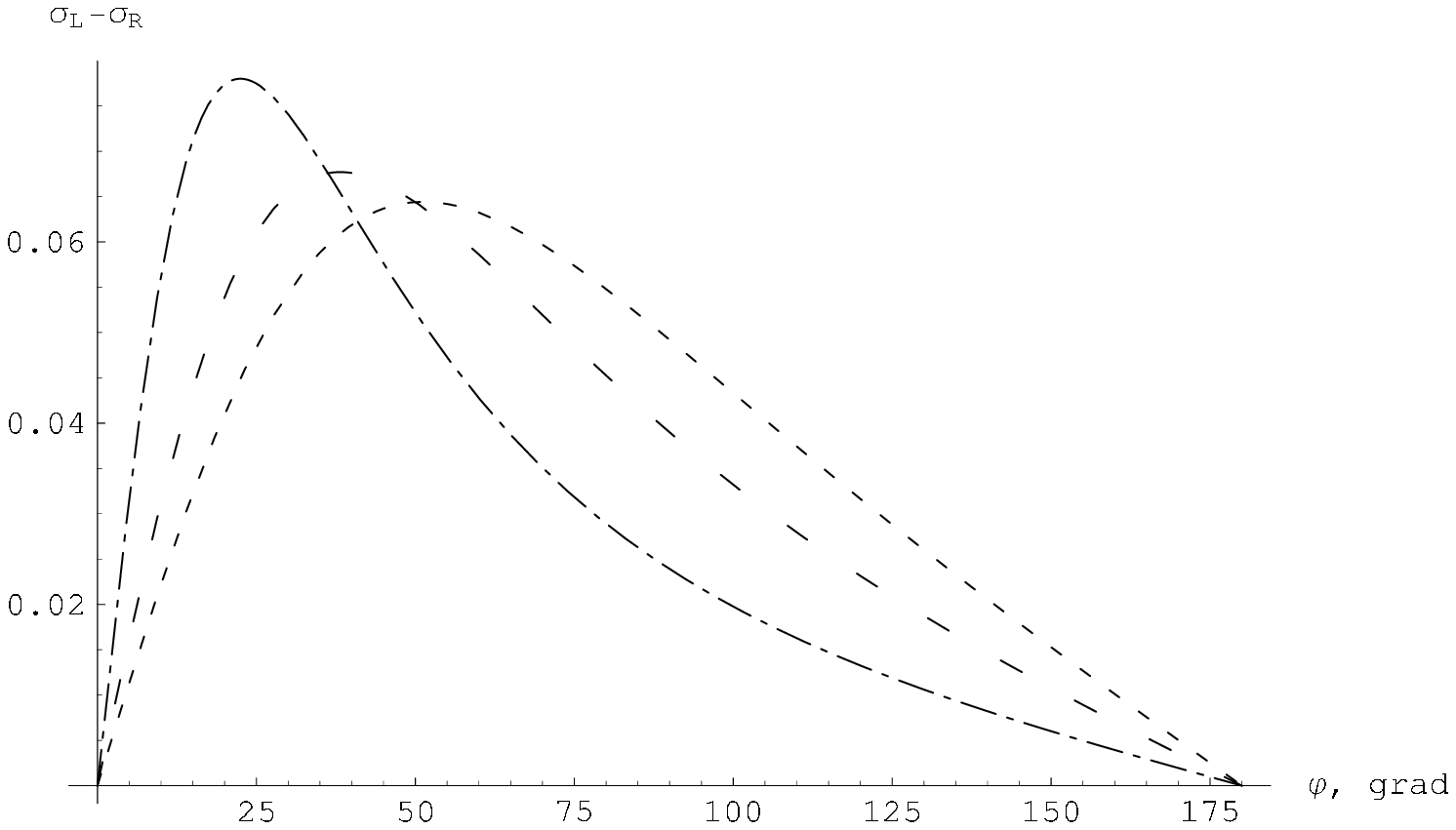}  }
\mbox{
   \epsfxsize=14cm
\epsfysize=8.4cm
 \epsffile{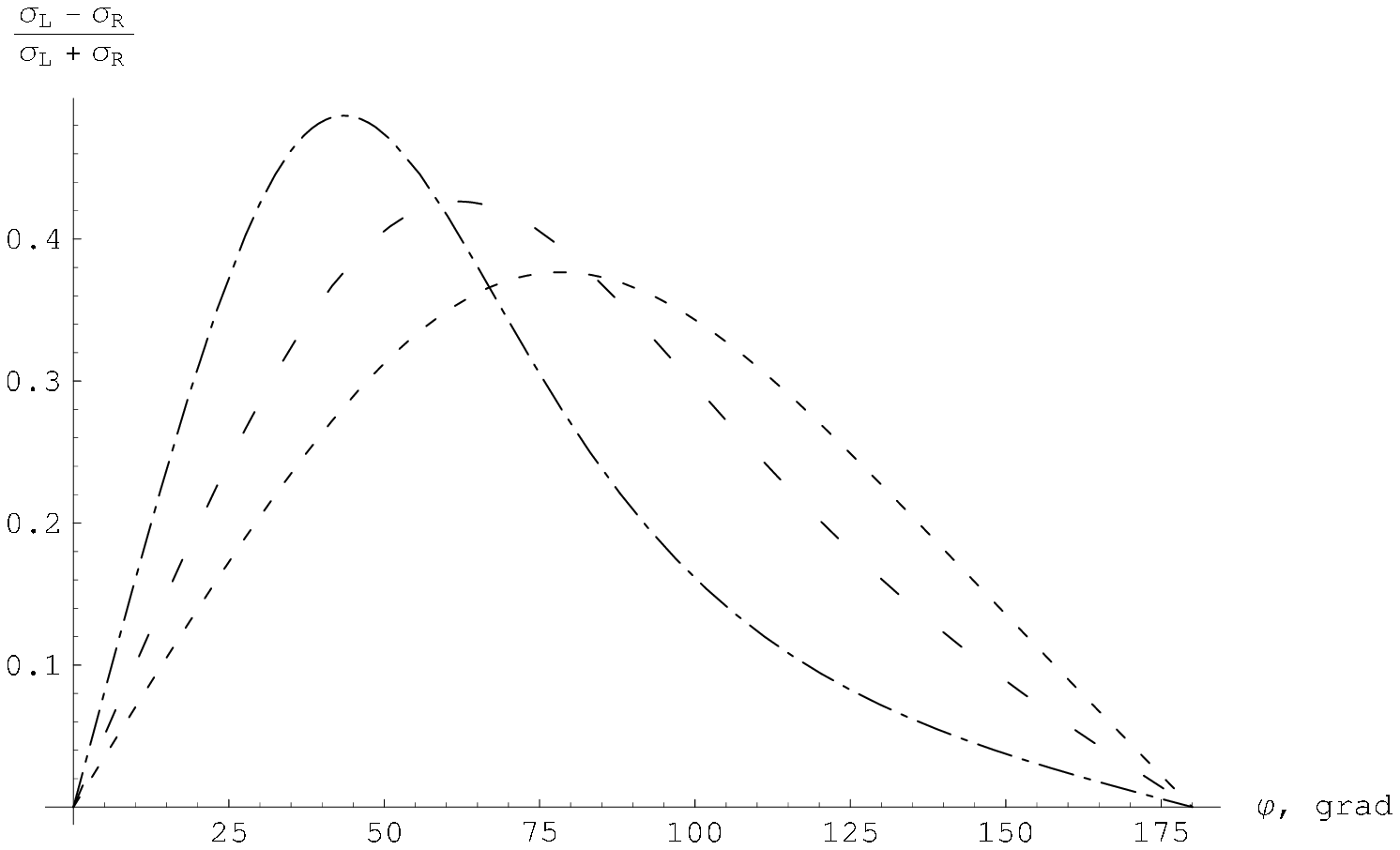}  }
{\caption{\label{res_6}The same for $E_e=6$GeV, $x_B=.3$, and $Q^2=1.5$GeV$^2$.
   }}
\end{figure}
\begin{figure}[ampl]
\mbox{
   \epsfxsize=14cm
\epsfysize=8.4cm
 \epsffile{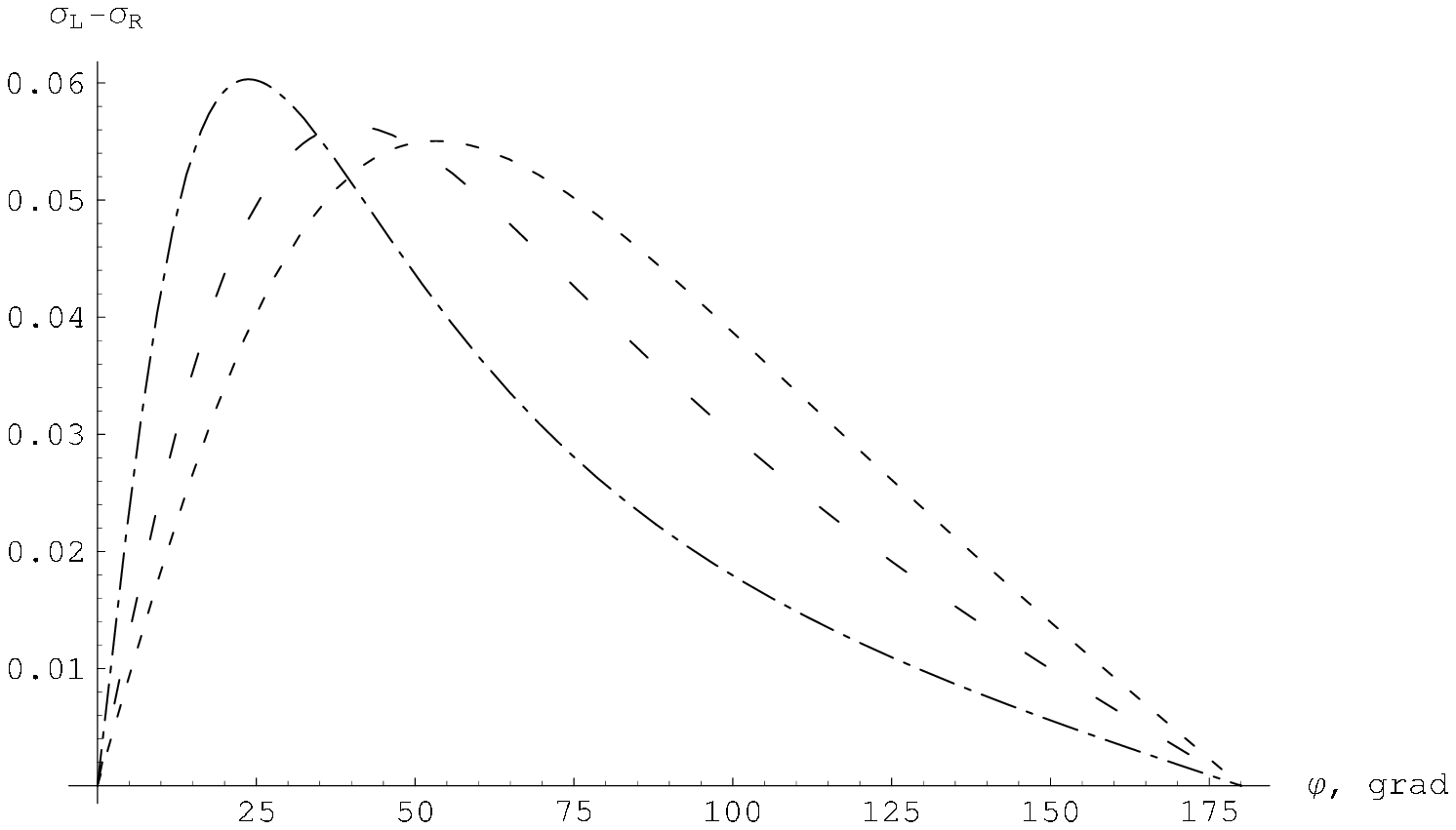}  }
\mbox{
   \epsfxsize=14cm
\epsfysize=8.4cm
 \epsffile{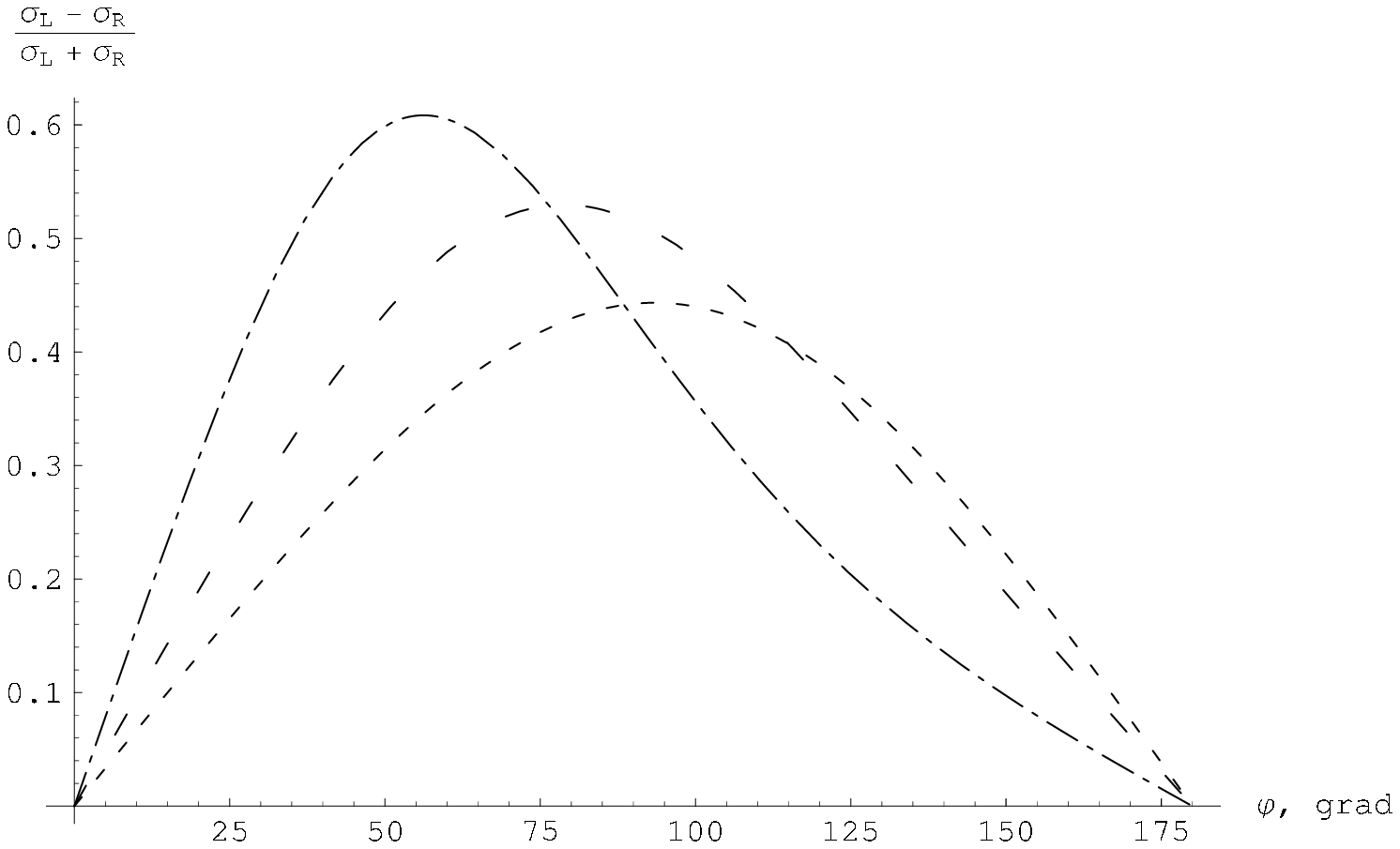}  }
{\caption{\label{res_7} The same for $E_e=6$GeV, $x_B=.3$, and $Q^2=2$GeV$^2$.
   }}
\end{figure}
\begin{figure}[ampl]
\mbox{
   \epsfxsize=14cm
\epsfysize=8.4cm
 \epsffile{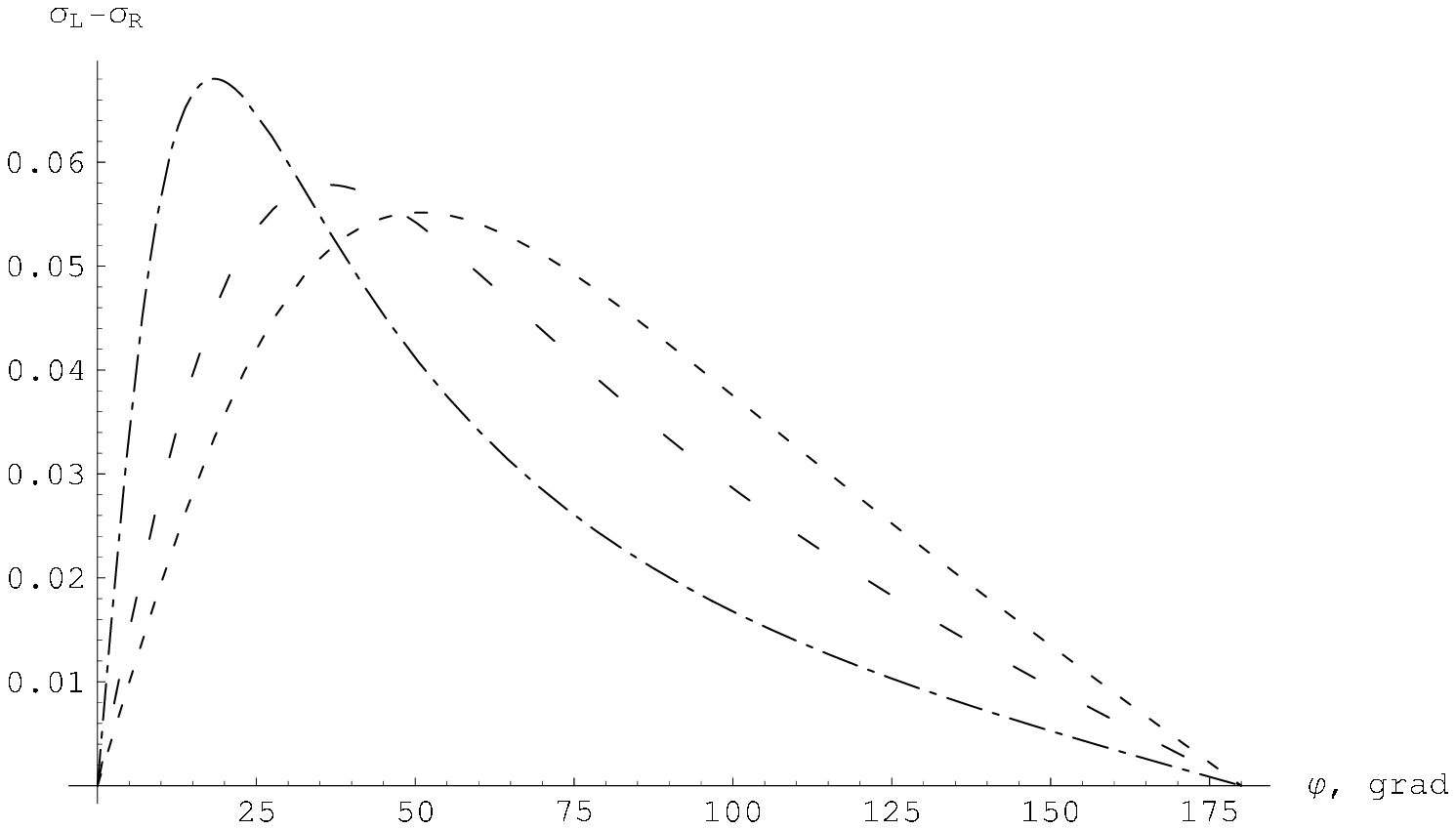}  }
\mbox{
   \epsfxsize=14cm
\epsfysize=8.4cm
 \epsffile{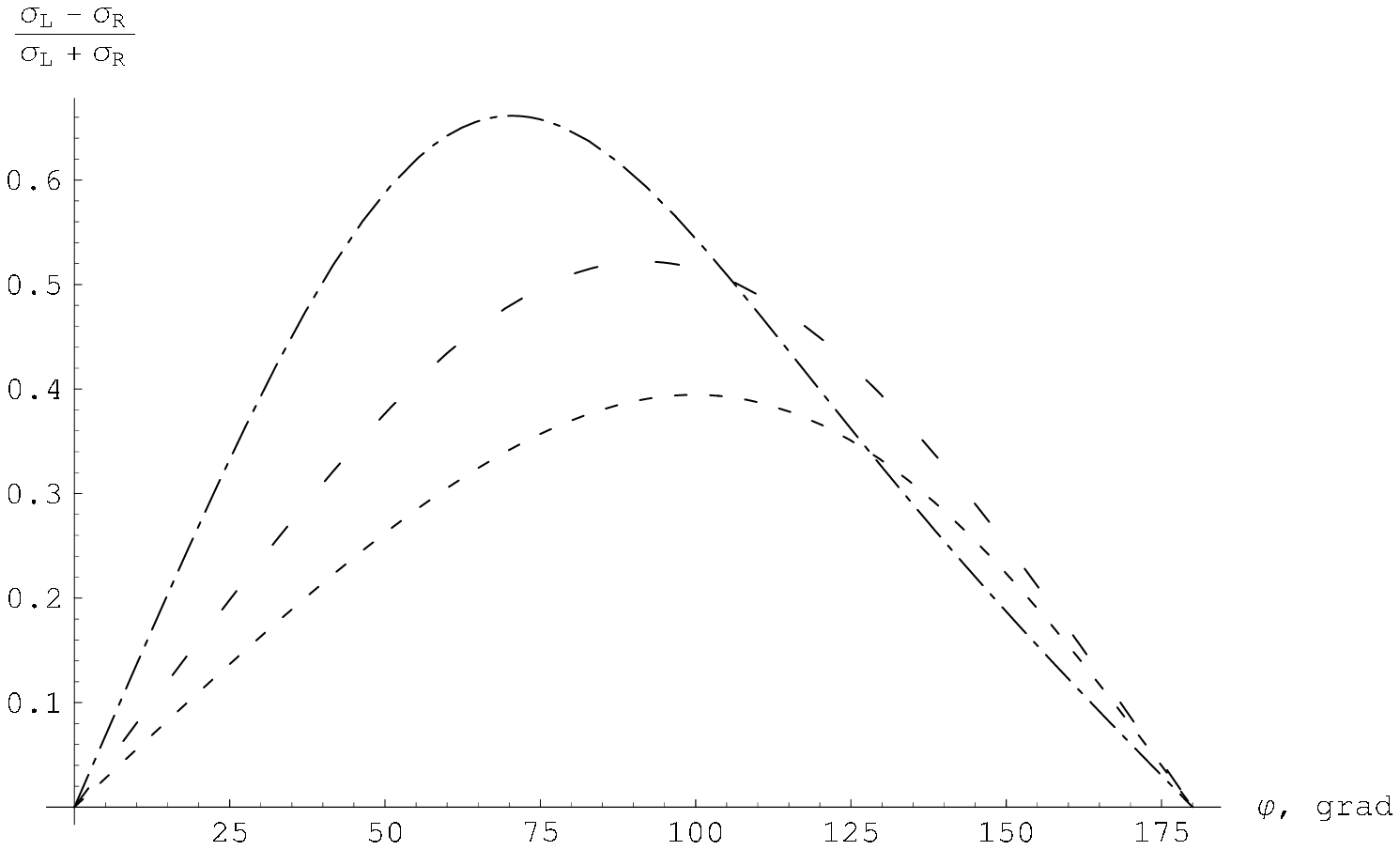}  }
{\caption{\label{res_8} The same for $E_e=6$GeV, $x_B=.3$, and $Q^2=2.5$GeV$^2$. 
   }}
\end{figure}
\begin{figure}[ampl]
\mbox{
   \epsfxsize=14cm
\epsfysize=8.4cm
 \epsffile{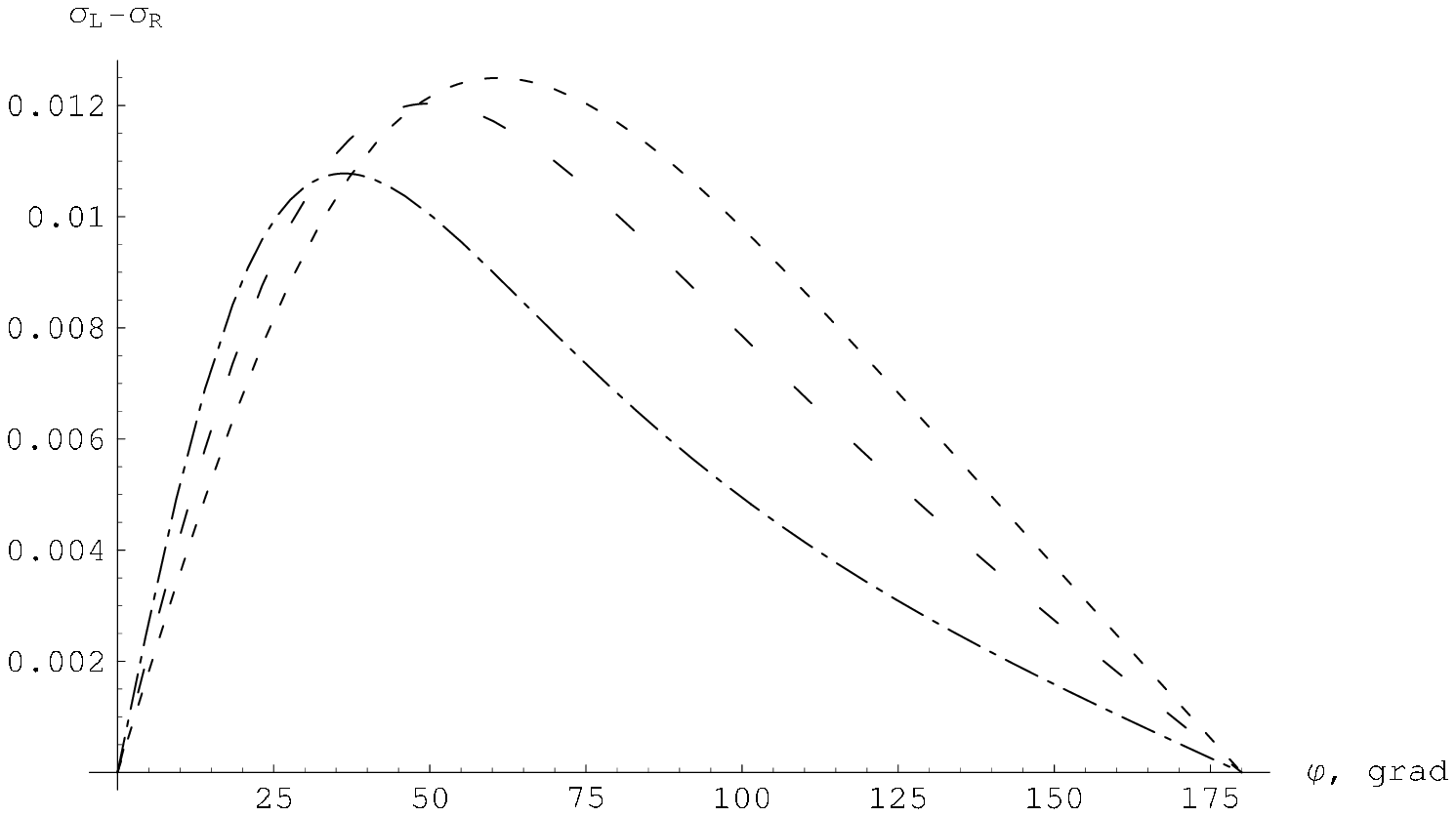}  }
\mbox{
   \epsfxsize=14cm
\epsfysize=8.4cm
  \epsffile{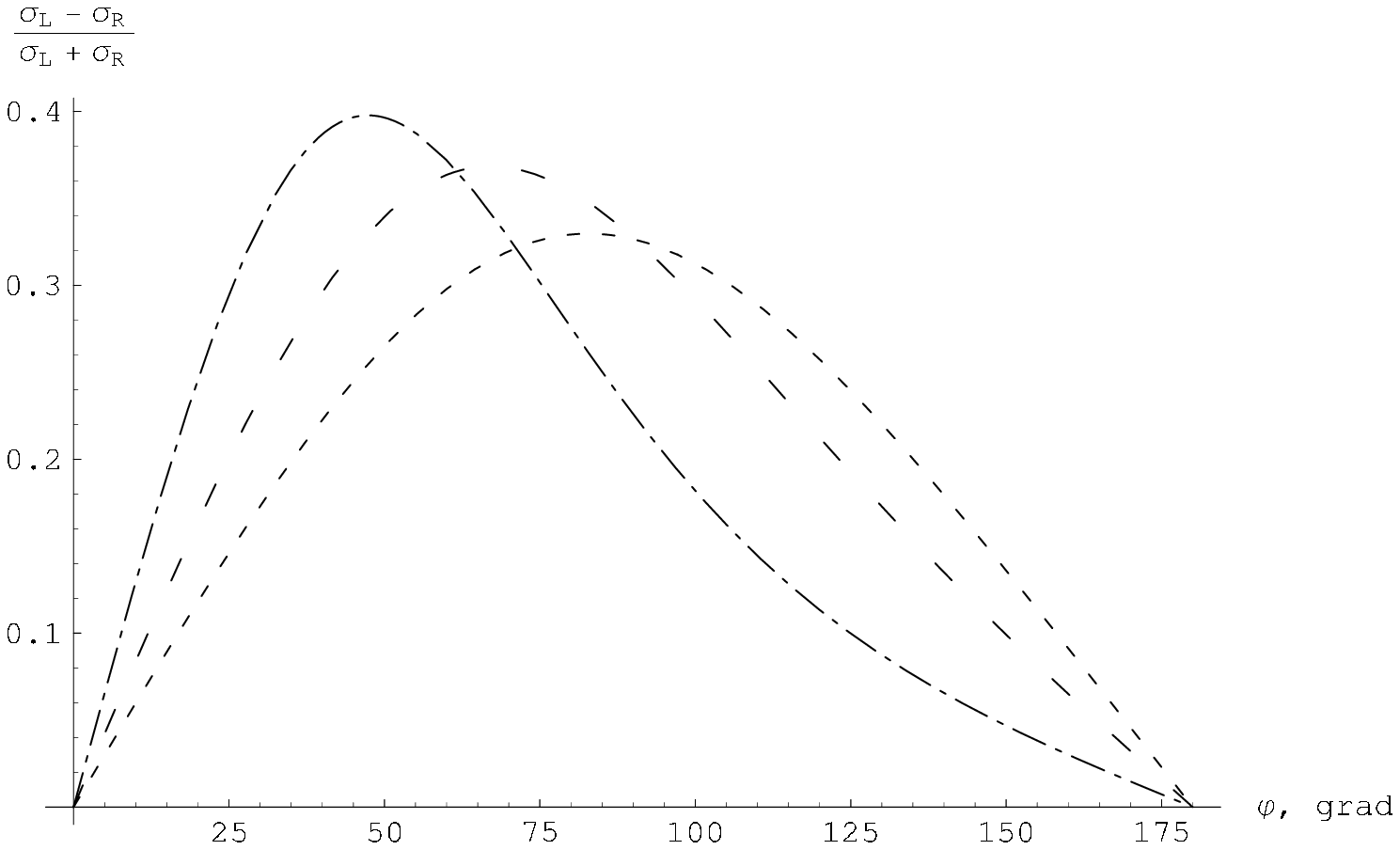}  }
{\caption{\label{res_9} The same for $E_e=11$GeV, $x_B=.3$, and $Q^2=2.5$GeV$^2$. 
   }}
\end{figure}

The imaginary part of the DVCS amplitudes
is directly proportional to the GPDs, evaluated along the line $x =
\xi$, and
measures in this way the `envelope functions'
$H(\xi, \xi, t)$, $E(\xi, \xi, t)$,
$\tilde{H}(\xi, \xi, t)$, and $\tilde{E}(\xi, \xi, t)$ (Eq.~(\ref{dvcs-ImT})).

Figures \ref{res_1} through \ref{res_9} show the results of our calculations for
the energies and kinematics accessible at Jefferson Lab. The values of transferred momentum squared run from $t=-.1$ to $t=-.5$ limited by
\begin{equation}
t_{min}=-4 M^2 \xi^2/(1-\xi^2)
\end{equation}
for each of kinematical sets.

One can see, that the interference term is larger at lower momentum transfer $t$ even though it is responsible 
for a smaller percentage of the total cross section.
As the initial energy of the electron $E_e$ goes up, the interference cross section value drops down by one order. The ratio of the interference term 
to the total cross section, nevertheless, stays the same: about 40 to 60 percents.

%%%%%%%%%%%%%%%%%%%%%%%%%%%%%%%%%%%%%%%%%%%%%%%%%%%%%%%%%%%%%%%%%%%%%%%%%%%%

\section{Twist-3 corrections for the DVCS process}
%\section{Definitions, link with ordinary parton distributions and nucleon form factors}
\label{chap3_1}

The twist-2 DVCS amplitude  (\ref{tvcs_munu}) is independent of $Q$. This means it has a scaling
behavior.
However, this amplitude is
not complete beyond the leading order in $Q$.
Although it is exactly
gauge invariant $q_\mu T^{\mu \nu }=0$ with respect to the virtual photon, the
electromagnetic gauge invariance is violated by the
real photon except in the forward direction $t=0$.
This violation of gauge invariance is a higher twist (twist-3) effect
compared to the leading order term $T^{\mu \nu }$.
Since the product  of final photon 4-momentum and the DVCS amplitude at twist-2 accuracy is proportional to the transverse component of the transferred momentum $\Delta _\perp$,
an improved DVCS amplitude has been
used  \cite{Gui98,Vdh99} to restore the gauge invariance in the
nonforward direction:
\begin{equation}
T^{\mu \nu }_{\mathrm{DVCS}}\, =\, T^{\mu \nu }_{\mathrm{DVCS-2}}\, +\,
{{P^{\nu }}\over {\left( P \cdot q'\right) }}\;
\left( \Delta _\perp \right) _{\lambda }\,
T^{\mu \lambda}_{\mathrm{DVCS-2}}\;.
\label{eq:dvcsgaugeinv}
\end{equation}
It creates a correction term of higher order in $Q$ to the twist-2
DVCS amplitude $T^{\mu \nu }_{\mathrm{DVCS-2}}$.

It is necessary to have an estimate of the effect of power
suppressed (higher twist) contributions to those observables in order to be able to extract the
twist-2 GPDs from DVCS observables at accessible values of the hard
scale $Q$.

The
first power correction to the DVCS amplitude of order ${\cal O}(1/Q)$ is called twist-3.

The twist-3 corrections to the DVCS
amplitude, which have been derived and calculated recently by several groups
\cite{Ani00,Pen00b,Bel00b,Rad00,Kiv01a,Kiv01b,Kiv01c,Rad01a,Bel00c,Bel01b}
using different approaches. In our study we used the notations and definitions from \cite{vander}.
The following section is a short overview of
these definitions.

\subsection{Kinematical variables of the DVCS process}
\label{chap5_3_1}

Let us start with the comparison of the kinematical variables used for the description of
twist-3 amplitude with the set of variables we used in the previous section. There is only a slight difference, but it is worth mentioning.

There are four generalized structure functions used for the description of DVCS amplitude. They are denoted by $H$, $E$, $\tilde{H}$ and $\tilde{E}$ and depend on three
variables. The light-cone momentum  fraction $x$ is defined by $k^+ = xP^+$,
where $k$ is the quark loop momentum and
$P$ is the average nucleon momentum ($P = (p + p')/2$, where $p$ ($p'$)
are the initial (final) nucleon four-momenta respectively). 
In the previous chapter we have used the initial momentum of the hadron $p$ as one of the basic 
light-cone momenta and up to twist-3 accuracy those momentum coincide as will be shown later. 
The skewedness variable $\xi$ is
defined by $\Delta^+ = - 2 \xi P^+$, where $\Delta = p' - p$ is the same as $r$ in the previous chapter, the
overall momentum transfer in the process, and where
$2 \xi \rightarrow x_B/(1 - x_B/2)$ in the Bjorken limit. 
The third variable entering the GPDs
is given by the Mandelstam invariant $t = \Delta^2$, exactly as it was before.

Using momenta $P$ and $\Delta$ instead of $p=P-\Delta/2$ and $p'=P+\Delta/2$ we can define the active quark momentum as
$k^+-\Delta^+/2=(x+\xi)P^+$ before the virtual photon impact and $k^++\Delta^+/2=(x-\xi)P^+$ afterwards.
The light-cone basis is chosen along the positive and negative $z$-direction (both $q^\mu$ and $P^\mu$ are collinear along the $z$-axis
and have opposite direction): $\tilde{p}^{\mu}=P^+/\sqrt{2}(1,0,0,1)$ and $n^\mu ==1/P^+ \cdot 1/\sqrt{2}(1,0,0,-1)$, satisfying $\tilde{p} \cdot \tilde{p}=0$, $n \cdot n=0$, $\tilde{p} \cdot n = 1$ and using the notation $P^+=P\cdot n$.

In this frame, the physical momenta entering the DVCS process
(\ref{eq:proc})
have the following decomposition \cite{Gui98}~:
\begin{eqnarray}
 &  & P^{\mu } \,=\, {1\over 2}\left( p^{\mu }+p'^{\mu }\right)
\,=\, \tilde{p}^{\mu }+{{\bar{m}^{2}}\over 2}\, n^{\mu }\; ,
\label{eq:dvcsp} \\
&  & q^{\mu }=-\left( 2\xi ^{'}\right) \,
\tilde{p}^{\mu }+\left( {{Q^{2}}\over {4\xi ^{'}}}\right) \, n^{\mu}\; ,
\label{eq:dvcsq} \\
&  & \Delta ^{\mu }\equiv p^{'\mu }-p^{\mu }
\,=\,-\left( 2\xi \right) \, \tilde{p}^{\mu }
+\left( \xi \, \bar{m}^{2}\right) \, n^{\mu }+\Delta^{\mu }_{\perp }\; ,
\label{eq:dvcsxi} \\
&  & q^{'\mu }\equiv q^{\mu }-\Delta ^{\mu } \,=\,-2\left( \xi ^{'}-\xi
 \right) \, \tilde{p}^{\mu }+\left( {{Q^{2}}\over {4\xi ^{'}}}-\xi \,
 \bar{m}^{2}\right) \, n^{\mu }-\Delta ^{\mu }_{\perp }\; ,
\label{eq:dvcsqp}
\end{eqnarray}
where $\Delta_\perp$ is the perpendicular component of the
momentum transfer $\Delta$
( i.e. $\tilde{p} \cdot \Delta_\perp = n \cdot \Delta_\perp = 0$ ),
and where the variables \( \bar{m}^{2} \), \( \xi ^{'} \) and \( \xi  \) are
given by
\begin{eqnarray}
&  & \bar{m}^{2}\,=\,{M}^{2}-{{\Delta ^{2}}\over 4}\; ,
\label{eq:dvcskin} \\
&  & 2\xi'=-q^+={{P\cdot q}\over {\bar{m}^{2}}}\,
\left[ -1+\sqrt{1+{{Q^{2}\,\bar{m}^{2}}\over {(P\cdot q)^{2}}}}\right] \;
\stackrel{Bj}{\longrightarrow }\; \frac{x_{B}}{1-\frac{x_{B}}{2}}\; ,\\
&  & 2\xi=-\Delta^+= 2\xi ^{'}\, {{Q^{2}-\Delta ^{2}}\over {Q^{2}
+\bar{m}^{2}(2\xi ^{'})^{2}}}\;
\stackrel{Bj}{\longrightarrow }\; \frac{x_{B}}{1-\frac{x_{B}}{2}}\; .
\label{eq:dvcskin3}
\end{eqnarray}

To twist-3 accuracy, Eqs.~(\ref{eq:dvcsp}-\ref{eq:dvcsqp}) reduce to
\begin{eqnarray}
P &=& \tilde{p}, \nonumber \\
\Delta &=& -2 \xi P + \Delta_\perp, \\
q &=&- 2 \xi P +\frac{Q^2}{4\xi} n,\nonumber \\
q'&=& \frac{Q^2}{4\xi} n - \Delta_\perp.\nonumber
\label{eq:kintw3}
\end{eqnarray}

Thus, the final (real) photon 4-momentum squared $q'^2->\Delta_\perp^2$ suggests the twist-3 accuracy $t=\Delta^2 \to \Delta_\perp^2 \to 0$.

\subsection{Twist-3 DVCS amplitude}
\label{chap5_3_3}

As a starting point we use the DVCS amplitude on the nucleon 
to the order ${\cal O}(1/Q)$ derived explicitly in a parton model approach \cite{Pen00b} and
in a light-cone expansion framework \cite{Bel00b}:
\begin{eqnarray}
&&H^{ \mu \nu} = \frac12 \int_{-1}^1 dx\quad \biggl\{ \left[(-g^{\mu
\nu})_\perp-\frac{P^\nu\Delta_\perp^\mu}{(P \cdot q')}
\right] \; n^\beta {\cal F}_\beta (x,\xi) \; C^+(x,\xi)
\label{eq:Tdvcstw3} \\
&&- \left[(-g^{\nu k})_\perp
-\frac{P^\nu\Delta_\perp^k}{(P \cdot q')}  \right]
i ({\epsilon_{k }}^{\mu})_\perp \; n^\beta \tilde{{\cal F}}_\beta (x,\xi)
\; C^-(x,\xi) \nonumber \\
&&- \frac{(q+4\xi P)^\mu}{(P \cdot q)}
\left[(-g^{\nu k})_\perp
-\frac{P^\nu\Delta_\perp^k}{(P \cdot q')}  \right]
\left\{ {\cal F}_k(x,\xi) \; C^+(x,\xi) \right. \nonumber \\
&&\left. ~~~~~~~~~~~~~~~~~~~~~~~~~~~~~~~~~~~~~~~~~~~~ - i (\epsilon_{k \rho})_\perp
\tilde{{\cal F}}^\rho (x,\xi) \; C^-(x,\xi)\right\} \biggr\} \, , \nonumber
\end{eqnarray}
where the functions ${\cal F}_\mu$ and $\tilde{{\cal F}}_\mu$ are given by:
\begin{eqnarray}
&& {\cal F}_\mu(x,\xi)= \frac{\Delta_\mu}{2 M\xi} \bar{N}(p'){} N(p)
E(x,\xi)- \frac{\Delta_\mu}{2\xi}\bar{N}(p') \hat{n} N(p)(H+E)(x,\xi)
\nonumber\\
&&~~~~+\int_{-1}^{1}du\
G_\mu(u,\xi)W_{+}(x,u,\xi)+i\epsilon_{\perp \mu k}
\int_{-1}^{1}du\  \tilde{G}^k (u,\xi)W_{-}(x,u,\xi)\, ,
\label{eq:F}
\\[4mm]
&&\tilde{{\cal F}}_\mu(x,\xi)=
\frac{\Delta_\mu}{2M}  \bar{N}(p'){\gamma_5} N(p)\tilde{E}(x,\xi)-
\frac{\Delta_\mu}{2\xi} \bar{N}(p') \hat{n} {\gamma_5} N(p)\tilde{H}(x,\xi) \nonumber \\
&&~~~~+\int_{-1}^{1}du\ \tilde{G}_\mu(u,\xi)W_{+}(x,u,\xi)+i \epsilon_{\perp \mu k}
\int_{-1}^{1}du\  G^k (u,\xi)W_{-}(x,u,\xi) \, .
\label{eq:Ft}
\end{eqnarray}
The following notations are used~:
\begin{eqnarray}
\label{eq:G}
&&G^\mu(u,\xi)=  \bar{N}(p'){\gamma^\mu_\perp} N(p)(H+E)(u,\xi)\nonumber \\[4mm]&&~~~~~~~~~~+
\frac{\Delta_\perp^\mu}{2\xi M}  \bar{N}(p'){} N(p)
\biggl[u\frac{\partial}{\partial u}+ \xi\frac{\partial}{\partial
\xi} \biggl] E(u,\xi) \nonumber \\[4mm]&&~~~~~~~~~~~~~~~~~~~~
-\frac{\Delta_\perp^\mu}{2\xi}
 \bar{N}(p')\hat{n} N(p)\biggl[u\frac{\partial}{\partial u}+
\xi\frac{\partial}{\partial \xi}\biggl] (H+E)(u,\xi) \, ,
\end{eqnarray}
\begin{eqnarray}
\label{eq:tG}
&&\tilde{G}^\mu (u,\xi)=\bar{N}(p'){\gamma^\mu_\perp\gamma_5} N(p) \tilde{H}(u,\xi)\nonumber \\[4mm]&&~~~~~~~~~~
+\frac{\Delta_\perp^\mu}{2M}  \bar{N}(p'){\gamma_5} N(p)
\biggl[1+u\frac{\partial}{\partial u}+\xi\frac{\partial}{\partial
\xi}\biggl] \tilde{E}(u,\xi) \nonumber\\[4mm]&&~~~~~~~~~~~~~~~~~~~~
-\frac{\Delta_\perp^\mu}{2\xi} \bar{N}(p')\hat{n}{\gamma_5} N(p)
\biggl[u\frac{\partial}{\partial u}+\xi\frac{\partial}{\partial
\xi}\biggl] \tilde{H}(u,\xi) \, .
\end{eqnarray}
The functions $W_{\pm}(x,u,\xi)$ are called \cite{Wan77}
Wandzura-Wilczek  kernels. They were
introduced in Ref.~\cite{Bel00b,Kiv01a} and are
defined as~:
\begin{eqnarray}
\label{eq:Wpm}
W_{\pm}(x,u,\xi)&=& \frac12\biggl\{
\theta(x>\xi)\frac{\theta(u>x)}{u-\xi}-
\theta(x<\xi)\frac{\theta(u<x)}{u-\xi} \biggl\} \nonumber
\\[4mm]&&\mskip-10mu \pm\frac12\biggl\{
\theta(x>-\xi)\frac{\theta(u>x)}{u+\xi}-
\theta(x<-\xi)\frac{\theta(u<x)}{u+\xi} \biggl\}.
\end{eqnarray}

We also
introduce the metric and totally antisymmetric tensors in the two
dimensional transverse plane ($\varepsilon_{0123}=+1$)~:
\begin{eqnarray}
\label{gt}
(-g^{\mu \nu})_\perp = -g^{\mu \nu}+ n^\mu \tilde{p}^{ \nu}+n^\nu
\tilde{p}^{ \mu},
\quad \epsilon^\perp_{\mu \nu}=
\epsilon_{\mu \nu \alpha\beta}n^\alpha \tilde{p}^{\beta} \, .
\end{eqnarray}
and  the coefficient functions
$C^{\pm}(x, \xi)$ are defined as~:
\begin{eqnarray}
C^\pm(x,\xi)=\frac{1}{x-\xi+i\varepsilon}\pm
\frac{1}{x+\xi-i\varepsilon}.
\label{eq:alf}
\end{eqnarray}

In the expression Eq.~(\ref{eq:Tdvcstw3})
for the DVCS amplitude to the twist-3 accuracy,
the first two terms correspond to the scattering of transversely
polarized virtual photons. Applying all the definitions to the first two terms of the expression Eq.~(\ref{eq:Tdvcstw3})
brings us to the twist-2 DVCS amplitude like we had in the previous chapter. The vector part
\begin{equation}
T_1=\frac{1}{2}(-g_{\perp}^{\mu\nu}-\frac{P^\nu\Delta_{\perp}^{\mu}}{(P \cdot q')})\Big(-\frac{1}{M}\bar{u}(p')u(p)C_{11}+\bar{u}(p')\hat{n}u(p)C_{12}\Big),
\end{equation}
with
\begin{equation}
C_{11}(\xi)=\int_{-1}^1 E(x,\xi)C^+(x,\xi)dx, 
\end{equation}
\begin{equation}
C_{12}(\xi)=\int_{-1}^1 (E(x,\xi)+H(x,\xi))C^+(x,\xi)dx, 
\end{equation}
and the axial-vector part
\begin{equation}
T_2=-\frac{i \epsilon_{\mu}^k}{2}(-g_{\perp}^{\nu k}-\frac{P^\nu \Delta_{\perp}^{\mu}}{(P \cdot q')})
\Big(-\frac{\xi}{M}\bar{u}(p')\gamma_5 u(p)C_{21}+\bar{u}(p')\hat{n}\gamma_5 u(p)C_{22}\Big),
\end{equation}
with
\begin{equation}
C_{21}(\xi)=\int_{-1}^1 \tilde{E}(x,\xi)C^-(x,\xi)dx, 
\end{equation}
\begin{equation}
C_{22}(\xi)=\int_{-1}^1 \tilde{H}(x,\xi)C^-(x,\xi)dx.
\end{equation}
This part of the amplitude,
containing $n_\beta \, {\cal F}^\beta$ and $n_\beta \,
\tilde{{\cal F}}^\beta$, depends
only on the twist-2 GPDs $H,E$ and $\tilde{H}, \tilde{E}$ and was
elaborated in Refs.~\cite{Gui98,Vdh99}.
Comparison with structure blocks of Eq.~(\ref{tvcs_munu}) gives us the idea of how the structure functions $T^a$ are
related to the structure functions parametrizing the tensor $E(x,\xi)$, vector $H(x,\xi)$, pseudoscalar $\tilde{E}(x,\xi)$, and axial-vector $\tilde{H}(x,\xi)$ transitions:
\begin{equation}
C_{11}(\xi) \leftrightarrow \frac{2}{1+\xi}T^a_K(\xi),
\end{equation}
\begin{equation}
C_{12}(\xi)-C_{11}(\xi) \leftrightarrow \frac{1}{(1+\xi)}T^a_F(\xi),
\end{equation}
\begin{equation}
C_{21}(\xi) \leftrightarrow \frac{1}{2(1+\xi)}T^a_P(\xi),
\end{equation}
\begin{equation}
C_{12}(\xi) \leftrightarrow -\frac{1}{2(1+\xi)}T^a_G(\xi).
\end{equation}
The generalized parton distributions coincide with the quark distributions at vanishing momentum
transfer and the first moments of the GPDs are related to
the elastic form factors
of the nucleon through model independent sum rules \cite{ji,ji2} :
\begin{eqnarray}
\int_{-1}^{+1}dx\, H^{q}(x,\xi ,t)\, &=&\, F_{1}^{q}(t)\, ,
\label{eq:ffsumruleh}\\
\int _{-1}^{+1}dx\, E^{q}(x,\xi ,t)\, &=&\, F_{2}^{q}(t)\, ,
\label{eq:ffsumrulee}\\
\int_{-1}^{+1}dx\, \tilde{H}^{q}(x,\xi ,t)\, &=&\, g_{A}^{q}(t)\, ,
\label{eq:ffsumruleht}\\
\int _{-1}^{+1}dx\,\tilde{E}^{q}(x,\xi ,t)\, &=&\, h_{A}^{q}(t)\,,
\label{eq:ffsumruleet}
\end{eqnarray}
while the second Mellin moment is closely related to the quark orbital momentum contribution to the proton spin \cite{Pen00b}.

The third term in Eq.~(\ref{eq:Tdvcstw3}) corresponds to
the contribution of the longitudinal polarization of the virtual
photon.
 Defining the polarization vector of the virtual photon as
\begin{eqnarray}
\varepsilon_L^\mu(q)=\frac{1}{Q}\biggl(
2\xi P^\mu+\frac{Q^2}{4\xi} n^\mu \biggr)\, ,
\label{eq:epsL}
\end{eqnarray}
we can easily calculate the DVCS amplitude for longitudinal
polarization of the virtual photon ($L\to T$ transition),
which is purely of twist-3~:
\begin{eqnarray}
(\varepsilon_L)_\mu \; H^{\mu\nu}=\frac{2\xi}{Q}\int_{-1}^1dx
\ \Biggl( {\cal F}_\perp^\nu\  C^+(x,\xi)- i\varepsilon_\perp^{\nu k}
\tilde{{\cal F}}_{\perp k}\  C^-(x,\xi)
\Biggr)\, .
\label{eq:LtoT}
\end{eqnarray}

It is therefore seen that this term depends only
on new `transverse' GPDs ${\cal F}_\perp^\mu$ and
${\tilde{\cal F}}_\perp^\mu$,
which can be related to the twist-2 GPDs
$H,E,\tilde{H}$ and $\tilde{E}$ with help of
Wandzura-Wilczek relations  given by Eqs.~(\ref{eq:F}-\ref{eq:tG}).
Twist-3 part of the amplitude could be divided into three separate terms. Two of them 
\begin{eqnarray}
T_3&=&-\frac{1}{2}\frac{q^\mu+4\xi P^\mu}{(P \cdot q)}(-g_{\perp}^{\nu k}-\frac{P^\nu\Delta_{\perp}^k}{(P \cdot q')}) \\&&
\Big(\bar{u}(p')u(p)\Delta_{\perp k}C_{31}+\bar{u}(p')\hat{n} u(p)\Delta_{\perp k}C_{32}\Big), \nonumber
\end{eqnarray}
and
\begin{eqnarray}
T_4&=&-\frac{i\epsilon_k^l}{2}\frac{q^\mu+4\xi P^\mu}{(P \cdot q)}(-g_{\perp}^{\nu k}-\frac{P^\nu\Delta_{\perp}^k}{(P \cdot q')}) \\&&
\Big(\bar{u}(p')\gamma_5u(p)\Delta_{\perp l} C_{42}+\bar{u}(p')\hat{n}\gamma_5 u(p)\Delta_{\perp l} C_{43}\Big), \nonumber
\end{eqnarray}
are similar to $T_1$ and $T_2$ and give no output to the cross section correction at twist-3 accuracy. The third one
\begin{eqnarray}
T_5&=&-\frac{1}{2}\frac{q^\mu+4\xi P^\mu}{(P \cdot q)}(-g_{\perp}^{\nu k}-\frac{P^\nu \Delta_{\perp}^k}{(P \cdot q')}) \\&&
\Big(\bar{u}(p')\gamma_k u(p)C_{33}+i \epsilon_k^l \bar{u}(p')\gamma_l\gamma_5 u(p)C_{41}\Big),\nonumber
\end{eqnarray}
with
\begin{eqnarray}\label{c33}
C_{33}(\xi)&=&\int_{-1}^1 du \Bigg\{(E(u,\xi)+H(u,\xi))\\
&&\int_{-1}^1dx \left(C^+(x,\xi)W^+(x,u,\xi)-C^-(x,\xi)W^-(x,u,\xi)\right)\Bigg\}, \nonumber
\end{eqnarray}
\begin{eqnarray}\label{c41}
C_{41}(\xi)&=&\int_{-1}^1 du \Bigg\{\tilde{H}(u,\xi) \\
&&\int_{-1}^1 dx\left(C^+(x,\xi)W^-(x,u,\xi)-C^-(x,\xi)W^+(x,u,\xi)\right)\Bigg\}, \nonumber
\end{eqnarray}
is the only one that needs to be calculated for the estimate of twist-3 corrections.

%%%%%%%%%%%%%%%%%%%%%%%%%%%%%%%%%%%%%%%%%%%%%%%%%%%%%%%%%%%%%%%%%%%%%%%%%%%%%%%%%%%%%%%%%%%%%%%%%%%%%%%%%%%%%%%%%%%%%%%%%%

\subsection{Twist-3 DVCS cross section}
 In the leading twist-2 approximation, the amplitude squared falls off as $1/Q^4$:

\begin{eqnarray}
\vert T_{twist 2}\vert^2&=&\frac{2 e^6}{Q^4  \xi^2 (1-\xi^2)}\Big((\beta^2+\xi^2) Q^2
-4\beta\xi(k_{\perp} \cdot \Delta_{\perp})\Big) 
  \nonumber \\&&
\Big(\vert C_{11}-(1-\xi^2)C_{12} \vert^2+\vert \xi^2 C_{21}-(1-\xi^2)C_{22} \vert^2\Big).
\end{eqnarray}
Next to the leading, twist-3, approximation is suppressed by $1/Q^2$:
\begin{eqnarray}
\vert T_{twist 3}\vert^2&=&\frac{8e^6(Q^2-2t)M^2}{Q^4(Q^2+t)^2(1-\xi^2)}\Big((\beta^2+3\xi^2) Q^2
-4\beta\xi(k_{\perp} \cdot \Delta_{\perp})\Big) \nonumber \\&&
\Big(-12 \xi^2 \vert C_{33}\vert^2+(4 \xi C_{33}+C_{41}\vert^2\Big).
\end{eqnarray}
Here we used the notation:
\begin{equation}
\beta\equiv \beta_{k}+\xi=\xi-4\xi^2\frac{k \cdot q'}{Q^2}.
\end{equation}

Coefficient functions $C_{33}$ and $C_{41}$ defined by Eqs.~(\ref{c33}) and (\ref{c41}) are affected by 
the Wandzura-Wilczek transformation. 
To better understand the properties of the WW transformations let us
 consider two limiting cases of the WW transformation: the forward
limit $\xi\to 0$ and the `meson' limit $\xi\to 1$. In the forward
limit we easily obtain:
\begin{eqnarray}
\nonumber \lim_{\xi\to 0}
\int_{-1}^1du\
W_+(x,u,\xi)\,\, f(u,\xi)&=&\, \theta(x\ge 0)\int_x^1 \,\,\frac{f(u,0)}{u}\
du\,\nonumber \\
&-& \,\theta(x\le 0)\int_{-1}^x \frac{f(u,0)}{u}\ du\,
,\\ \lim_{\xi\to 0} \int_{-1}^1du\
W_-(x,u,\xi)\,\, f(u,\xi)&=&0\, .
\label{fwd} \end{eqnarray}
From the second expression it follows that the $W_-$ kernel is a `genuine non-forward' object as it disappears in the forward limit.

In the limit $\xi\to 1$ the generalized parton distributions have
properties of meson distribution amplitudes \cite{Bal96,Ball98}. In this limit the WW
transformations have the form: 
\begin{eqnarray} \nonumber \lim_{\xi\to 1}
\int_{-1}^1du
W_\pm(x,u,\xi) f(u,\xi)&=&\frac{1}{2} \biggl\{ \int_{-1}^x
\frac{du}{1-u} f(u,1) \pm \int_{x}^1 \frac{du}{1+u}f(u,1)\biggr\}.
\label{meson}\end{eqnarray}

In Refs.~\cite{Rad00,Kiv01c,Rad01a} it was
demonstrated that the twist-3 skewed parton distributions in the
WW approximation exhibit discontinuities at the points
$x=\pm\xi$. Using general properties of the WW transformation (\ref{skachki}) and
Eqs.~(\ref{eq:F},\ref{eq:Ft}) one obtains:
\begin{eqnarray} \nonumber &&\lim_{\delta\to 0}
\int_{-1}^1du (W_\pm(\xi+\delta,u,\xi)-W_\pm(\xi-\delta,u,\xi)) f(u,\xi)=\frac{1}{2} {\cal P}\int_{-1}^1 \frac{f(u,\xi)}{u-\xi}du,\\&&
\lim_{\delta\to 0}\int_{-1}^1du(W_\pm(-\xi+\delta,u,\xi)-W_\pm(-\xi-\delta,u,\xi)) f(u,\xi)=\\&&
~~~~~~~~~~~~~~~~~~~~~~~~~~~~~~~ ~ ~~~~~~~~~~~~~~~~~~~~~~~~~~~~~~~~~~~\pm\frac{1}{2} {\cal P}\int_{-1}^1 \frac{f(u,\xi)}{u+\xi}du.\nonumber
\label{skachki}
\end{eqnarray}
Here ${\cal P}$ means an integral in the sense of {\it
principal value}. We see that for a very wide class of functions
$f(u,\xi)$, the discontinuity of the corresponding WW transforms
is nonzero. This feature of the WW transformation may lead to the
violation of the factorization for  the twist-3 DVCS amplitude.

Fortunately, some combinations of the distributions
${\cal F}_\mu$ and $\tilde{{\cal F}}_\mu$  are free of
discontinuities due to a certain  symmetry of the Eqs.~(\ref{eq:F}) and
(\ref{eq:Ft}).  For example:
\begin{eqnarray}
\label{comb1} {\cal F}_\mu(x,\xi)-i\varepsilon_{\perp\mu\rho}
\tilde{{\cal F}}_\rho(x,\xi)\, , \end{eqnarray}
 is free of the discontinuity at
$x=\xi$ while its `dual' combination:
\begin{eqnarray}
\label{comb2}
 {\cal F}_\mu(x,\xi)+ i\varepsilon_{\perp\mu\rho} \tilde{{\cal F}}_\rho(x,\xi)\, , \end{eqnarray}
has no discontinuity at
$x=- \xi$. The cancellation of
discontinuities in these particular combinations of the GPDs
ensures the factorization of the twist-3 DVCS amplitude on the
nucleon.

One of the non-trivial properties of the generalized parton
distributions is the polynomiality of their Mellin moments which
follows from the Lorentz invariance of nucleon matrix elements \cite{ji,ji2,Ji01}:
\begin{eqnarray}
\label{pc}
\int_{-1}^1 dx\ x^N\ H^q(x,\xi)&=&h_0^{q(N)}+h_2^{q(N)}\ \xi^2+\ldots+ h_{N+1}^{q(N)}\
\xi^{N+1}\, ,\\
\nonumber
\int_{-1}^1 dx\ x^N\ E^q(x,\xi)&=&e_0^{q(N)}+e_2^{q(N)}\ \xi^2+\ldots+ e_{N+1}^{q(N)}\
\xi^{N+1}\, .
\end{eqnarray}
The time reversal invariance requirement \cite{Man98a,Ji98b} leaves only even powers of the skewedness parameter $\xi$.
 This fact implies that the highest power
of $\xi$ is $N+1$ for odd $N$ (singlet GPDs ) and $N$ for even $N$ (nonsinglet GPDs).
Furthermore, there is a relation between  the highest power of
$\xi$ for the singlet functions $H^q$ and $E^q$ due to the fact that the nucleon has spin $1/2$ \cite{ji,ji2,Ji98b}:
\begin{eqnarray}
\label{HE}
e_{N+1}^{q(N)}=-h_{N+1}^{q(N)}\, .
\end{eqnarray}
The polynomiality conditions (\ref{pc}) strongly restrict the
class of functions of two variables
$H^q(x,\xi)$ and $E^q(x,\xi)$. In the previous section the polynomiality conditions was implemented by using 
 the double distributions
\cite{compton,npd,Mul94}.
In this case the generalized distributions
are obtained as a one-dimensional section (Eq.~\ref{evo-71}) of the two-variable
double distributions
$F(x,y)$ (Eq.~\ref{dd}) \cite{Rad01b}.

It is easy to check that the GPDs obtained by reduction from the
double distributions satisfy the polynomiality conditions
(\ref{pc}) but always lead to $h_{N+1}^{q(N)}=e_{N+1}^{q(N)}=0$, {\em i.e.}
the highest power of $\xi$ for the singlet GPDs is absent. In other words the
parametrization of the singlet GPDs in terms of double distributions is not
complete. It can be completed by adding the so-called D-term to
Eq.~(\ref{evo-71}) \cite{Pol99b}:
\begin{equation}
{\cal F}_\zeta(X)=\int^1_0dx\int_0^{1-x}\delta(x+\zeta y-X) F(x,y)dy \pm \theta
\left[1-\frac{X^2}{\zeta^2}\right]\ D\left(\frac{X}{\zeta}\right).
\label{addingDterm}
\end{equation}
Note that for both tensor $E^q(x,\xi)$ and vector $H^q(x,\xi)$ GPDs the
absolute value of the D-term is
the same and has an  opposite sign.

For the quark helicity dependent GPDs $\tilde{H}(x,\xi)$ and $\tilde{E}(x,\xi)$  the D-term is absent.

\section{Conclusion}

In this chapter, we give the predictions for the DVCS cross section for the kinematics reachable by CEBAF at Jefferson Lab. 
The DVCS amplitude was estimated using the model for the non-forward parton distributions from \cite{musatov}. 

The results for the DVCS part of the total cross section are close to those obtained in \cite{vander}.

As was pointed out, the contribution of the interference term to the differential cross section is small compared to that of the Bethe-Heitler part alone.
This gives the idea of how difficult it could be to get the information on real and imaginary parts of the DVCS amplitude from the experiment.
However, it is a well known fact that the interference term for the cross section of DVCS with the electrons of opposite 
helicity will give us an estimate of the imaginary part of the DVCS amplitude, while the asymmetry of electron-positron cross sections will  reveal the real part of
the amplitude. 

Moreover, at sufficiently large $\Theta^{\gamma\gamma}_{CM}$ angles the DVCS part of the total cross section dominates, but those angles correspond to the large values of $|t|$.
That is why it is also necessary to create a ``non-zero-$t$'' model of the non-forward distributions to better estimate the DVCS input from the generalized parton distributions.

%% file: conclu/conclu.tex
In this dissertation two aspects of deeply virtual Compton scattering were studied. 

One aspect considers the small-$x$ DVCS where the energy of the incoming
virtual photon is very large in comparison to its virtuality.
The DVCS in this region can be described by the BFKL pomeron. 
At the same time these kinematics $x \sim 10^{-2} - 10^{-4}, Q^2 \geq 6$GeV$^2$, and $-t \sim 1 - 5$GeV$^2$ could be accessed at HERA \cite{zeus}.

It is shown that at large momentum transfer the coupling of the BFKL 
pomeron to the nucleon is equal to the Dirac form factor of the nucleon. 
It allows us to calculate the DVCS cross section and to investigate $t$-dependence of the amplitude in a  model-independent way.
These approximate calculations of the DVCS cross section in the Regge regime in QCD are very timely
because all the other predictions for the small-$x$ DVCS rely on some model assumptions.

The algorithm for the calculation of the photon impact factor was elaborated. It allows also to estimate the meson production processes amplitude. 

The study showed that the $t$-dependence of the BFKL pomeron is very important and it should be possible to detect at $t<2$GeV$^2$.

Unfortunately, this approximation contains some uncertainties.
There is no evidence that the next $\sim(\alpha_s \ln x)^3$ term in the BFKL series is as small as we assume.
 Also, we are unable to fix the $\alpha_s$ in the leading logarithmic approximation.  

Thus, the first part of this study is devoted to the exploration of the deeply virtual Compton scattering process in the Regge regime.
It gives an estimation for the cross section in soon to be accessible kinematics without any model assumptions. 
On the other hand, it does not reveal any information about the structure of the targeted nucleon. 

To study the internal structure of the nucleon, we used the concept of Generalized Parton Distributions within the quantum chromodynamics framework.
Those distribution functions are accessible through deeply virtual exclusive reactions and mainly through the deeply virtual Compton scattering at
the kinematical region of high virtuality $Q^2$ and low momentum transfer $t$. 

In our study we analyzed the leading twist-2 approximation of DVCS amplitude
 and prepared the algorithm for the estimation of the next-to-leading order, 
twist-3 term of the amplitude. In our analysis we neglected the 
$\Delta_{\perp}^2$ (twist-3 approximation) 
which is natural for the parton model.

DVCS is regarded as the cleanest tool to access the underlying GPDs. Unfortunately, virtual Compton scattering is always in competition with the dominating BH process. 

In our study we created a computer program for precise calculation of the Bethe-Heitler process in the required kinematics.

To exclude the unnecessary input from BH process, the study of asymmetry with two electrons of opposite helicities was invoked. 
It was shown that the main input in this case comes from the interference DVCS-BH term, which simultaneously filters DVCS observables 
and magnifies them by the Bethe-Heitler magnitude.

Numerical algorithms for calculating the DVCS amplitude in twist-2 and twist-3 approximations were developed. 
The results for the DVCS cross section at different kinematical regions accessible on CEBAF at Jefferson Lab in twist-2 approximation were presented.
\begin{figure}[t]
\hspace{4cm}
\mbox{
\epsfxsize=6cm
\epsfysize=6cm
\epsffile{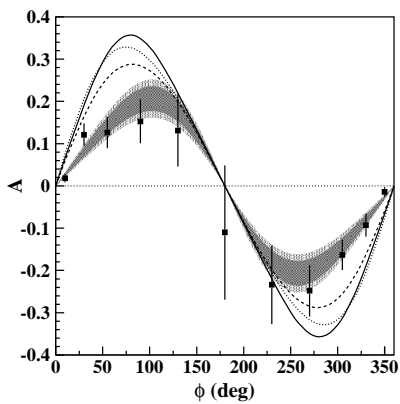}}
\caption{$\phi$ dependence of the beam spin asymmetry A \cite{stepa}.
$Q^2=1.25$(GeV/c)$^2$, $x_B=0.19$, and $-t=0.19$ (GeV/c)$^2$.} 
\label{asym1}
\end{figure}
\begin{figure}[t]
\hspace{4cm}
\mbox{
\epsfxsize=6cm
\epsfysize=6cm
\epsffile{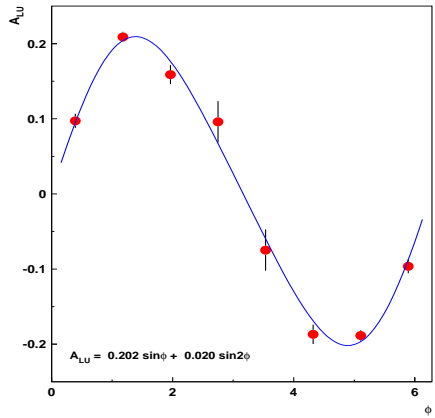}}
\caption{ The new data 
sample at 5.75 GeV \cite{smith} averaged over the entire 
kinematic range 
of the data.} 
\label{asym2}
\end{figure}
The first measurements of the beam
spin asymmetry in the DVCS
regime were performed at CEBAF at Jefferson Lab \cite{stepa,smith}. As was expected, 
the interference of the DVCS and Bethe-Heitler processes provide a 
clear asymmetry (Eq.~(\ref{fla:asym})). 
The results of our study based on models from \cite{musatov} agree in 
sign, and in magnitude to predictions made before ~\cite{Vdh99,vander} and to the experimental results represented on Figs. \ref{asym1} and \ref{asym2}.

It proves that DVCS could be a good tool to access GPDs at relatively low
energies and momentum transfers.

With the upcoming CEBAF upgrade to 12 GeV
\cite{ref:clas++} at Jefferson Lab further measurements at higher beam 
energy are expected. Construction of new detectors, increase in beam energy, 
and an improvement 
of polarization, luminosity, acceptance, and precision will broaden the 
range of the $Q^2$ and $x_B$. This will make possible a better understanding 
of 
nucleon structure. 

From the theoretical side of the study, higher order perturbative 
QCD corrections as well as new models of $t$-dependence of GPDs  are expected.

%% file: GPD/GPD-appendix.tex
 
\chapter{\label{A} Bethe-Heitler cross-section in DVCS regime}

This program was created on the request of experimentalists at Jefferson Lab.
It calculates the cross-section of the Beth-Heitler process at DVCS kinematics.
The cross-section is calculated in nb$/$GeV$^4$ with respect to the initial and final electron energy and azimuthal and out-of-the-plane angles of the final electron and photon.

This program uses the Dirac $F_1^p(t)$ and Pauli $F_2^p(t)$ form factors of the proton as a nucleon form factors.

\begin{verbatim}
c     Subroutine BH(E0,Ee,ae,ag,fe,fg,cs) calculates the cross 
c     section of the e+p->e'+p'+gamma  reaction, which is dif- 
c     ferential with respect to Q**2, Bjorken x, t and out-of- 
c     lepton-plane angle f for the process where the photon is 
c     emitted from the initial or final lepton ( Bethe-Heitler 
c     process).
c     Input parameters include:
c     E0 - initial electron energy, Gev
c     Ee - final electron energy, Gev
c     ae - azimuthal angle of final electron, rad
c     ag - azimuthal angle of emitted photon, rad
c     fe - out-of-plane electron angle, rad
c     fg - out-of-plane photon angle, rad
c     Outcoming parameter is 
c     cs - Bethe-Heitler cross section, nb/Gev**4
c     
      Program BetheHeitler
      Pi=3.141592663
      E0=5.75
      Ee=3.75
      ag=20.*Pi/180.
      fe=Pi
      fg=0.
      do 1 i=-24,24
      ae=i*Pi/180.-0.001
      call BH(E0,Ee,ae,ag,fe,fg,cs)
      print 2,ae*180/Pi,cs
 1    end do
 2    format(' ag=',F10.5,' cs=',1pE10.4)
      end
      Subroutine BH(E0,Ee,ae,ag,fe,fg,cs)
      Real M,k1q2
      Pi=3.141592663
      M=.938
      x=E0*Ee*(1.-Cos(ae))/((E0-Ee)*M)
      y=(E0-Ee)/E0
      Eg=-(-E0*(Ee*(1-Cos(ae))-M)-Ee*M)/(-E0*(1-Cos(ag))-M+
     +Ee*(1-Cos(ae)*Cos(ag)-Cos(fe-fg)*Sin(ae)*Sin(ag)))
      if(Eg.lt.0) return
      Q2=2*E0*Ee*(1-Cos(ae))
      t=2*(-E0*(Ee*(1-Cos(ae))+Eg*(1-Cos(ag)))+
     +Ee*Eg*(1-Cos(ae)*Cos(ag)-Cos(fe-fg)*Sin(ae)*Sin(ag)))
      k1q2=Eg*(1-Cos(ag))/(2*Ee*(1-Cos(ae)))
      elch=2.*Sqrt(Pi/137.)
      vph=(x*y**2)/(512.*Pi**4*Q2**2*Sqrt(1+(4*M**2*x**2)/Q2))
      bhsq=-((2*(2*M**2*(1+8*k1q2**2+t**2/Q2**2+
     +k1q2*(4+4*t/Q2))*x**2*y**2+t*(4-2*(2+x+4*k1q2*x+t*x/Q2)*y+
     +((t**2*x**2+2*Q2*t*x*(1+2*k1q2*x)+Q2**2*(2+x**2+8*
     *k1q2**2*x**2+4*k1q2*x*(1+x)))*y**2)/Q2**2))*
     *F1(t)**2)/(x**2*y**2)+4*t*(1+8*k1q2**2+t**2/Q2**2+k1q2*
     *(4+4*t/Q2))*F1(t)*F2(t)+(t*(M**2*(1+8*k1q2**2+t**2/
     /Q2**2+k1q2*(4+4*t/Q2))*x**2*y**2+t*(-2+(2+x+4*k1q2*x+
     +t*x/Q2)*y-(1+2*k1q2*x+t*x/Q2)*y**2))*F2(t)**2)/
     /(M**2*x**2*y**2))/(k1q2*t**2*(1+2*k1q2+t/Q2))
      cs=elch**6*bhsq*vph/(2.56942E-6)
      return
      end
      Function F1(t)
      f1=1/(1-t/.706)**2*(1-2.79*t/(4*.938**2))/(1-t/(4*.938**2))
      end
      Function F2(t)
      f2=1/(1-t/.706)**2*1.79/(1-t/(4*.938**2))
      end
\end{verbatim}